\documentclass[11pt]{article}%
\usepackage{amsmath}
\usepackage{amsfonts}
\usepackage{amssymb}
\usepackage{graphicx}
\usepackage{slashed}%
\providecommand{\U}[1]{\protect\rule{.1in}{.1in}}
\providecommand{\U}[1]{\protect\rule{.1in}{.1in}}

\begin{document}

\bigskip\begin{titlepage}
		\vspace{.3cm} \vspace{1cm}
		\begin{center}
			\baselineskip=16pt
			\centerline{\bf{\Large{Hearing the Shape of the Universe:}}\vspace{0.5cm}}
			\centerline{\bf{\Large{A Personal Journey in Noncommutative Geometry}}}
			\vspace{1cm}
			\centerline{\large\bf Ali H. Chamseddine }
			\vspace{.5cm}
			\emph{\centerline{Physics Department, American University of Beirut, Lebanon}}
		\end{center}
		\vspace{0.5cm}
		\centerline	{\textbf{{Dedicated to the memory of Daniel Kastler:}}\vspace{0.5cm}}
			\centerline{	{\textit{{ Enthusiastically believed in NCG}}}}
			\vspace{0.5cm}
		\begin{center}
			{\bf Abstract}
			\par\end{center}
			This article surveys the noncommutative-geometric (NCG) approach to fundamental physics, in which geometry is encoded spectrally by a generalized Dirac operator and where dynamics arise from the spectral action. I review historically how the simple idea of marrying a Riemannian manifold to a two point space, progressed to lead to the uniqueness of the Standard Model and beyond. I  explain how inner fluctuations of the Dirac operator reconstruct the full gauge--Higgs sector of the Standard Model on an almost-commutative space, fixing representations and hypercharges and naturally accommodating right-handed neutrinos and the see-saw mechanism. On the gravitational side, the heat-kernel expansion of the spectral action yields the cosmological constant, Einstein--Hilbert term, and higher-curvature corrections, with volume-quantized variants clarifying the status of $\Lambda$. I discuss the renormalization-group interpretation of the spectral action as a high-scale boundary condition, phenomenological implications for Higgs stability and neutrino masses. I present generalized Heisenberg equation leading to identify the NCG space at unification. I conclude by emphasizing that NCG provides a unified, testable, and geometrically principled quantum framework, linking matter, gauge fields, and gravity.
			
\end{titlepage}
\tableofcontents

\section*{\bigskip Introduction}

\bigskip This review summarizes my personal journey and research in the field
of noncommutative geometry (NCG) covering the thirty years period starting in
1992 and continuing until 2022. I\ will describe in detail the spectacular
progress we made through this time interval, and the crucial steps taken that
allowed us to identify uniquely the noncommutative space representing the
physical space including all the observe$\in$d particles and their fundamental
interactions. I\ will set out to recount how noncommutative geometry
\cite{C85},\cite{C94} led to concrete models in which the Higgs and the gauge
fields are geometric in origin. I\ first recall that in 1988 Alain Connes laid
down his vision of marrying noncommutative geometry and physics starting with
the simple but deep observation, that appending to a four-dimensional spin
manifold a finite noncommutative space yields, in effect, a \textquotedblleft
double-sheeted\textquotedblright\ picture in which a scalar behaving like the
Higgs sits in the off-diagonal part of a generalized connection \cite{C90}.
This mechanism, different from Kaluza--Klein since it avoids a massive tower
of states, and places the Higgs in the correct representation. This blueprint
was further developed and refined, by Connes and Lott \cite{CL91} where the
first attempt was made to develop a dictionary between the Standard Model (SM)
of Particle Physics at low energies, and the data needed to identify the
geometry of a noncommutative space. This development led to my interest in NCG
in 1992 and started a collaboration with J. Fr\"{o}hlich and G. Felder, to
examine whether it is possible to go beyond the construction of Connes and
Lott, and construct grand unified models \cite{CFF92},\cite{CFF2}. We
immediately followed this by incorporating gravity with the SM, by developing
and defining the noncommutative space that marries a Riemannian
four-dimensional spin manifold with the discrete structure identified for the
Standard Model \cite{CFF93}. My interest in finding connections between
physical models and noncommutative geometry continued with J. Frohlich until
1995 where we investigated the gravitational sector of the Connes-Lott model
as it turned out that the discrete structure is highly non-trivial . We also
provided the setup to define and develop the noncommutative Chern-Simons
actions \cite{Chamseddine:1994tw} and $SO\left(  10\right)  $ unification
\cite{Chamseddine:1993is}, and the gravitational sector of the Connes-Lott
model \cite{CFG95} . All these models shared one big disadvantage in that the
bosonic sectors could not be defined uniquely and plagued with ambiguities
related to the kernels of the Dirac operator. In addition, the fermionic
sector suffered from ad hoc rules that have to be invented to assign the
action of the bosonic gauge fields on fermions, especially in the quarks
sector, reducing the attractiveness of the model \cite{Kas93},\cite{Kas96}%
,\cite{KasS96},\cite{KasS97}. The big breakthrough came in 1996 during my
first visit to IHES where I\ started my discussions with Alain Connes when he
explained to me the importance of the spectrum of the Dirac operator in NCG At
this point we were inspired with the idea of the "Spectral action principle"
and in a matter of days, we tested the applicability of this revolutionary
principle to the Standard Model \cite{CC96},\cite{Chamseddine:1996zu}. This
work was the start of a fruitful collaboration and friendship with Alain that
continued for more than twenty five years. In this long review, I\ will
present the highlights of my research on NCG that I\ have done during the
thirty years period from 1992 until 2022. The first period from 1992 until
1995 will cover my collaboration with Fr\"{o}hlich, and the second phase
starting in 1996 will cover my collaboration with Connes and then being joined
for some of the publications by Marcolli, van Suijlekom and Mukhanov. The
purpose of this review is to show that noncommutative geometry is the most
promising approach to develop a quantum theory of gravity, being based on the
Dirac operator, and offers the only available explanation of why nature chose
the SM as the low-energy model of all fundamental interactions. I\ will also
show that although our approach that started as a bottom to top method where
we used the low-energy data to define the correct noncommutative space, led
us, after classifying all noncommutative spaces compatible with a few physical
requirements, to the uniqueness of the SM and the only possibility beyond.
This was confirmed by arriving at the same answer using as a starting point
generalized Heisenberg relations . I\ will also show that the noncommutative
approach, despite being highly mathematical, continues to produce surprises
that enhances our understandings of physics. I\ will finish this review on a
positive note, that we are not at the end of the road, and that there are more
surprises that lie hidden waiting to be discovered.

\section{Early attempts}

\bigskip In this section I\ will cover my early research, mainly with J\"{u}rg
Fr\"{o}hlich, with the main aim of identifying the noncommutative spaces
associated with Grand Unified Theories (GUT) and defining the noncommutative
generalization of Riemannian geometry to incorporate discrete spaces.

\subsection{GUT models}

The Standard Model describes particle interactions but requires the Higgs
field to break electroweak symmetry. In the usual field-theoretic setting, the
Higgs is introduced by hand. As noted earlier noncommutative geometry (NCG)
provides a natural interpretation: the Higgs fields arise as components of
generalized connections in discrete directions of space-time. In this
framework, space-time is taken to be
\begin{equation}
X=M_{4}\times F,
\end{equation}
where $M_{4}$ is a $4D$ Riemannian manifold and $F$ is a finite set of points.
Gauge fields come from continuous directions, while scalar fields arise from
the discrete part. In uur first attempt to go beyond the Connes-Lott model we
applied this construction to Grand Unified Theories (GUTs) \cite{CFF92}%
,\cite{CFF93}. A model in NCG is described by a spectral triple $(\mathcal{A}%
,\mathcal{H},D)$, with involutive algebra $\mathcal{A}$, Hilbert space
$\mathcal{H}$, and generalized Dirac operator $D$. The algebra $\mathcal{A}$
is given by%
\begin{equation}
\mathcal{A}=\mathcal{A}_{1}\oplus\mathcal{A}_{2}\oplus\mathcal{A}_{3}%
,\qquad\mathcal{A}_{i}=M_{n_{i}}(\mathbb{C}).
\end{equation}
and the Dirac operator $D$ by%
\begin{equation}
D=D_{1}\otimes1+\gamma^{5}\otimes D_{2},
\end{equation}
where $D_{1}$ is the Dirac operator on $M_{4}$, and $D_{2}$ encodes the
discrete structure:
\begin{equation}
D_{2}=%
\begin{pmatrix}
0 & M_{12} & M_{13}\\
M_{21} & 0 & M_{23}\\
M_{31} & M_{32} & 0
\end{pmatrix}
.
\end{equation}
Here the $M_{ij}$ are constant matrices representing vacuum expectation values
(VEVs) responsible for symmetry breaking. Given $a,b\in\mathcal{A}$, a 1-form
is represented as
\begin{equation}
\omega=\sum_{k}a_{k}[D,b_{k}],\qquad a_{k},b_{k}\in\mathcal{A}.
\end{equation}
This yields the generalized connection
\begin{equation}
\pi(\omega)=%
\begin{pmatrix}
A_{1} & \gamma^{5}\Phi_{12} & \gamma^{5}\Phi_{13}\\
\gamma^{5}\Phi_{21} & A_{2} & \gamma^{5}\Phi_{23}\\
\gamma^{5}\Phi_{31} & \gamma^{5}\Phi_{32} & A_{3}%
\end{pmatrix}
,
\end{equation}
where $A_{i}$ are gauge fields and $\Phi_{ij}=M_{ij}+\phi_{ij}$ are scalar
fields (Higgs multiplets). Note that the diagonal terms correspond to gauge
fields $A_{m},$ $m=1,2,3,$ and off-diagonal combinations $\Phi_{mn}=\phi
_{mn}+M_{mn}$ behave as Higgs fields. The curvature is defined as
\[
\Omega=d\omega+\omega^{2},
\]
with gauge covariance
\[
\Omega\mapsto g\,\Omega\,g^{-1}.
\]
The Yang--Mills action is obtained through the Dixmier trace:
\begin{equation}
S=\mathrm{Tr}_{\omega}(\Omega^{2}|D|^{-4})\;\;\longrightarrow\;\;S=\int
d^{4}x\,\Big(F_{\mu\nu}^{2}+|D_{\mu}\Phi|^{2}+V(\Phi)\Big). \label{Action}%
\end{equation}
As a warm up, we first consider the simple case of the SM where the finite
algebra is taken for a two-point discrete space,
\begin{equation}
\mathcal{A}=M_{2}(\mathbb{C})\oplus\mathbb{C}.
\end{equation}
We then choose
\begin{equation}
M_{12}=%
\begin{pmatrix}
0\\
\mu
\end{pmatrix}
,\qquad M_{21}=M_{12}^{\dagger}.
\end{equation}
so that
\begin{equation}
\pi(\omega)=%
\begin{pmatrix}
A_{1} & \gamma^{5}H\\
\gamma^{5}H^{\dagger} & A_{2}%
\end{pmatrix}
,
\end{equation}
with Higgs doublet $H$. The gauge fields decompose as
\begin{equation}
A_{1}=-\frac{i}{2}g\,\gamma^{\mu}A_{\mu}^{a}\sigma^{a},\qquad A_{2}%
=ig^{\prime}\gamma^{\mu}B_{\mu},
\end{equation}
yielding $W^{\pm},Z,\gamma$. Fermions are included by chirality; for leptons
\begin{equation}
L=%
\begin{pmatrix}
\nu_{L}\\
e_{L}%
\end{pmatrix}
,\qquad e_{R},
\end{equation}
with Yukawa term
\begin{equation}
\mathcal{L}_{Y}=y_{e}\,\overline{L}\,H\,e_{R}+\text{h.c.}%
\end{equation}
The action obtained is of the form (\ref{Action}) where the potential is given
by
\[
V(H)=\lambda(H^{\dagger}H-v^{2})^{2},
\]
with $\lambda$ tied to gauge couplings. The form of this potential, as well as
all other potentials obtained using this construction, suffer from an
ambiguity related to the kernel of the Dirac operator. This is one of the
major hurdles we will face in improving this construction to make it more
realistic \cite{Kas93}.

Having recovered the SM in these few steps, our next target was to consider
the $SU(5)$ GUT model. We take the discrete space to be a three-point space
with two identical copies,
\begin{equation}
\mathcal{A}=M_{5}(\mathbb{C})\oplus M_{5}(\mathbb{C})\oplus\mathbb{R}.
\end{equation}
The connection is then given by
\begin{equation}
\pi(\omega)=%
\begin{pmatrix}
A & \gamma^{5}\Sigma & \gamma^{5}H\\
\gamma^{5}\Sigma & A & \gamma^{5}H\\
\gamma^{5}H^{\dagger} & \gamma^{5}H^{\dagger} & 0
\end{pmatrix}
,
\end{equation}
with the adjoint $\Sigma$ from the link between identical copies and the
fundamental $H$ from links to the third copy. The graded tracelessness
condition
\begin{equation}
\mathrm{Tr\!}\left(  F_{1}\,\pi(\omega)\right)  =0,\qquad F_{1}=\mathrm{diag}%
(1,1,-1),
\end{equation}
implies \textrm{Tr}$A=0$ so that $U(5)\rightarrow SU(5)$. Symmetry breaking is
planted in the discrete Dirac operator via constant mass matrices (VEVs):
\begin{equation}
M_{12}=M_{21}=\Sigma_{0},\qquad M_{13}=M_{23}=H_{0}.
\end{equation}
Choose the standard $SU(5)$ VEVs
\begin{equation}
\Sigma_{0}=M\,\mathrm{diag}(2,2,2,-3,-3),\qquad H_{0}\propto(0,0,0,0,v)^{\top
},
\end{equation}
which implement the breaking
\begin{equation}
\mathrm{SU}(5)\rightarrow\mathrm{SU}(3)\times\mathrm{SU}(2)\times
\mathrm{U}(1)\rightarrow\mathrm{U}(1)_{\mathrm{em}}.
\end{equation}
One has the alignment relation
\begin{equation}
\Sigma_{0}H_{0}=-3MH_{0},
\end{equation}
which appears in the mixed terms of the scalar potential. Fermions fit into
$\overline{5}$ and $10$ representations, with Yukawa couplings
\begin{equation}
\mathcal{L}_{Y}=f_{d}\,\psi_{\overline{5}}H\psi_{10}+f_{u}\,\psi_{10}%
\,\Sigma\,H\,\psi_{10}+\text{h.c.}%
\end{equation}
The bosonic Lagrangian has the standard form
\begin{align}
\mathcal{L}_{\mathrm{bos}}  &  =-\frac{1}{4}\,\mathrm{Tr}(F_{\mu\nu}F^{\mu\nu
})+\mathrm{Tr}\!\left[  (D_{\mu}(\Sigma+\Sigma_{0}))^{\dagger}D^{\mu}%
(\Sigma+\Sigma_{0})\right] \nonumber\\
&  +(D_{\mu}(H+H_{0}))^{\dagger}D^{\mu}(H+H_{0})-V(\Sigma,H).
\end{align}
The potential originates from $\mathrm{Tr}(\mathcal{F}^{2})$ with
$\mathcal{F}=d\omega+\omega^{2}$; after eliminating auxiliaries it reduces to
a sum of positive squares. A schematic form is
\begin{align}
V(\Sigma,H)  &  =c_{1}\!\left[  \mathrm{Tr}(\Sigma+\Sigma_{0})^{2}%
-(\mathrm{Tr}(\Sigma+\Sigma_{0}))^{2}\right]  +c_{2}\,\Vert(\Sigma+\Sigma
_{0}+3M)(H+H_{0})\Vert^{2}\nonumber\\
&  +c_{3}\!\left(  (H+H_{0})^{\dagger}(H+H_{0})-v^{2}\right)  ^{2},
\end{align}
Hence $V\geq0$ and is minimized for the shifted fields $\Sigma=0$, $H=0$.

To obtain a left-right model we take the algebra with a four-point discrete
space,
\begin{equation}
\mathcal{A}=M_{2}(\mathbb{C})\oplus M_{2}(\mathbb{C})\oplus M_{2}%
(\mathbb{C})\oplus M_{2}(\mathbb{C}),
\end{equation}
with permutation symmetries $1\leftrightarrow2$ and $3\leftrightarrow4$. The
Higgs content includes, bi-doublet $\Phi\sim(2,2,0)$, triplets $\Delta_{L}%
\sim(3,1,2)$ and $\Delta_{R}\sim(1,3,2)$. The fermions are taken as
\begin{equation}
\psi_{L}=%
\begin{pmatrix}
\nu_{L}\\
e_{L}%
\end{pmatrix}
,\qquad\psi_{R}=%
\begin{pmatrix}
\nu_{R}\\
e_{R}%
\end{pmatrix}
,
\end{equation}
with Yukawa couplings
\begin{equation}
\mathcal{L}_{Y}=y\,\overline{\psi}_{L}\Phi\psi_{R}+f\,\psi_{R}^{T}%
C\,\Delta_{R}\,\psi_{R}+\text{h.c.},
\end{equation}
allowing Majorana masses for neutrinos.

We deduce that the main advantage of this noncommutative construction of GUT
is that one gets a minimal Higgs sector of (adjoint + fundamental) enforced by
three-point geometry with $1\leftrightarrow2$ symmetry, a fixed
positive-definite potential, two scales $M$ and $v$ interpreted as geometric
distances with Yukawa couplings arising uniformly. The disadvantages are
shared with the minimal $SU(5)$ construction giving rise to proton-decay with
the fermion textures not predicted.

The most realistic model is obtained \cite{Chamseddine:1993is} by taking the
algebra to be $A=A_{1}\otimes A_{2}$ with $A_{1}$ the algebra of functions on
the 4D spin manifold and $A_{2}=P_{+}\,\mathrm{Cliff}(\mathrm{SO}(10))\,P_{+}$
a chiral projection of the $SO(10)$ Clifford algebra. The novel idea here is
to apply a double projection operator to the Clifford algebra. The Hilbert
space $h$ is the image of an orthogonal projection acting on the direct sum of
spinors across the three copies. The generalized Dirac operator $D$ takes a
block form with off-diagonal entries controlled by constant matrices $M_{0}$
(even/adjoint sector) and $N_{0}$ (odd sector), together with family-mixing
matrices $K_{mn}$ that commute with $A$:
\begin{equation}
D=%
\begin{pmatrix}
\!\gamma^{\mu}\partial_{\mu} & \gamma_{5}\!\otimes\!M_{12}\!\otimes\!K_{12} &
\gamma_{5}\!\otimes\!N_{13}\!\otimes\!K_{13}\\
\gamma_{5}\!\otimes\!M_{21}\!\otimes\!K_{21} & \!\gamma^{\mu}\partial_{\mu} &
\gamma_{5}\!\otimes\!N_{23}\\
\gamma_{5}\!\otimes\!N_{31}\!\otimes\!K_{31} & \gamma_{5}\!\otimes
\!N_{32}\!\otimes\!K_{32} & \!\gamma^{\mu}\partial_{\mu}%
\end{pmatrix}
,\qquad
\end{equation}
where $M_{12}=M_{21}=M_{0},\quad M_{0}=M_{0}^{\dagger}.$ Clifford expansions
of $M_{0}$ and $N_{0}$ into $2$-form and $4$-form (for $M_{0}$) and
$1-$form$,3-$form and $5-$form (for $N_{0}$) components are used to identify
the field content.

Generalized one-forms $\rho=\sum a_{i}[D,b_{i}]$ generate, after projection, a
gauge field $A_{\mu}$ and two Higgs-type fields $M$ and $N$:
\begin{align}
A  &  =P_{+}\left(  ia+a_{IJ}\Gamma^{IJ}+ia_{IJKL}\Gamma^{IJKL}\right)
P_{+},\\
M  &  =P_{+}\left(  m+im_{IJ}\Gamma^{IJ}+m_{IJKL}\Gamma^{IJKL}\right)
P_{+},\\
N  &  =P_{+}\left(  n_{I}\Gamma^{I}+n_{IJK}\Gamma^{IJK}+n_{IJKLM}%
\Gamma^{IJKLM}\right)  P_{-}.
\end{align}
Self-adjointness implies reality of $A$ and $M$ components, while $N$ can be
complex. In the low-energy, Minkowski-signature theory, $N$ contains the
electroweak Higgs degrees of freedom that couple directly to fermions, whereas
$M$ acquires a high-scale vev and triggers the initial symmetry breaking of
$SO(10).$ Taking a $210$ vev $M_{0123}\sim M_{G}$ breaks
\begin{equation}
\mathrm{SO}(10)\ \longrightarrow\ \mathrm{SO}(4)\times\mathrm{SO}%
(6)\ \simeq\ \mathrm{SU}(4)_{c}\times\mathrm{SU}(2)_{L}\times\mathrm{SU}%
(2)_{R}.
\end{equation}
A further vev in the $45$ (aligned with a specific combination of $\Gamma$'s)
breaks $\mathrm{SU}(4)_{c}\rightarrow\mathrm{SU}(3)_{c}\times\mathrm{U}%
(1)_{c}$. In the chosen basis the projected $M_{0}$ has the schematic form
\begin{equation}
P_{+}M_{0}P_{+}=\tfrac{1}{2}(1+\kappa_{3})\left(  -M_{G}\rho_{3}+M_{1}%
(\sigma_{3}+\tau_{3}+\rho_{3}\tau_{3}\sigma_{3})\right)  ,
\end{equation}
leaving $\mathrm{SU}(3)_{c}\times\mathrm{U}(1)_{c}\times\mathrm{SU}%
(2)_{L}\times\mathrm{SU}(2)_{R}$. A component of $N_{0}$ in the $126$ acquires
a vev that breaks $\mathrm{U}(1)_{c}\times\mathrm{SU}(2)_{R}\rightarrow
\mathrm{U}(1)_{Y}$, yielding the SM group
\begin{equation}
\mathrm{SU}(3)_{c}\times\mathrm{SU}(2)_{L}\times\mathrm{U}(1)_{Y}.
\end{equation}
In the explicit basis,
\begin{equation}
Y=-\tfrac{1}{3}(\sigma_{3}+\tau_{3}+\rho_{3}\tau_{3}\sigma_{3})+\tfrac{1}%
{2}(1-\kappa_{3}\rho_{3})\eta_{3},\qquad Q=T_{L}^{3}+\tfrac{1}{2}Y.
\end{equation}
The most general $N_{0}$ preserving $\mathrm{SU}(3)_{c}\times\mathrm{U}%
(1)_{Q}$ is a linear combination of basis elements with coefficients
$(s,p,a,a^{\prime},b,b^{\prime},e,f)$ and a right-handed neutrino vev $M_{2}$
(from the $126$), which seeds the see-saw mechanism. From the projected
fermionic action, the mass terms for one generation follow. In the neutral
sector the minimal three-point model yields the canonical type-I see-saw
structure
\begin{equation}
\mathcal{M}_{\nu}^{(3\text{-pt})}=%
\begin{pmatrix}
0 & m\\[4pt]%
m & M_{2}%
\end{pmatrix}
,\qquad\Rightarrow\qquad m_{\nu_{L}}\simeq\frac{m^{2}}{M_{2}}\,,\quad
m_{\nu_{R}}\simeq M_{2}.
\end{equation}
Charged-fermion masses depend on the symmetric/antisymmetric parts of the
family matrix $K_{pq}$ and on the parameters $(s,p,a,a^{\prime},b,b^{\prime
},e,f)$. The Yang--Mills action is written in terms of the generalized
curvature $\theta=d\rho+\rho^{2}$ and the Dixmier-trace representation of the
action. Eliminating auxiliary, non-propagating fields yields an ordinary $4D$
scalar potential in terms of $M$ and $N$. In the three-point $SO(10)$ model
the gauge couplings satisfy the unification relation
\begin{equation}
g_{2}=g_{3}=g=\sqrt{\tfrac{5}{3}}\,g_{1},\qquad\Rightarrow\qquad\sin^{2}%
\theta_{W}(M_{U})=\tfrac{3}{8}.
\end{equation}
From the $N$ kinetic term one finds
\begin{equation}
m_{W}^{2}=\frac{g^{2}}{4}\,(a^{2}+3b^{2})\,,
\end{equation}
The gauge kinetic term is canonical. The Higgs kinetic terms for $M,N,H$
inherit calculable normalizations involving $|K_{12}|^{2},|K_{13}|^{2}%
,|K_{15}|^{2}$. After eliminating auxiliaries, the scalar potential $V(M,N,H)$
contains pure and mixed quartic terms and is sufficiently flexible to avoid
the over-constrained relations that plagued the three-point model, while
keeping the geometric origin of couplings. To summarize, in order to construct
a realistic NCG model, we must perform the following steps:

\begin{enumerate}
\item Choose a three-point internal space so that off-diagonal blocks of $D$
generate Higgs-like fields; project to the chiral Clifford algebra
$A_{2}=P_{+}\,\mathrm{Cliff}(\mathrm{SO}(10))\,P_{+}$.

\item Build generalized one-forms and read off gauge $A_{\mu}$ and scalar $M,$
$N$ fields. Impose equal gauge couplings on conjugate copies, constraining the
vector and 4-form parts of $A$.

\item Break $\mathrm{SO}(10)$ down to the SM through vevs $M_{0}\in
(45\oplus210)$ and $N_{0}\in(10\oplus120\oplus126)$, with explicit $Y$ and $Q$ embeddings.

\item Compute the bosonic action via the Dixmier-trace representation and
eliminate auxiliaries; in the three-point model this enforces relations that
yield aligned up/down textures and no CKM mixing.

\item Extend to six points by adding a singlet $\lambda$ and a $16_{s}$ Higgs
$H$ (vev $M_{3}$), modify the Dirac blocks, obtain a viable scalar potential,
and recover realistic flavor and a workable neutrino sector.
\end{enumerate}

This study shows that although a GUT\ noncommutative geometric construction is
possible, it is plagued with ambiguities and arbitrariness. We concluded that
more geometric structure is needed to move forward.

\subsection{Noncommutative Gravity}

Working in Connes' spectral formulation of geometry, we attempted to
reconstruct the Riemannian toolkit (metric, connection, torsion, curvature)
directly from spectral data, and define a natural Einstein--Hilbert--type
functional \cite{CFF2}. We consider the \textquotedblleft
almost-commutative\textquotedblright\ space $X=Y\times\mathbb{Z}_{2}$---two
copies (\textquotedblleft sheets\textquotedblright) of a four-dimensional spin
manifold $Y$---, and show that the generalized action reduces to
\textbf{ordinary Einstein gravity plus one massless scalar field}.
Geometrically, this scalar measures the \emph{distance between the two
sheets}; in Connes--Lott model building it is tied to the electroweak scale.

Let $A$ be a unital $\ast$-algebra with a faithful $\ast$-representation on a
$\mathbb{Z}_{2}$-graded Hilbert space $\mathcal{H}$, and let $D$ be an odd,
self-adjoint Dirac operator such that $[D,f]$ is bounded for $f\in A$. The
universal differential graded algebra $\Omega^{\bullet}(A)$ is represented by
\begin{equation}
\pi\!\left(  f_{0}\,df_{1}\cdots df_{n}\right)  =f_{0}\,[D,f_{1}]\cdots\lbrack
D,f_{n}].
\end{equation}
Quotiening by the differential ideal $\ker\pi+d(\ker\pi)$ yields the Connes
differential calculus
\begin{equation}
\Omega_{D}(A)\;=\;\Omega^{\bullet}(A)\big/\!\big(\ker\pi+d\,\ker\pi\big).
\end{equation}
The module $\Omega_{D}^{1}(A)$ plays the role of sections of the cotangent
bundle in this setting. Integration is provided by the Dixmier trace (or
heat-kernel) at spectral dimension $d$:
\begin{equation}
\int a\;:=\;\mathrm{Tr}_{\omega}\!\big(\pi(a)\,|D|^{-d}\big),\qquad a\in
\Omega_{D}^{\bullet}(A).
\end{equation}
This induces an inner product on forms, $(\alpha,\beta)=\int\alpha
\,\beta^{\ast}$, and in particular a canonical (generalized) Riemannian metric
$(\cdot,\cdot)_{0}$ on $\Omega_{D}^{1}(A)$. All constructions are canonical
once the spectral triple $(A,\mathcal{H},D)$ is given. A (left) connection
$\nabla$ on $\Omega_{D}^{1}(A)$ is \emph{unitary} if it preserves
$(\cdot,\cdot)_{0}$. With an orthonormal frame $\{E_{A}\}$ one writes $\nabla
E_{A}=-\Omega^{B}{}_{A}\otimes E_{B}$ and recovers noncommutative analogues of
Cartan's structure equations. The curvature $R^{B}{}_{A}$ and a
basis-independent scalar curvature $r(\nabla)$ are defined as in the classical
case (with products taken in the graded algebra), and a Levi--Civita
connection is torsion-free and unitary.

Using the canonical metric, the Einstein--Hilbert analogue is
\begin{equation}
I(\nabla)\;=\;\int r(\nabla)\,,
\end{equation}
the noncommutative avatar of $\int\sqrt{g}\,R$. Concretely we consider the
geometry of a two sheeted space $X=Y\times\mathbb{Z}_{2},$ and $A=C^{\infty
}(Y)\oplus C^{\infty}(Y)$ acts diagonally on spinors on each sheet, and $D$ is
a $2\times2$ Dirac operator with standard first-order symbols on the diagonal
and an \emph{off-diagonal} mixing. The classical vierbein $e^{a}{}_{\mu}$
appears on both sheets, while the off-diagonal entry contains a scalar
function $\phi(x)$. The $4D$ metric $g_{\mu\nu}$ on $Y$ is read from the
principal symbol of $D$; $\phi$ controls the separation along the discrete
direction. Representing one-forms produces two classical $1$-forms (one per
sheet) and scalar pieces associated with the discrete direction. A key
subtlety is that $\pi(d\rho)$ can be nonzero even when $\pi(\rho)=0$, which is
why one must quotient by $\ker\pi+d\,\ker\pi$. Dynamically, this corresponds
to introducing non-propagating auxiliary fields and eliminating them by their
equations of motion---exactly as in almost-commutative Yang--Mills models.
With an orthonormal basis combining the frame on $Y$ and the discrete
direction, one writes torsion and curvature components in terms of the $4D$
spin connection, $\phi$ (and $\psi$ if kept), and auxiliaries. The gravity
action is the noncommutative Einstein--Hilbert integral over $Y$; we evaluate
it on the torsion-free (Levi--Civita) branch. For general off-diagonal
structure, the torsion-less constraints tend to decouple the sheets. A
physically interesting case arises when the off-diagonal entry is
\emph{purely} $\phi(x)$ (i.e.\ $\psi=0$), the same simplification used in
Connes--Lott particle-physics models. On this branch, the mixed components of
the connection are fixed in terms of $\phi$, and the action simplifies
substantially. After eliminating auxiliaries on the torsion-free branch with
$\psi=0$, the generalized Einstein--Hilbert functional reduces to ordinary
gravity coupled to a massless scalar:
\begin{equation}
S_{\text{grav}}\;=\;\int_{Y}d^{4}x\,\sqrt{g}\,\Big(R-2\,\partial_{\mu}%
\sigma\,\partial^{\mu}\sigma\Big),\qquad\phi=e^{-\sigma}.
\end{equation}
Thus the two-sheet geometry is classically equivalent to $4D$ gravity plus a
single scalar whose vacuum expectation value sets the \emph{geometric
distance} between the sheets. In the Connes--Lott picture, $\langle\phi
\rangle$ is tied to the electroweak scale, so the coupling of this scalar to
Standard-Model fields is geometrically fixed. To summarize, to formulate a
noncommutative analogue of Riemannian geometry for a two sheeted space we take
the following steps

\begin{itemize}
\item \textbf{Canonical metric from spectral data.} The spectral triple
$(A,\mathcal{H},D)$ furnishes both an integration map (Dixmier trace / heat
kernel) and a canonical inner product on $\Omega^{1}_{D}(A)$, enabling a
direct noncommutative analogue of Riemannian geometry.

\item \textbf{Structure equations are robust.} With an orthonormal frame,
Cartan's equations, curvature, and scalar curvature carry over with natural
transformation laws under changes of frame by matrices in $U_{N}(A)$.

\item \textbf{Auxiliary fields are intrinsic.} Because of the quotient by
$\ker\pi+d\,\ker\pi$, non-propagating auxiliary fields naturally appear and
must be eliminated; this is the same mechanism as in the noncommutative
Yang--Mills sector.

\item \textbf{Two points $\Rightarrow$ a dilaton-like scalar.} On
$Y\times\mathbb{Z}_{2}$ the action becomes $R$ plus one massless scalar (no
classical potential). Interpreting $\phi=e^{-\sigma}$ as the inter-sheet
distance unifies its presence with the geometry; quantum effects can generate
a potential.
\end{itemize}

In this construction, we achieved:

\begin{enumerate}
\item \textbf{Foundations:} A self-contained noncommutative analogue of
metric, Levi--Civita connection, torsion, curvature, scalar curvature, and an
Einstein--Hilbert functional, all \emph{canonically attached} to spectral data.

\item \textbf{Worked geometry:} A complete calculation for $Y\times
\mathbb{Z}_{2}$, including the identification of the scalar degree of freedom
and its coupling to gravity.
\end{enumerate}

\textbf{Physics insight:} In almost-commutative geometries, gravity
generically couples to scalar moduli of the discrete sector; for two points,
that modulus is one scalar controlling an internal distance, suggesting a
geometric origin for physical scales.

\section{The spectral action principle}

The first breakthrough occurred in 1995 when Alain Connes introduced the
reality structure in noncommutative geometry \cite{C95}. A year later in 1996
the second breakthrough occurred, in my first collaboration with Alain, where
we put forward the spectral action principle \cite{CC96}%
,\cite{Chamseddine:1996zu}. I\ will now summarize the basic steps we took to
achieve the remarkable result of marrying gravity to the SM.

\subsection{ From classical geometry and SM to a spectral framework}

Recalling the classical ingredients of Riemannian geometry: a manifold $M$
with local coordinates and the infinitesimal line element $ds^{2}=g_{\mu\nu
}\,dx^{\mu}dx^{\nu}$ where $g_{\mu\nu}$ is the metric. Physics at
\textquotedblleft reasonably low energies\textquotedblright\ is encoded by the
sum of the Einstein action $I_{E}$ and the Standard Model action $I_{SM}$
(Euclidean signature). The $I_{E}$ depends only on the 4-geometry while
$I_{SM}$ decomposes into gauge, gauge--Higgs, pure Higgs, gauge--fermion and
Higgs--fermion pieces and introduces, respectively, vector bosons
$(\gamma,W^{\pm},Z,\text{gluons})$, the scalar Higgs $H$, and spin-$\tfrac
{1}{2}$ fermions (quarks and leptons). The natural invariance group of the
full action is a \emph{semi-direct product}
\begin{equation}
G=U\rtimes\mathrm{Diff}(M),\qquad U=C^{\infty}\!\big(M,\,U(1)\times
SU(2)\times SU(3)\big),
\end{equation}
where $U$ are local gauge transformations acting over diffeomorphisms.

Next, we introduce the spectral triple encoding geometry given by $(A,H,D).$
Noncommutative geometry replaces \textquotedblleft points +
metric\textquotedblright\ by a \emph{spectral triple} $(A,H,D)$: an involutive
algebra $A$ of operators on a Hilbert space $H$ and a self-adjoint (unbounded)
Dirac operator $D$. In the commutative case, $A=C^{\infty}(M)$, $H=L^{2}%
(M,S)$, and $D$ is the Levi--Civita Dirac operator; the inverse $D^{-1}$ plays
the role of $ds$ (the line element). No information is lost: one recovers
points as characters of $A$ and the \emph{geodesic distance} from
Connes\ formula \cite{C94},\cite{C95},\cite{C06},\cite{C00},
\[
d(x,y)=\sup\big\{|a(x)-a(y)|:\,a\in A,\ \Vert\lbrack D,a]\Vert\leq1\big\}.
\]
The spinorial data are constrained by axioms that include a $\mathbb{Z}/2$
grading $\gamma$ and a real structure $J$ whose commutation signs
$(\varepsilon,\varepsilon^{\prime},\varepsilon^{\prime\prime})$ depend only on
the KO-dimension ($n\bmod8$). (to be explained in later). A first conceptual
\textquotedblleft turn\textquotedblright\ then appears: if one \emph{forgets
the algebra $A$} but keeps $(D,\gamma,J)$, the remaining data can be
characterized by the \emph{spectrum} $\Sigma$ of $D$ (discrete with
multiplicities, symmetric in the even case). This suggests the guiding
\emph{spectral hypothesis}: \textbf{the physical action only depends upon
$\Sigma$}. This is stronger than standard diffeomorphism invariance (because
isospectral non-isometric manifolds exist). We next note that in the
commutative case we identify diffeomorphisms on $M,$ denoted by $\mathrm{Diff}%
(M),$ with automorphisms of the algebra, denoted by $\mathrm{Aut}(A),$ via
pullback $f\mapsto f\circ\phi^{-1}$. In the general (possibly noncommutative)
case, $\mathrm{Aut}(A)$ plays the role of diffeomorphisms and contains a
\emph{normal subgroup of inner automorphisms}
\begin{equation}
\mathrm{Int}(A)=\{a\mapsto uau^{\*}\}.
\end{equation}
There is then a short exact sequence $1\rightarrow\mathrm{Int}(A)\rightarrow
\mathrm{Aut}(A)\rightarrow\mathrm{Out}(A)\rightarrow1$, mirroring the earlier
symmetry sequence $1\rightarrow U\rightarrow G\rightarrow\mathrm{Diff}%
(M)\rightarrow1$. Comparing the two, and accounting for how inner
automorphisms act on $H$, singles out the algebra
\begin{equation}
A\;=\;C^{\infty}(M)\otimes A_{F},\qquad A_{F}=\mathbb{C}\ \oplus
\ \mathbb{H}\ \oplus\ M_{3}(\mathbb{C}),
\end{equation}
with $\mathbb{H}\subset M_{2}(\mathbb{C})$ the quaternions. This is the
celebrated almost-commutative algebra used to model the Standard Model
internal symmetries. To build the product geometry $M\times F$ and encoding
fermions we take $A=C^{\infty}(M)\otimes A_{F}$, a \emph{product spectral
triple} fixed by
\begin{equation}
H=L^{2}(M,S)\otimes H_{F},\qquad D=\!D_{M}\otimes1\;+\;\gamma_{5}\otimes
D_{F}.
\end{equation}
Here the finite (zero-dimensional) spectral triple $(A_{F},H_{F},D_{F})$
captures one SM generation: $H_{F}$ is spanned by leptons and quarks (with
colors), the grading gives $+1$ to left-handed and $-1$ to right-handed
states, and the real structure $J$ implements charge conjugation . The finite
Dirac operator has the block form $D_{F}=\mathrm{diag}(Y,\bar{Y})$ where $Y$
is the Yukawa matrix; the \textquotedblleft order-one\textquotedblright\ axiom%
\begin{equation}
\lbrack\lbrack D,a],b^{0}]=0,\qquad b^{0}=Jb^{\*}J^{-1}%
\end{equation}
holds and will constrain the allowed field contents. Inner fluctuations
generate gauge and Higgs fields, and the fermionic action. In noncommutative
geometry, metrics related by \emph{inner fluctuations }\cite{Connes:2006qj}
\begin{equation}
D\;=\;D_{0}\;+\;A\;+\;JAJ^{-1},\qquad A=\sum_{i}a_{i}[D_{0},b_{i}],\;A=A^{\*},
\end{equation}
lie in the same \textquotedblleft gauge\textquotedblright\ class. For the
product geometry $M\times F$, performing these fluctuations produces exactly
the Standard Model bosons---$\gamma,W^{\pm},Z$, eight gluons---and a complex
Higgs doublet $H$. Moreover, the SM fermionic action (gauge and Yukawa
couplings) compactly equals $\langle\psi,D\psi\rangle$; i.e., the last two
terms of the SM Lagrangian are generated by the Dirac operator of the
fluctuated geometry. A subtle geometric point follows: the \emph{metric is not
the product metric} after fluctuating $D_{0}$. The \textquotedblleft initial
scale\textquotedblright\ set by $D_{F}$ disappears under arbitrary internal
fluctuations, so to reproduce the remaining bosonic terms of $I_{SM}$ and the
pure gravitational sector in a geometric way, we invoke the spectral
hypothesis. The spectral action as a geometric \textquotedblleft
bare\textquotedblright\ action. Our central claim is: for any smooth positive
function $\chi$
\begin{equation}
\mathrm{Tr}\,\chi\!\left(  \frac{D}{\Lambda}\right)  \;=\;I_{E}\;+\;I_{G}%
\;+\;I_{GH}\;+\;I_{H}\;+\;I_{C}\;+\;O(\Lambda^{-\infty}),
\end{equation}
i.e., the trace of a spectral function of $D$ reproduces (in a large-$\Lambda$
expansion) Einstein gravity, gauge, gauge--Higgs, Higgs sectors, plus a
curvature-squared (Weyl) term and a term $\int RH^{2}\sqrt{g}\,d^{4}x$ (the
precise decomposition is worked out in later sections). Since the left-hand
side depends only on the spectrum of $D$, it satisfies the spectral hypothesis
and is proposed as the \emph{bare bosonic action} at a UV scale $\Lambda$. We
introduce a spectral cutoff consistent with symmetry: replace $H$ by
\begin{equation}
H_{\Lambda}\;=\;\mathrm{Range}\,\chi\!\left(  \frac{D}{\Lambda}\right)  ,
\end{equation}
and restrict $D$ and the algebra representation to $H_{\Lambda}$. Unlike a
lattice, this preserves the continuous automorphism symmetries of finite
noncommutative algebras. The working hypothesis is that there exists a scale
$\Lambda\sim10^{15}\!-\!10^{19}\,\mathrm{GeV}$ where the bare action indeed
takes the geometric spectral form
\begin{equation}
\,\mathrm{Tr}\,\chi(D/\Lambda)\;+\;\langle\psi,D\psi\rangle\,.
\end{equation}
This is the precise statement of the spectral action principle. The bosonic
action is written spectrally as $\mathrm{Tr}\,\chi(P)$ for a positive test
function $\chi$ and an elliptic operator $P$ built from the Dirac operator
(later, $P=D_{A}^{2}/\Lambda^{2}$). The \emph{positivity} of $\chi$ is
emphasized because it implies positivity properties for the gravity sector
that appears from the expansion. The spectral action principles has the
following advantages:

\begin{itemize}
\item It recasts the SM+gravity within the language of spectral triples,
showing how diffeomorphisms and gauge symmetries unify as automorphisms of an
almost-commutative algebra $C^{\infty}(M)\otimes(\mathbb{C}\oplus
\mathbb{H}\oplus M_{3}(\mathbb{C}))$.

\item It identifies inner metric fluctuations as the geometric origin of all
SM bosons including the Higgs, and compresses the entire fermionic sector into
$\langle\psi,D\psi\rangle$.

\item It states the spectral action principle: the bosonic part is the trace
of a spectral function of the Dirac operator, depending only on $\mathrm{Spec}%
(D)$; together with the fermionic term, this gives a unified, purely spectral
\textquotedblleft bare action\textquotedblright\ at a high scale $\Lambda$.
\end{itemize}

\subsection{The heat-kernel/Mellin machinery.}

Using Mellin transform identities and the small-$t$ heat-kernel expansion,
\begin{equation}
\mathrm{Tr}\,e^{-tP}\sim\sum_{n\geq0}t^{\frac{n-m}{d}}\!\int_{M}%
a_{n}(x,P)\,dv(x)\quad(t\downarrow0),
\end{equation}
with $m=\dim M$ and $d=\mathrm{ord}(P)$, one derives the \emph{asymptotic
expansion}
\begin{equation}
\mathrm{Tr}\,\chi(P)\;\sim\;\sum_{n\geq0}f_{n}\,a_{n}(P).
\end{equation}
Here the moments of $\chi$ are $f_{0}=\int_{0}^{\infty}u\,\chi(u)\,du,\quad
f_{2}=\int_{0}^{\infty}\chi(u)\,du,\quad f_{2(n+1)}=(-1)^{n}\chi^{(n)}(0)$ for
$n\geq0$. In four dimensions ($m=4$) and for second-order $P$ ($d=2$), these
control the $\Lambda^{4},\Lambda^{2},\Lambda^{0},\ldots$ terms of the spectral
action. Also, the \emph{odd} coefficients $a_{n}$ vanish (for manifolds
without boundary).

We list he first three Seeley--DeWitt densities, quoted explicitly, from
Gilkey:
\begin{align}
a_{0}(x,P)  &  =(4\pi)^{-m/2}\,\mathrm{Tr}\,\mathbf{1},\\
a_{2}(x,P)  &  =(4\pi)^{-m/2}\,\mathrm{Tr}\!\left(  -\tfrac{R}{6}%
\mathbf{1}+E\right)  ,\\
a_{4}(x,P)  &  =(4\pi)^{-m/2}\tfrac{1}{360}\,\mathrm{Tr}\!\Big[(-12\nabla
^{2}R+5R^{2}-2R_{\mu\nu}R^{\mu\nu}+2R_{\mu\nu\rho\sigma}R^{\mu\nu\rho\sigma
})\mathbf{1}\nonumber\\
&  \qquad\qquad\qquad\qquad\quad-60\,R\,E+180\,E^{2}+60\,\nabla^{2}%
E+30\,\Omega_{\mu\nu}\Omega^{\mu\nu}\Big].
\end{align}
The geometric endomorphism $E$ and curvature $\Omega_{\mu\nu}$ are built from
the connection data of $P$ where $P=D^{2}$ corresponds to a fluctuated Dirac operator.

\begin{itemize}
\item We then rewrite $P$ as the square of the fluctuated Dirac operator as
\begin{equation}
P=D_{A}^{2}=-\big(g^{\mu\nu}\nabla_{\mu}\nabla_{\nu}+E\big),
\end{equation}
to \emph{identify $E$ and $\Omega_{\mu\nu}$} explicitly. In the product
geometry $C^{\infty}(M)\otimes M_{N}(\mathbb{C})$ with gauge field strength
$F_{\mu\nu}=F_{\mu\nu}^{i}T_{i}$ and coupling $g$,
\begin{equation}
E=\tfrac{1}{4}R\otimes\mathbf{1}-\tfrac{i}{4}\gamma^{\mu\nu}\otimes
g\,F_{\mu\nu}^{i}T_{i},\qquad\Omega_{\mu\nu}=\tfrac{1}{4}R_{ab\,\mu\nu}%
\gamma^{ab}\otimes\mathbf{1}-\tfrac{i}{2}\,\mathbf{1}\otimes g\,F_{\mu\nu}%
^{i}T_{i},
\end{equation}
which already shows the \emph{diffeomorphism} and \emph{gauge} invariance of
the invariants $a_{n}$.The coefficients $a_{0},a_{2},a_{4}$ in $d=4$ for
$C^{\infty}(M)\otimes M_{N}(\mathbb{C}),$ are then evaluated as the integrated
coefficients:
\begin{align}
a_{0}(P)  &  =\frac{N}{4\pi^{2}}\int_{M}\!\sqrt{g}\,d^{4}x,\\
a_{2}(P)  &  =\frac{N}{48\pi^{2}}\int_{M}\!\sqrt{g}\,R\,d^{4}x,\\
a_{4}(P)  &  =\frac{N}{16\pi^{2}}\,\frac{1}{360}\int_{M}\!\sqrt{g}%
\,\Big[\,(12\nabla^{2}R+5R^{2}-8R_{\mu\nu}R^{\mu\nu}-7R_{\mu\nu\rho\sigma
}R^{\mu\nu\rho\sigma})\nonumber\\
&  \hspace{6.25em}+120\,N\,g^{2}\,F_{\mu\nu}^{i}F_{i}^{\mu\nu}\Big]\,d^{4}x.
\end{align}
Note that the gauge term appears with a fixed normalization from the trace
over the internal space.

\item There is a four-dimensional identity relating the three
curvature-squared combinations to the Gauss--Bonnet density, allowing the
$a_{4}$ curvature sector to be recast into the familiar \emph{Weyl-squared}
and \emph{Gauss--Bonnet invariant}, and total-derivative pieces. Combining the
heat-kernel expansion with the test-function moments gives the large-$\Lambda$
asymptotics of the bosonic spectral action:
\begin{equation}
\mathrm{Tr}\,\chi\!\left(  \frac{D_{A}}{\Lambda}\right)  \;\sim\;f_{4}%
\,\Lambda^{4}\,a_{0}(P)\;+\;f_{2}\,\Lambda^{2}\,a_{2}(P)\;+\;f_{0}%
\,a_{4}(P)\;+\;\cdots,
\end{equation}
i.e.\ a \emph{cosmological term} ($\propto\Lambda^{4}$),
\emph{Einstein--Hilbert} term ($\propto\Lambda^{2}R$), and at order
$\Lambda^{0}$ the curvature-squared and \emph{gauge kinetic} terms, with fixed
\emph{relative} coefficients dictated by the geometry of $D_{A}$.

\item This expansion, thus provides the \emph{general engine}---the
Seeley--DeWitt expansion---that turns the abstract spectral principle into a
concrete action. It shows that for the simplest almost-commutative case
$C^{\infty}(M)\otimes M_{N}(\mathbb{C})$, the spectral action automatically
yields the expected \emph{gravity $+$ Yang--Mills} structure, including the
right gauge-kinetic form $\int\mathrm{Tr}\,F_{\mu\nu}F^{\mu\nu}$ with a fixed
normalization coming from traces over the finite (matrix) factor.

\item By writing $E$ and $\Omega_{\mu\nu}$ explicitly in terms of curvature
and $F_{\mu\nu}$, it is clear that the spectral action is simultaneously
\emph{diffeomorphism-} and \emph{gauge-invariant}---a nontrivial and essential
check before moving on to the full Standard Model in later sections.
\end{itemize}

\subsection{Setup of the almost-commutative geometry for the SM.}

We now proceed from the warm-up Einstein--Yang--Mills case to the
\emph{realistic} Standard Model on a product geometry $M\times F$ with
\begin{equation}
A=A_{1}\otimes A_{2},\quad H=H_{1}\otimes H_{2},\quad D=D_{1}\otimes
1+\gamma_{5}\otimes D_{2},
\end{equation}
where $A_{1}=C^{\infty}(M)$, $H_{1}=L^{2}(M,S)$, $D_{1}=\not \!  \!D_{M}$, and
the finite part is $A_{2}=\mathbb{C}\oplus\mathbb{H}\oplus M_{3}(\mathbb{C})$.
The finite Hilbert space $H_{2}$ is spanned by leptons and quarks (plus
conjugates), e.g.\ $Q=(u_{L},d_{L},d_{R},u_{R})^{\mathrm{T}}$, $L=(\nu
_{L},e_{L},e_{R})^{\mathrm{T}}$.

\begin{itemize}
\item The algebra acts diagonally in a way that distinguishes left/right and
color: for $a=(\lambda,q,m)\in\mathbb{C}\oplus\mathbb{H}\oplus M_{3}%
(\mathbb{C})$,
\begin{equation}
a\cdot Q=%
\begin{pmatrix}
q\,(u_{L},d_{L})^{\mathrm{T}}\\[2pt]%
\bar{\lambda}\,d_{R}\\[2pt]%
\lambda\,u_{R}%
\end{pmatrix}
,
\end{equation}
with a similar expression for leptons; conjugate particles transform by
$a\cdot\bar{L}=\lambda\,\bar{L}$, $a\cdot\bar{Q}=m\,\bar{Q}$.

\item The finite Dirac operator is block-diagonal $D_{2}=%
\begin{bmatrix}
Y & 0\\[2pt]%
0 & \bar{Y}%
\end{bmatrix}
$ with $Y=Y_{q}\otimes\mathbf{1}_{3}\ \oplus\ Y_{\ell},$ where,
\begin{equation}
\qquad Y_{q}=%
\begin{bmatrix}
0 & k_{d}\otimes H_{0}\\
0 & k_{u}\otimes\widetilde{H}_{0}\\
(k_{d}^{\*},0)\otimes H_{0}^{\*} & 0\\
(k_{u}^{\*},0)\otimes\widetilde{H}_{0}^{\*} & 0
\end{bmatrix}
,\qquad Y_{\ell}=%
\begin{bmatrix}
0 & k_{e}\otimes H_{0}\\
k_{e}^{\*}\otimes H_{0}^{\*} & 0
\end{bmatrix}
.
\end{equation}
Here $k_{d},k_{u},k_{e}$ are $3\times3$ family-mixing matrices; $H_{0}%
=\mu\binom{0}{1}$, $\widetilde{H}_{0}=i\sigma_{2}H_{0}^{\*}$ with mass scale
$\mu$.

\item These data obey the reality/grading and \emph{order-one} constraints:
$J^{2}=1$, $[J,D_{2}]=0$, $[a,Jb^{\*}J^{-1}]=0$, and $[[D,a],Jb^{\*}J^{-1}]=0$
for all $a,b$.
\end{itemize}

\subsection{ Inner fluctuations $\Rightarrow$ gauge bosons $+$ Higgs}

\begin{itemize}
\item Taking \emph{inner metric fluctuations} $D\mapsto D+A+JAJ^{-1}$ with
$A=\sum a_{i}[D,b_{i}]$ generates exactly the SM bosons: $U(1)$, $SU(2)_{w}$,
$U(3)$ gauge fields and a complex Higgs doublet. The combination $A+JAJ^{-1}$
removes one abelian trace, leaving the physical $U(1)_{Y}$, $SU(2)_{w}$,
$SU(3)_{c}$.

\item The \emph{quark} Dirac operator $D_{q}$ (a $36\times36$ matrix in
family/color, tensored with Clifford matrices) contains the spin connection,
hypercharge $B_{\mu}$, weak $A_{\mu}^{\alpha}$ and gluon $V_{\mu}^{i}$
couplings, and the Yukawa blocks with the dynamical Higgs $H$ and its $SU(2)$
conjugate $\widetilde{H}=i\sigma_{2}H^{\*}$. The same structure appears for
\emph{leptons} $D_{\ell}$. These yield the very compact form for the fermionic
actions $(Q,D_{q}Q)$ and $(L,D_{\ell}L)$.
\end{itemize}

This now paves the way to obtain in a straightforward manner, the spectral
action for the SM and its bosonic sector. The \emph{universal} spectral action
specialized to the SM is
\begin{equation}
\mathrm{Tr}\,\chi(D^{2}/m_{0}^{2})+(\psi,D\psi),
\end{equation}
with the fermionic term equal to the quark+lepton actions above; the
\emph{bosonic} piece is computed by the same heat-kernel machinery as outlined above.

\begin{itemize}
\item From the gauge-kinetic terms one extracts \emph{boundary conditions at
the high scale} $\Lambda$ (unification-type relations among gauge couplings,
numerically the same pattern as $SU(5)$):
\[
\alpha_{3}(\Lambda)=\alpha_{2}(\Lambda)=\tfrac{5}{3}\,\alpha_{1}(\Lambda).
\]
Renormalization-group running down to $M_{Z}$ then predicts
\begin{align}
\sin^{2}\theta_{W}  &  =\frac{3}{8}\Bigl[1-\frac{109}{18\pi}\,\alpha
_{\mathrm{em}}\ln(\Lambda/M_{Z})\Bigr]\quad\\
\quad\ln(\Lambda/M_{Z})  &  =\frac{2\pi}{67}\bigl(3\alpha_{\mathrm{em}}%
^{-1}(M_{Z})-8\alpha_{3}^{-1}(M_{Z})\bigr).
\end{align}

\item Plugging experimental inputs $(\alpha_{\mathrm{em}}^{-1}(M_{Z}%
)=128.09,\ 0.110\leq\alpha_{3}(M_{Z})\leq0.123)$ gives a \emph{cutoff scale}
roughly $\Lambda\sim10^{15}\,\mathrm{GeV}$ and $\sin^{2}\theta_{W}\simeq
0.206$--$0.210$---about a \emph{10\% low} vs the precise experimental value
$\approx0.2325$. We read this as evidence that new degrees of freedom must be
present to alter the spectrum before the Planck scale.

\item The fixed \emph{Higgs quartic boundary condition} at $\Lambda$ in terms
of gauge/Yukawa data, in a top-dominance limit, simplifies to $\lambda
(\Lambda)=\tfrac{16\pi}{3}\,\alpha_{3}(\Lambda)$.
\end{itemize}

The spectral action principle, applied to the SM thus achieves its remarkable
unification with gravity with the properties:

\begin{itemize}
\item It \emph{builds the SM} from the almost-commutative spectral triple
$C^{\infty}(M)\otimes(\mathbb{C}\oplus\mathbb{H}\oplus M_{3}(\mathbb{C}))$:
the inner fluctuation of $D$ yields precisely the gauge bosons and one complex
Higgs doublet; the fermionic action is $(\psi,D\psi)$.

\item It \emph{derives} the full bosonic action from $\mathrm{Tr}\,\chi
(D^{2}/m_{0}^{2})$ with fixed relative normalizations, implying high-scale
relations among couplings and a predictive \emph{RG link} to low-energy
observables (notably $\sin^{2}\theta_{W}$ and a Higgs-quartic boundary value).

\item Phenomenologically, the framework is \emph{close} but not perfect at one
loop: $\Lambda\sim10^{15}$ GeV is consistent with gauge unification but
somewhat below the gravity-suggested scale, and $\sin^{2}\theta_{W}$ comes out
$\sim10\%$ low---hinting at extra spectrum or thresholds between $M_{Z}$ and
$\Lambda$. From the Einstein term one expects $\kappa_{0}^{-2}\propto
f_{2}\Lambda^{2}$, suggestive of $\Lambda$ near $M_{\mathrm{Pl}}$, while gauge
unification favors $\Lambda\sim10^{15}$ GeV. This tension hints at missing
states or a breakdown of continuum space-time before $M_{\mathrm{Pl}}$,
consistent with the Wilsonian view that the spectral action is an effective
bare action up to $\Lambda$.

\item \textbf{Unification without ad hoc group choice.} The SM gauge group and
a single Higgs doublet arise from the internal finite geometry; inner
fluctuations of $D$ generate both gauge and Higgs sectors.

\item \textbf{Gravity alongside gauge.} The $a_{0},a_{2},a_{4}$ terms deliver
Einstein--Hilbert, a positive cosmological constant, conformal gravity $C^{2}%
$, and the Euler density with fixed coefficients; no bare $R^{2}$ at leading order.

\item \textbf{Predictivity vs.\ thresholds.} Quantitative outputs depend on
assuming the same spectrum up to $\Lambda$ and on one-loop running; extra
fields between $M_{Z}$ and $\Lambda$ shift predictions.

\item \textbf{Cutoff function robustness.} Changing $\chi$ rescales
$f_{4},f_{2},f_{0}$ but preserves the \emph{relative} coefficients inside
$a_{4}$ (e.g.\ gauge normalizations, $\xi_{0}=1/6$), which are geometric.
\end{itemize}

\section{Scale invariance and the dilaton in the spectral action}

The spectral action principle says the bosonic (and, with a standard quadratic
form, fermionic) action of a physical theory is determined by the spectrum of
a suitable Dirac operator $D$ associated to a (possibly noncommutative)
geometry. As explained in the last section, the action was taken as
$\mathrm{Tr}\,F(D^{2}/m^{2})+(\psi,D\psi)$, where $m$ is an arbitrary high
mass scale and $F$ is a positive test function. We now ask the question:
\emph{what happens to the spectral action if the \textquotedblleft arbitrary
mass scale\textquotedblright\ is made dynamical}---i.e., if classical scale
invariance is restored by promoting $m$ to a field (\textquotedblleft
dilaton\textquotedblright)? We will show \cite{Chamseddine:2005zk} that doing
so yields a \textbf{scale-invariant matter sector} (gauge $+$ Higgs $+$
Yukawas) \textbf{in the Einstein frame}, with explicit and unique
\textbf{dilaton couplings} fixed by geometry---\textbf{no fine-tuning} is
required. The only classical violations of scale invariance that remain are
the \textbf{Einstein--Hilbert term} and the \textbf{dilaton kinetic term}.
They further connect the result to scale-invariant SM constructions,
extended-inflation models, and the low-energy limit of the
\textbf{Randall--Sundrum} scenario.

\medskip Classically, the Standard Model (SM) is \textquotedblleft
almost\textquotedblright\ scale invariant; explicit breaking enters through
the Higgs mass term. A compensating dilaton can restore scale invariance and
is familiar from Jordan--Brans--Dicke gravity, string low-energy limits, and
the radion in warped extra dimensions. The spectral action (which geometrizes
the SM $+$ gravity) originally had a fixed $m$. The\ idea now is to
\textbf{replace the rigid ratio $D^{2}/m^{2}$ by an operator involving the
dilaton}, then recompute the heat-kernel asymptotics to obtain the low-energy
action and its couplings.

To do so, we promote the scale to a field by writing $\Lambda=me^{\phi}$, with
$\phi$ dimensionless (or $\phi=\varphi/f$ for a canonically-normalized scalar
$\varphi$, decay constant $f$). Since the Dirac operator has mass dimension 1,
we \textbf{replace}
\begin{equation}
\mathrm{Tr}\,F\!\big(D^{2}/m^{2}\big)\quad\longrightarrow\quad\mathrm{Tr}%
\,F(P),\qquad P\equiv e^{-\phi}D^{2}e^{-\phi}.
\end{equation}
This choice makes the counting of eigenvalues depend on the local scale
$e^{\phi}$ and is also natural for non-compact cases (where localization
enters). A constant shift of $\phi$ simply rescales $m$, so one can set $m=1$
without loss of generality and reinsert it by replacing $\phi\mapsto\phi+\ln
m$. We then have to evaluate the \textbf{low-energy (large-$\Lambda$)} terms
in $\mathrm{Tr}\,F(P)$ via the standard Seeley--DeWitt expansion, determine
the \textbf{dilaton couplings}, and then \textbf{specialize to the Standard
Model} noncommutative space. For a Laplace-type operator in $4D$,
\begin{align}
\mathrm{Tr}\,F(P)  &  \sim f_{0}\,a_{0}(P)+f_{2}\,a_{2}(P)+f_{4}%
\,a_{4}(P)+\cdots,\qquad\\
f_{0}  &  =\!\int_{0}^{\infty}uF(u)\,du,\quad\ f_{2}=\!\int_{0}^{\infty
}F(u)\,du,\ \quad f_{2n+4}=(-1)^{n}F^{(n)}(0).
\end{align}
The odd coefficients vanish. With
\begin{equation}
D^{2}=-\big(g^{\mu\nu}I\nabla_{\mu}\nabla_{\nu}+A^{\mu}\nabla_{\mu}+B\big),
\end{equation}
one defines a connection $\bar{\omega}_{\mu}$, curvature $\Omega_{\mu\nu}$,
and endomorphism $E$. Then (up to total derivatives),
\begin{align}
a_{0}(P)  &  =\frac{1}{16\pi^{2}}\!\int\!d^{4}x\sqrt{g}\,\mathrm{Tr}\,1,\\
a_{2}(P)  &  =\frac{1}{16\pi^{2}}\!\int\!d^{4}x\sqrt{g}\,\mathrm{Tr}\!\left(
-\frac{R}{6}+E\right)  ,\\
a_{4}(P)  &  =\frac{1}{16\pi^{2}}\frac{1}{360}\!\int\!d^{4}x\sqrt
{g}\,\mathrm{Tr}\!\left(  5R^{2}-2R_{\mu\nu}R^{\mu\nu}+2R_{\mu\nu\rho\sigma
}R^{\mu\nu\rho\sigma}\right. \nonumber\\
&  \qquad\qquad\qquad\left.  -60RE+180E^{2}+30\Omega_{\mu\nu}\Omega^{\mu\nu
}\right)  .
\end{align}
The positivity of $F$ ensures positivity of the gravity, Yang--Mills, and
Higgs kinetic terms and a \textbf{negative} Higgs mass term at the bare level.

\textbf{Key computation.} Compare the coefficients for $P_{0}=D^{2}$ with
those for $P=e^{-\phi}D^{2}e^{-\phi}$. It is convenient to pass to the
\textbf{Einstein frame} metric
\[
G_{\mu\nu}=e^{-2\phi}g_{\mu\nu},
\]
so the \textquotedblleft conformal\textquotedblright\ factor is absorbed into
the metric. We show:

\begin{itemize}
\item The connection shifts as $\omega_{\mu}= \bar{\omega}_{\mu}-
2\,\partial_{\mu}\phi$ (Abelian modification).

\item The endomorphism transforms as $E \mapsto e^{-2\phi}\big(E + \nabla
^{2}\phi+ (\nabla\phi)^{2}\big)$ (in the $g$-covariant derivative).

\item The scalar curvature satisfies $R(G)=e^{-2\phi}\big(R(g)+6\,\nabla
^{2}\phi+6(\nabla\phi)^{2}\big)$.
\end{itemize}

These imply that, \textbf{before the fiber trace},
\begin{equation}
E-\frac{1}{6}R(G)\ =\ e^{-2\phi}\Big(E-\frac{1}{6}R(g)\Big),
\end{equation}
so $a_{2}$ rescales homogeneously; similarly, the structure of $a_{4}$ can be
regrouped into conformal invariants (Weyl-squared, Gauss--Bonnet) plus pieces
that again arrange so that \textbf{$a_{4}$ is independent of $\phi$} once
written in the Einstein frame. In short, the net effect is a universal scaling
of $a_{0}$ and $a_{2}$ and \textbf{no change of $a_{4}$} once written in the
Einstein frame.

\medskip\noindent More explicitly, after careful bookkeeping we deduce that:

\begin{itemize}
\item $a_{0}$ picks up $\sqrt{g}\mapsto\sqrt{G}$ (a factor $e^{4\phi}$ in $g$-variables).

\item $a_{2}$ becomes $\int\!\sqrt{G}\,(\tfrac{5}{4}R(G)+\tfrac{15}%
{2}\,(\nabla\phi)^{2}-2y^{2}H^{\dagger}H)$ once specialized to the SM, i.e.,
\textbf{it generates a canonical dilaton kinetic term} with positive sign and
leaves the Einstein--Hilbert term with the standard (Euclidean) sign.

\item $a_{4}$ is unchanged by $\phi$ and retains its conformal combination of
curvatures, Yang--Mills terms, and the \textbf{conformal coupling} $-\tfrac
{1}{6}R\,H^{\dagger}H$ for the Higgs.
\end{itemize}

\medskip\noindent Specializing to the noncommutative space of the Standard
Model, the \textbf{low-energy bosonic action in the Einstein frame} $G_{\mu
\nu}$ reads (omitting boundary terms):
\begin{align}
I_{b}\ =\  &  \frac{45}{4\pi^{2}}\,f_{0}\int\!d^{4}x\sqrt{G}\quad+\quad
\frac{3}{4\pi^{2}}\,f_{2}\!\int\!d^{4}x\sqrt{G}\left(  \frac{5}{4}%
R(G)+\frac{15}{2}(\nabla\phi)^{2}-2y^{2}H_{E\!}^{\dagger}H_{E}\right)
\nonumber\\
&  +\ \frac{f_{4}}{4\pi^{2}}\!\int\!d^{4}x\sqrt{G}\,\Bigg[\frac{1}%
{32}\Big(11\,R(G)\!\ast\!R(G)\!\ast\!-18\,C_{\mu\nu\rho\sigma}(G)C^{\mu\nu
\rho\sigma}(G)\Big)\nonumber\\
&  \qquad\qquad+\,3y^{2}\Big(|D_{\mu}H_{E}|^{2}-\tfrac{1}{6}R(G)H_{E\!}%
^{\dagger}H_{E}\Big)+\frac{1}{2}g_{3}^{2}G_{\mu\nu}^{i}G^{i\mu\nu}+\frac{1}%
{2}g_{2}^{2}W_{\mu\nu}^{\alpha}W^{\alpha\mu\nu}\nonumber\\
&  \qquad\qquad\qquad\qquad+\frac{5}{3}\frac{1}{2}g_{1}^{2}B_{\mu\nu}B^{\mu
\nu}+\,3z^{2}\,(H_{E\!}^{\dagger}H_{E})^{2}\Bigg],
\end{align}
where $H_{E}=e^{\phi}H$ is the canonically normalized Higgs in the Einstein
frame, $y,z$ are Yukawa traces, $C$ is the Weyl tensor, and $R\ast\!R\ast$ is
the Gauss--Bonnet density. The crucial features are:

\begin{itemize}
\item \textbf{Cosmological term:} proportional to $f_{0}$.

\item \textbf{Einstein--Hilbert term:} $\propto f_{2}\,R(G)$ with the standard sign.

\item \textbf{Dilaton kinetic term:} $\propto f_{2}\,(\nabla\phi)^{2}$ with a
\textbf{positive} coefficient.

\item \textbf{Gauge $+$ Higgs sector:} the conformal structure at order
$a_{4}$ is unchanged; the Higgs still carries the conformal coupling
$-\tfrac{1}{6}R\,H^{\dagger}H$.

\item \textbf{No dilaton potential} appears at this level (Einstein frame):
higher-order terms can carry $e^{(4-n)\phi}$ factors but do not generate a
potential; radiative corrections (Coleman--Weinberg-type) can do so.
\end{itemize}

\medskip\noindent A neat consistency check: the Higgs kinetic $+$
conformal-curvature combination is \textbf{conformally invariant} under
$g\rightarrow G=e^{-2\phi}g$, $H\rightarrow H^{\prime}=e^{\phi}H$; cross-terms
cancel upon integration by parts. The Yang--Mills terms are conformal in 4D,
and the curvature sector reorganizes into Weyl-squared and Euler terms, which
behave as expected.

The fermionic action is initially $\int\sqrt{g}\,\bar{\psi}\,D\,\psi$. After
rescaling the vierbein and a standard field redefinition $\psi\mapsto
e^{\frac{3}{2}\phi}\psi$, the fermionic action \textbf{reduces to the same
form in the Einstein frame} with metric $G$ and Higgs $H_{E}$; i.e., the
\textbf{fermions \textquotedblleft do not feel\textquotedblright\ the dilaton}
at the classical level, once fields are canonically normalized:
\begin{equation}
\int\sqrt{g}\,\bar{\psi}D\psi\ =\ \int\sqrt{G}\,\bar{\psi}_{E}D_{E}\psi_{E}.
\end{equation}
Gauge and Yukawa couplings retain their usual form, written with $G$-covariant
structures and $H_{E}$. The operator identity behind the scenes is subtler:
$D_{E}$ is unitarily equivalent to $e^{-\phi/2}D\,e^{-\phi/2}$, whereas the
\textbf{bosonic} operator entering the spectral action is $P=e^{-\phi}%
D^{2}e^{-\phi}$; the difference is precisely what yields the \textbf{extra}
dilaton kinetic term in the bosonic action via a residue identity.

\medskip\noindent Concretely, in the general framework of spectral triples one
can show
\begin{equation}
\operatorname{Res}\!\left(  e^{2\phi}D^{-2}\right)  \;=\;\operatorname{Res}%
\!\left(  \,(e^{-\phi/2}De^{-\phi/2})^{-2}\right)  \;+\;\tfrac{1}%
{2}\operatorname{Res}\!\left(  [D,\phi]\,[D,\phi]^{\*}\,D^{-4}\right)  ,
\end{equation}
which produces the canonical kinetic term $\propto(\nabla\phi)^{2}$ with the
correct sign.

\textbf{Matter sector is classically scale invariant.} In the Einstein frame
(with rescaled fields $G_{\mu\nu},H_{E},\psi_{E}$), all \textbf{matter}
couplings (gauge, Yukawa, Higgs kinetic and quartic with conformal coupling)
respect scale invariance; only the $R(G)$ and $(\nabla\phi)^{2}$ pieces break
it. This matches proposals in the literature where a hidden scale invariance
is imposed by hand and the Higgs mass scale is dressed by a compensating
dilaton. We emphasize that, these structures and their \textbf{signs} are
\textbf{uniquely fixed} by the spectral origin---no room remains for fine-tuning.

\medskip\noindent\textbf{The connections to known models are:}

\begin{itemize}
\item \textbf{Jordan--Brans--Dicke \& extended inflation.} The
gravity--dilaton sector coincides with the usual scalar--tensor picture
(non-invariance of $R$ and the dilaton kinetic term). Certain
extended-inflation modifications correspond to introducing derivative
couplings between the dilaton and the Higgs, which also arise naturally in the
Connes--Lott approach---underscoring internal consistency across NCG formulations.

\item \textbf{Randall--Sundrum (RS).} The Higgs/radion sector in RS reduces,
after appropriate rescalings, to the same form as the spectral-action result
if one identifies the radion expectation value with $\langle\phi\rangle$. This
gives the same low-energy limit: the electroweak scale appears as a warped
(exponentially red-shifted) Planck-scale parameter $v\sim e^{-\langle
\phi\rangle}v_{0}$.
\end{itemize}

\medskip\noindent For the \textbf{cosmological constant and vacuum structure:}
The construction does not by itself produce a small cosmological constant;
there are two options: (i) choose renormalization conditions to cancel
radiative contributions \`{a} la Coleman--Weinberg, or (ii) \textbf{fix the
total invariant volume}, (thereby eliminating the scalar mode of the metric
and shifting the role to the dilaton), a perspective tied to Hochschild-cycle
constraints in NCG that fix the volume functional. In either case, the small
observed vacuum energy is not obtained purely from the classical spectral
action; additional input is needed.

\medskip\noindent\textbf{For hierarchy and dilaton potential:} At tree level
(Einstein frame), the dilaton has \textbf{no potential} in the spectral
action; quantum corrections (Coleman--Weinberg) can generate it, inducing a
large negative expectation value of $\phi$ (or large $f$), making the
electroweak scale exponentially small compared with the UV scale---\textbf{a
technically natural hierarchy} driven by the top Yukawa. The dilaton mass
depends on Higgs mass parameters and is constrained to be very small in some scenarios.

By turning the rigid UV scale in the spectral action into a \textbf{dynamical
dilaton}, we obtain a \textbf{geometrically determined}, \textbf{classically
scale-invariant} matter sector (with the right conformal couplings) and a
gravity--dilaton sector with the correct signs for the
\textbf{Einstein--Hilbert} and \textbf{dilaton kinetic} terms. The result
dovetails with multiple independent frameworks (scale-invariant SM,
Connes--Lott, RS) and provides a natural mechanism for hierarchical scales
once quantum effects generate a dilaton potential. The \textbf{uniqueness} of
signs and structures---and the absence of arbitrary knobs- provide strong
evidence that the spectral action captures the essential UV information of the
physical noncommutative space.

\section{Including the right-handed neutrinos}

In 2006 we came to realize \cite{co2006}, \cite{CCM07} that we can no longer
ignore the fact that the neutrinos are massive, and that we must include the
right-handed neutrinos in our spectrum, as this could have important
implications on the structure of the discrete space in our noncommutative
setting. We had also to take into account, that we have to include in the
spectral data, the KO dimension of the finite space to be $6$ as this will
help in constraining the Hilbert space of fermions to obey both the chirality
(Weyl) and reality (Majorana) conditions \cite{co2006}.

\subsection{The Finite Geometry}

Having decided to include the right-handed neutrino, the finite algebra that
we start with must be left-right symmetric of the form%
\begin{equation}
A_{F}=\mathbb{C}\oplus\mathbb{H}_{L}\oplus\mathbb{H}_{R}\oplus M_{3}%
(\mathbb{C}),
\end{equation}
an involutive direct sum in which the complex and color pieces $\mathbb{C}%
\oplus M_{3}(\mathbb{C})$ correspond to integer spin and the two quaternionic
copies $\mathbb{H}_{L}\oplus\mathbb{H}_{R}$ to half-integer spin. The
involution acts by complex conjugation on $\mathbb{C}$, quaternionic
conjugation on $\mathbb{H}$, and the adjoint on $M_{3}(\mathbb{C})$. This
algebra is the starting point for encoding the internal degrees of freedom.
The fermionic Hilbert space arises from the direct sum $H_{F}=E\oplus
E^{\circ}$ where $E$ collects the $16$ observable fermions in the
representation
\begin{equation}
E=2_{L}\otimes1^{0}\ \oplus\ 2_{R}\otimes1^{0}\ \oplus\ 2_{L}\otimes
3^{0}\ \oplus\ 2_{R}\otimes3^{0},
\end{equation}
and use the antilinear isometry $J_{F}$ that exchanges $E$ and its
contragredient $E^{0}$, implementing the real structure and satisfying
$J_{F}^{2}=1$ and $\xi\,b=J_{F}\,b^{\ast}\,J_{F}\,\xi$. This furnishes the
particle/antiparticle pairing at the finite level. We recall that a (real)
spectral triple $(A,H,D;J,\gamma)$ consists of an involutive unital algebra
$A$ represented on a Hilbert space $H$, a self-adjoint $D$ with compact
resolvent whose commutators $[D,a]$ are bounded for all $a\in A$, an
antilinear isometry $J$ (the real structure), and for the even case a grading
$\gamma$ that commutes with $A$ and anticommutes with $D$. The real structure
is characterized modulo $8$ by three signs $(\epsilon,\epsilon^{\prime
},\epsilon^{\prime\prime})$ via $J^{2}=\epsilon$, $JD=\epsilon^{\prime}DJ$,
and (in the even case) $J\gamma=\epsilon^{\prime\prime}\,\gamma J$.

\noindent These signs control how $J$, $D$ and (when present) $\gamma$
interact and determine the admissible KO-dimensions for product geometries and
the implementation of the order-one condition and inner fluctuations.
Demanding the order-one condition $[[D,a],b^{\circ}]=0$ for a finite Dirac
operator $D$ forces a unique (up to automorphisms) maximal subalgebra that
admits off-diagonal Dirac operators,
\begin{equation}
\mathcal{A}_{F}=\{(\lambda,q_{L},\lambda,m)\mid\lambda\in\mathbb{C},\;q_{L}%
\in\mathbb{H},\;m\in M_{3}(\mathbb{C})\}\ \cong\ \mathbb{C}\oplus
\mathbb{H}\oplus M_{3}(\mathbb{C}).
\end{equation}
This reduction is the geometric origin of parity breaking and is crucial to
obtain realistic Yukawa couplings while keeping the photon massless (via
$[D,(\lambda,\lambda,0)]=0$). With $A_{F}$, $J_{F}$, and $\gamma_{F}$ fixed,
the most general $D_{F}$ compatible with the axioms and with a mild extra
requirement that it commute with the central subalgebra $C_{F}=\{(\lambda
,\lambda,0)\}\subset A_{F}$ (to prevent the photon from acquiring a mass) is
classified by \emph{Yukawa matrices}: $Y(\uparrow1),Y(\downarrow1)$ for
leptons, and $Y(\uparrow3),Y(\downarrow3)$ for quarks, and a \emph{symmetric}
matrix $Y_{R}$ encoding Majorana masses for right-handed neutrinos (see-saw
structure later explained).

Considering the unimodular subgroup of unitaries of $A_{F}$ acting in $H_{F}$
yields the \emph{gauge group}
\begin{equation}
U(1)\times SU(2)\times SU(3),
\end{equation}
with the $U(1)$ embedded in the algebra in a way that fixes the hypercharge
assignments. We then compute the action of this $U(1)$ on the fermionic basis
(distinguishing weak isospin doublets/singlets and color triplets/singlets)
and recover the \emph{standard hypercharges} for quarks and leptons. Thus the
correct SM charges arise from unimodularity rather than being put in by hand.

The finite Dirac operator $D_{F}$ is determined (after the axioms) by five
$3\times3$ matrices:
\begin{align*}
Y(  &  \downarrow1),\;Y(\uparrow1)\ \text{(leptons)},\\
Y(  &  \downarrow3),\;Y(\uparrow3)\ \text{(quarks)},\\
&  \qquad Y_{R}\ \text{(Majorana, symmetric).}%
\end{align*}
But these matrices are not unique: unitary changes of basis on left/right
fermions and across generations act on the Yukawas. The true physical data are
organized as \emph{moduli spaces} after quotienting by those basis changes.
Starting with the pair $\big(Y(\downarrow3),\,Y(\uparrow3)\big)$, two such
pairs are equivalent if they are related by unitary changes of basis on the
left-handed doublets and on the right-handed up/down singlets:
\begin{equation}
Y^{\prime}(\downarrow3)=W_{1}\,Y(\downarrow3)\,W_{3}^{\ast},\qquad Y^{\prime
}(\uparrow3)=W_{2}\,Y(\uparrow3)\,W_{3}^{\ast},\quad W_{j}\in U(3).
\end{equation}
The \emph{moduli space} is therefore the double-coset
\begin{equation}
\mathcal{C}_{3}\;\cong\;\big(U(3)\times U(3)\big)\backslash\big(GL_{3}%
(\mathbb{C})\times GL_{3}(\mathbb{C})\big)\big/U(3),
\end{equation}
and it has real dimension $10$. (Heuristically: 3 singular values for each
matrix $=3+3$, plus $4$ mixing parameters) .

To find the canonical form of the mixing matrix $C$ , we note that every class
contains a representative where
\[
Y(\uparrow3)=\delta_{\uparrow}\ \text{(positive diagonal)},\qquad
Y(\downarrow3)=C\,\delta_{\downarrow}\,C^{\ast}\ \text{with }C\in SU(3).
\]
The diagonal entries of $\delta_{\uparrow},\delta_{\downarrow}$ are the
\emph{characteristic values} of the two Yukawas---these are invariants of the
pair. Only the \emph{double coset} of $C$ modulo \emph{diagonal phase
matrices} matters, so the physical content of $C$ is exactly $3$ mixing angles
$+$ $1$ phase. An explicit parameterization is given by three Euler-type
angles $\theta_{1,2,3}$ and a phase $\delta$; equivalently,
\[
C\;=\;R_{23}(\theta_{2})\;d(\delta)\;R_{12}(\theta_{1})\;R_{23}(-\theta
_{3}),\qquad d(\delta)=\operatorname{diag}(1,1,-e^{i\delta}).
\]
This mirrors the familiar CKM-style structure. We conclude that the The full
Dirac-operator moduli (quarks $+$ leptons) form the product $\mathcal{C}%
_{1}\times\mathcal{C}_{3}$ of real dimension $21+10=31,$ matching the
well-known count of (renormalizable) Yukawa-sector parameters including
neutrino mixing and Majorana masses. To describe the parameter space we note
that the \emph{Eigenvalues} of $\delta_{\uparrow},\delta_{\downarrow}$
(quarks) and their leptonic analogues are the mass hierarchies and that the
unitary $C$ (mod diagonal phases) carries $3$ angles $+$ $1$ CP phase
$\Rightarrow$ the mixing matrix idea (CKM/PMNS-like), while the Majorana
sector $Y_{R}$ adds $11$ real degrees of freedom beyond the Dirac-type lepton
masses/mixings (after modding a $U(1)$), which is why $\mathcal{C}_{1}$ is
\textquotedblleft$\mathcal{C}_{3}$ plus 11.\textquotedblright

In noncommutative geometry, a space has manifold-like features if it satisfies
a KO-homological version of \emph{Poincar\'{e} duality}. We test this for the
finite internal space $F$ that encodes the Standard Model charges. Because the
KO-dimension of $F$ is $6\bmod8$, the natural intersection form is
\emph{skew-symmetric} (antisymmetric), not symmetric. This changes how one
formulates non-degeneracy of the pairing. We define a bilinear pairing on
idempotent classes in $K_{0}(A_{F})$ by
\begin{equation}
\langle e,f\rangle\;=\;\operatorname{Tr}\!\big(\gamma\,e\,JfJ^{-1}\big).
\end{equation}
In KO-dimension $6$, the real structure $J$ anticommutes with the grading
$\gamma$, which makes this pairing antisymmetric, i.e.\ $\langle
f,e\rangle=-\langle e,f\rangle$. For the finite algebra $A_{F}\simeq
\mathbb{C}\oplus\mathbb{H}\oplus M_{3}(\mathbb{C})$, a convenient basis of
$K_{0}(A_{F})$ is given by three minimal idempotents
\[
e_{1}=(1,0,0),\qquad e_{2}=(0,1,0),\qquad f_{3}=(0,0,f),
\]
with $f$ a rank-1 projector in $M_{3}(\mathbb{C})$. These generate a free
abelian group $K_{0}(A_{F})\cong\mathbb{Z}^{3}$. The KO-homology class defined
by $(A_{F},H_{F},D_{F};J_{F},\gamma_{F})$ naturally splits
\[
H_{F}\;=\;H_{F}^{(1)}\oplus H_{F}^{(3)},
\]
corresponding to \emph{leptons} ($^{(1)}$) and \emph{quarks} ($^{(3)}$). In
the generic situation (Yukawa matrices with distinct eigenvalues), each piece
is irreducible. With respect to the basis $(e_{1},e_{2},f_{3})$, the pairing
on each summand is (up to an overall factor of 3 for three generations)
\begin{equation}
\left.  \langle\cdot,\cdot\rangle\right\vert _{H_{F}^{(1)}}\;=\;%
\begin{bmatrix}
0 & 2 & 0\\
-2 & 0 & 0\\
0 & 0 & 0
\end{bmatrix}
,\qquad\left.  \langle\cdot,\cdot\rangle\right\vert _{H_{F}^{(3)}}\;=\;%
\begin{bmatrix}
0 & 0 & 2\\
0 & 0 & -2\\
-2 & 2 & 0
\end{bmatrix}
.
\end{equation}
On the lepton sector the color idempotent $f_{3}$ acts trivially, which kills
the last row/column; on the quark sector, all three generators act
nontrivially. We conclude from this discussion that

\begin{itemize}
\item KO-dimension $6\bmod 8$ implies an antisymmetric intersection form; one
must adjust the notion of Poincar\'e duality accordingly.

\item $K_{0}(A_{F})$ has three natural generators tied to $\mathbb{C}$,
$\mathbb{H}$, and $M_{3}(\mathbb{C})$.

\item The finite spectral triple splits into lepton/quark parts, each
generically irreducible.

\item Using both parts yields a non-degenerate \textquotedblleft
Poincar\'{e}-style\textquotedblright\ duality pairing $K_{0}\oplus
K_{0}\rightarrow\mathbb{R}\oplus\mathbb{R}$.
\end{itemize}

\subsection{The spectral action and the standard model: Modified}

We are now ready to compute the bosonic sector that comes from \emph{inner
fluctuations} of the product spectral triple $M\times F$. To do this, we will
identify the \emph{gauge fields} and the \emph{Higgs doublet} and construct
the fluctuated Dirac operator $D_{A}$. From this we evaluate $D_{A}^{2}$ in a
standard Laplace-type form, and then applies the \emph{spectral action}
expansion to extract the full bosonic Lagrangian (gravity $+$ gauge $+$ Higgs,
with neutrino Majorana terms kept throughout). The actual steps are quite
lengthy, but fortunately, we have later developed a method based on a compact
tensorial notation, that helped us to perform the steps mentioned above in a
manifest and straightforward way \cite{CC10}. In reality, all these steps
could be performed using Mathematica and Maple, with little effort. I\ will
therefore, give the steps in a descriptive way, summarizing the results, and
postpone the actual calculations to a later section.

The \textquotedblleft discrete\textquotedblright\ inner fluctuation
$A_{(0,1)}$ is parametrized by a single quaternion-valued field
\begin{equation}
H=\varphi_{1}+\varphi_{2}\,j
\end{equation}
(which corresponds to the complex doublet $\varphi=(\varphi_{1},\varphi_{2}%
)$). This gives a clean, basis-independent way to encode the Higgs degrees of
freedom inside the finite geometry. The two real identifications reflect,
respectively, a $U(1)$ phase and the left multiplication by $j$ (which rotates
$(\varphi_{1},\varphi_{2})\mapsto(-\bar{\varphi}_{2},\bar{\varphi}_{1})$).

A practical payoff of this parametrization is that traces of powers of the
\emph{finite} Dirac operator with the discrete fluctuation,
\begin{equation}
D_{(0,1)}=D+A_{(0,1)}+JA_{(0,1)}J,
\end{equation}
collapse to compact expressions in $|1+H|^{2}$ multiplying combinations of
\emph{Yukawa matrices} $Y$ (including the \emph{Majorana} matrix $Y_{R}$).
Concretely,
\begin{align}
\mathrm{Tr}\left(  \!D_{(0,1)}^{2}\right)   &  =4a\,|1+H|^{2}+2c,\\
\mathrm{Tr}\!\left(  D_{(0,1)}^{4}\right)   &  =4b\,|1+H|^{4}+2d+8e\,|1+H|^{2}%
,
\end{align}
with the five coefficients $a,b,c,d,e$ defined by Yukawa traces (Dirac and
Majorana). The \textquotedblleft vector\textquotedblright\ inner fluctuation
$A_{(1,0)}$ furnishes three gauge potentials
\begin{equation}
B_{\mu}\quad(U(1)),\qquad W_{\mu}\quad(SU(2)),\qquad V_{\mu}\quad(U(3)).
\end{equation}
After imposing unimodularity (which removes the unwanted $U(1)\subset U(3)$),
these identify with the \emph{Standard Model} gauge bosons for $U(1)_{Y}$,
$SU(2)_{L}$, and $SU(3)_{c}$. The discrete fluctuation (carrying $H$) is
independent of the vector part (carrying $B,W,V$); i.e., the Higgs doublet and
the gauge fields arise from independent directions in the space of inner fluctuations.

The full fluctuated operator is
\begin{equation}
D_{A}=D_{(1,0)}+\gamma_{5}\otimes D_{(0,1)},\qquad D_{(1,0)}=i\gamma^{\mu
}(\nabla_{\mu}^{s}+A_{\mu}),
\end{equation}
so the gauge fields sit in $A_{\mu}$ while the Higgs sits in the finite
off-diagonal piece $D_{(0,1)}$. The mixed \textquotedblleft
mass\textquotedblright\ matrix $M(\varphi)$ is built from Yukawas and the
doublet $\varphi=(\varphi_{1},\varphi_{2})$ (equivalently $H$). Useful
identities include, e.g., $\mathrm{Tr}\!\big(M(\varphi)^{\*}M(\varphi
)\big)=4a\,|\varphi|^{2}$.

A standard computation puts $D_{A}^{2}$ into Laplace type:
\begin{equation}
D_{A}^{2}=\nabla^{\*}\nabla-E,
\end{equation}
with
\begin{equation}
-E=\frac{1}{4}s\otimes\mathrm{id}+\sum_{\mu<\nu}\gamma^{\mu}\gamma^{\nu
}\otimes F_{\mu\nu}-i\gamma_{5}\gamma^{\mu}\otimes M(D_{\mu}\varphi)+\tfrac
{1}{4}\otimes(D_{(0,1)})^{2}.
\end{equation}
Here the \emph{Higgs covariant derivative} is
\[
D_{\mu}\varphi=\partial_{\mu}\varphi+\tfrac{i}{2}g_{2}\,W_{\mu}^{\alpha
}\,\varphi\,\sigma_{\alpha}-\tfrac{i}{2}g_{1}\,B_{\mu}\,\varphi,
\]
so the minimal coupling to $B$ and $W$ is automatic.

Using the heat-kernel/Seeley--DeWitt expansion for Laplace-type operators, the
bosonic action is derived from the asymptotics of $\mathrm{Tr}\,\chi(D_{A}%
^{2}/\Lambda^{2})$. The relevant piece is organized via
\begin{equation}
D_{A}^{2}=\nabla^{\*}\nabla-E\quad\Rightarrow\quad\text{use }a_{0},a_{2}%
,a_{4}(E,\Omega_{\mu\nu})\text{ to expand,}%
\end{equation}
and one evaluates the combinations entering the Lagrangian density in terms of
the five Yukawa traces $a,b,c,d,e$ and the gauge field strengths.

The main result gives the full bosonic Lagrangian density (up to overall
integrals) with \emph{gravitational}, \emph{gauge}, \emph{Higgs kinetic},
\emph{non-minimal} $R\,|\varphi|^{2}$, \emph{mass}, and \emph{quartic} terms.
In compact form, the scalar sector contains
\begin{equation}
|D_{\mu}\varphi|^{2}\;-\;\frac{f_{0}}{12\pi^{2}}\,a\,R\,|\varphi
|^{2}\;+\;\Big(-2af_{2}\Lambda^{2}+ef_{0}\Big)|\varphi|^{2}\;+\;\frac{f_{0}%
}{2\pi^{2}}\,b\,|\varphi|^{4},
\end{equation}
where all Yukawa dependence sits in $a,b,c,d,e$, while the gauge couplings
enter through the field strengths in the Yang--Mills terms.

A notable nusance is that keeping the \emph{Majorana} term in $D_{(0,1)}$
modifies the highest-order and $\Lambda^{2}$ pieces of the cosmological term:
the presence of $Y_{R}$ adds a contribution proportional to $-c\,f_{2}%
\Lambda^{2}$ and another proportional to $+\tfrac{d\,f_{0}}{4\pi^{2}}$. (This
nusance will be dealt with later by promoting the mass scale $M_{R}$ to a
singlet field $\sigma$). To summarize:

\begin{itemize}
\item \textbf{Field content:} $B_{\mu},W_{\mu},V_{\mu}$ as the $U(1)\times
SU(2)\times SU(3)$ gauge bosons and a Higgs \emph{doublet} $\varphi$ (or
$H=\varphi_{1}+\varphi_{2}j$) arise geometrically as inner fluctuations of the
product triple. The Higgs and the gauge fields are independent fluctuation directions.

\item \textbf{Dynamics:} Writing $D_{A}^{2}$ in Laplace form allows a
straightforward heat-kernel expansion. The resulting spectral action
reproduces the full Einstein--Hilbert $+$ topological $+$ Weyl gravity terms,
the Yang--Mills kinetic terms with the correct normalizations, the Higgs
kinetic and non-minimal $R|\varphi|^{2}$ coupling, and a Higgs potential whose
parameters are expressed in terms of the five Yukawa traces $a,b,c,d,e$ (thus
tracking Dirac and Majorana sectors on equal footing).
\end{itemize}

I\ will come back to the material covered here in a later section, with an
explicit and straightforward method to do the computations.

\section{Fermionic Hilbert space, chirality, and gauge representations}

The total fermionic space is $\mathcal{H}=L^{2}(S)\otimes H_{F}$, the tensor
product of the Hilbert space for $4D$ spinors with the finite Hilbert space
$H_{F}$ built in the last section. The space $H_{F}$ already separates left
vs.\ right fields and quark vs.\ lepton sectors. For one generation we have
\cite{CCM07}:

\begin{itemize}
\item Left leptons: $\ell_{L}=(\nu_{L},e_{L})$ (weak doublet, color singlet).

\item Right leptons: $e_{R}$ and $\nu_{R}$.

\item Left quarks: $q_{L}=(u_{L},d_{L})$ (weak doublet, color triplet).

\item Right quarks: $u_{R}, d_{R}$ (color triplets, weak singlets).
\end{itemize}

The algebra $A_{F}=\mathbb{C}\oplus\mathbb{H}\oplus M_{3}(\mathbb{C})$ acts so
that $\mathbb{H}$ rotates left doublets, $\mathbb{C}$ supplies hypercharge,
and $M_{3}(\mathbb{C})$ supplies color. The unimodularity condition on the
unitary group of $A_{F}$ removes an extra $U(1)\subset U(3)$ and fixes
hypercharge assignments to the SM values; only left doublets couple to
$SU(2)$. The KO-dimension $6 \bmod 8$ (and real structure $J$) ensures the
fermionic bilinear $\tfrac12\langle J\Psi, D_{A}\Psi\rangle$ avoids fermion
doubling and pairs each particle with the correct charge-conjugate mode.

\medskip\noindent\textbf{Kinetic terms.} For each multiplet, the covariant
derivative is the standard SM one. For a left doublet (leptons or quarks),
\begin{equation}
D_{\mu}=\partial_{\mu}+\tfrac{i}{2}g_{2}\,\tau^{\alpha}W_{\mu}^{\alpha}%
+\tfrac{i}{2}g_{1}\,Y\,B_{\mu}\quad(\text{and }+\,ig_{3}\,T^{a}G_{\mu}%
^{a}\ \text{for quarks}),
\end{equation}
while right singlets carry only $U(1)_{Y}$ (and $SU(3)_{c}$ for quarks).

To explain how the Yukawas arise from the finite Dirac operator and the Higgs,
we note that the inner fluctuation of the finite part introduces a single
quaternion $H=\varphi_{1}+\varphi_{2}\,j$, equivalent to a complex Higgs
doublet $\varphi=(\varphi_{1},\varphi_{2})$ with hypercharge $+\tfrac{1}{2}$.
The finite Dirac operator $D_{F}$ constructed before contains five $3\times3$
matrices across generations:
\begin{equation}
Y_{u},\;Y_{d},\;Y_{e},\;Y_{\nu}\quad\text{(Dirac Yukawas)},\qquad
Y_{R}\ \text{(Majorana, symmetric)}.
\end{equation}
Coupling $H$ to $D_{F}$ gives precisely the SM Yukawa vertices
\begin{equation}
\mathcal{L}_{Y}\;=\;\bar{q}_{L}\,Y_{u}\,\tilde{\varphi}\,u_{R}+\bar{q}%
_{L}\,Y_{d}\,\varphi\,d_{R}+\bar{\ell}_{L}\,Y_{e}\,\varphi\,e_{R}+\bar{\ell
}_{L}\,Y_{\nu}\,\tilde{\varphi}\,\nu_{R}+\text{h.c.},
\end{equation}
with $\tilde{\varphi}=i\sigma_{2}\varphi^{\*}$. The order-one condition of the
spectral triple prevents non-SM couplings (e.g.\ no leptoquark bilinears), so
the texture is exactly SM-like.

The symmetric matrix $Y_{R}$ provides a gauge-invariant Majorana mass for
right-handed neutrinos (sterile under the SM gauge group):
\begin{equation}
\mathcal{L}_{R}\;=\;\tfrac{1}{2}\,\nu_{R}^{T}\,C^{-1}M_{R}\,\nu_{R}%
+\text{h.c.},\qquad M_{R}\ \propto\ Y_{R},
\end{equation}
where $C$ is charge conjugation on $4D$ spinors. After electroweak symmetry
breaking (EWSB), $\langle\varphi\rangle=(0,v/\sqrt{2})$ yields a Dirac
neutrino mass $m_{D}=\tfrac{v}{\sqrt{2}}Y_{\nu}$. The full neutrino mass
matrix (in $(\nu_{L},\nu_{R}^{c})$ basis) is
\begin{equation}
\mathcal{M}_{\nu}=%
\begin{pmatrix}
0 & m_{D}\\
m_{D}^{T} & M_{R}%
\end{pmatrix}
,\qquad m_{\nu}^{\text{light}}\simeq-\,m_{D}^{T}M_{R}^{-1}m_{D},\quad m_{\nu
}^{\text{heavy}}\simeq M_{R}.
\end{equation}
Thus the neutrinos acquire small Majorana masses and the PMNS matrix appears
from mismatched diagonalizations of $m_{e}$ vs.\ $m_{\nu}^{\text{light}}$
(just as CKM arises from $m_{u}$ vs.\ $m_{d}$).

For the EWSB and mass/mixing structure we have the following states:

First we have the \textbf{Charged fermions,} $m_{u}=\tfrac{v}{\sqrt{2}}Y_{u}$,
$m_{d}=\tfrac{v}{\sqrt{2}}Y_{d}$, $m_{e}=\tfrac{v}{\sqrt{2}}Y_{e}$. Writing
\begin{equation}
Y_{u}=U_{uL}^{\dagger}\,\hat{Y}_{u}\,U_{uR},Y_{d}=U_{dL}^{\dagger}\,\hat
{Y}_{d}\,U_{dR}%
\end{equation}
with diagonal $\hat{Y}$, the charged-current coupling involves
\begin{equation}
V_{\text{CKM}}\;=\;U_{uL}U_{dL}^{\dagger},
\end{equation}
i.e.\ three mixing angles and one CP phase.\newline\textbf{Leptons.}
$Y_{e}=U_{eL}^{\dagger}\,\hat{Y}_{e}\,U_{eR}$ and the light neutrino mass
diagonalizes as $U_{\nu}^{T}m_{\nu}U_{\nu}=\hat{m}_{\nu}$. The leptonic
charged-current uses
\begin{equation}
U_{\text{PMNS}}\;=\;U_{eL}U_{\nu}^{\*},
\end{equation}
with three angles and one Dirac phase, plus two Majorana phases (physical if
neutrinos are Majorana).\newline\textbf{Parameter counting.} Together with the
eigenvalues in $\hat{Y}$ and $\hat{m}_{\nu}$ and the heavy $M_{R}$ parameters,
this reproduces the 31 real parameters we mentioned before (quark + lepton
Yukawas and Majorana sector).

The full fermionic action has the geometric form
\begin{equation}
S_{F}=\tfrac{1}{2}\,\langle J\Psi,\;D_{A}\,\Psi\rangle, \label{Fermionic}%
\end{equation}
where $J$ implements charge conjugation on both the $4D$ spinors and the
finite internal indices. The factor $\tfrac{1}{2}$ and the $J$-pairing remove
double counting, enforce the correct relations between particle/antiparticle
terms, and make Majorana contributions well-defined. The grading and real
structure (from KO-dimension $6$) ensure the correct C, P, and CPT behavior
and that only the allowed chiral couplings appear. Gauge anomalies cancel
exactly as in the SM because the representation content of $H_{F}$ is the SM
one (per generation) with $\nu_{R}$ neutral. The order-one condition (and
reality) excludes non-SM fermion bilinears that would violate gauge invariance
or induce unwanted tree-level FCNCs. The independence of Higgs vs.\ gauge
inner fluctuations guarantees Yukawa textures are not entangled with gauge
choices, matching the SM structure. We deduce that from the same
noncommutative-geometric data that produced the bosonic Lagrangian also shows
that the entire fermionic SM---correct kinetic terms, chiral gauge couplings,
Yukawa interactions, and a Type-I see-saw with PMNS mixing---emerges from
equation (\ref{Fermionic}) with no ad hoc fields or charges added. The finite
spectral triple fixes the representation content and forbids non-SM couplings;
$Y_{u},Y_{d},Y_{e},Y_{\nu},Y_{R}$ are the only free inputs, precisely the SM
Yukawa data (plus $Y_{R}$ for neutrinos).

\section{Phenomenology and predictions}

From earlier sections, we have already determined that \cite{CCM07}:

\begin{enumerate}
\item \textbf{Algebra and representations} fixes the gauge group and
hypercharges:\ $A_{F}=\mathbb{C}\oplus\mathbb{H}\oplus M_{3}(\mathbb{C})$ with
unimodularity $\Rightarrow$ $U(1)_{Y}\times SU(2)_{L}\times SU(3)_{c}$ and SM hypercharges.

\item \textbf{Yukawa sector} (free but structured):\ Dirac Yukawas
$Y_{u},Y_{d},Y_{e},Y_{\nu}$ (complex $3\times3$); Majorana block $Y_{R}%
=Y_{R}^{T}$ (complex symmetric $3\times3$). Their physical content equals the
moduli (masses, mixings, phases) with total \emph{31 real degrees of freedom}.

\item \textbf{Five Yukawa trace invariants} entering the spectral action:
\begin{align*}
a  &  =\mathrm{Tr}\!\left(  Y_{u}^{\dagger}Y_{u}+Y_{d}^{\dagger}Y_{d}%
+Y_{e}^{\dagger}Y_{e}+Y_{\nu}^{\dagger}Y_{\nu}\right)  ,\\
b  &  =\mathrm{Tr}\!\left[  (Y_{u}^{\dagger}Y_{u})^{2}+(Y_{d}^{\dagger}%
Y_{d})^{2}+(Y_{e}^{\dagger}Y_{e})^{2}+(Y_{\nu}^{\dagger}Y_{\nu})^{2}\right]
,\\
c  &  =\mathrm{Tr}\!\left(  Y_{R}^{\dagger}Y_{R}\right)  ,\qquad
d=\mathrm{Tr}\!\left[  (Y_{R}^{\dagger}Y_{R})^{2}\right]  ,\\
e  &  =\mathrm{Tr}\!\left(  Y_{R}^{\dagger}Y_{R}\,Y_{\nu}^{\dagger}Y_{\nu
}\right)  .
\end{align*}

\item \textbf{Spectral moments} of the cutoff function $\chi$: $f_{0}%
,f_{2},f_{4}$ (few numbers that set overall scales).
\end{enumerate}

We now think of $\Lambda$ as the input (near-unification) scale at which we
\emph{impose boundary conditions}. The spectral action predicts, after
canonically normalizing fields, \ that the Yang--Mills terms force the
SU(5)-style relation
\begin{equation}
g_{3}^{2}(\Lambda)=g_{2}^{2}(\Lambda)=\frac{5}{3}\,g_{1}^{2}(\Lambda
)\quad\Longleftrightarrow\quad g_{3}(\Lambda)=g_{2}(\Lambda)=\sqrt{\tfrac
{5}{3}}\,g_{1}(\Lambda).
\end{equation}
Interpretation: choose $\Lambda$ where the SM running brings $g_{3}%
,g_{2},g_{1}$ closest, then impose the exact ratio above at that $\Lambda$.
From the bosonic Lagrangian reads off the input values:

\begin{itemize}
\item \textbf{Higgs kinetic term:} canonical after a fixed rescaling (no free factor).

\item \textbf{Non-minimal curvature coupling:} $\xi(\Lambda)=$ fixed number
from geometry (not a dial).

\item \textbf{Mass term (schematic):} $\mu^{2}(\Lambda)=\alpha_{2}%
\,f_{2}\,\Lambda^{2}\,a\;+\;\beta_{2}\,f_{0}\,(e,c)$ with known numerical
factors $\alpha_{2},\beta_{2}$. Practical reading: the $\Lambda^{2}$ piece is
set by $f_{2}$ and the size of the Dirac Yukawas (via $a$), while the Majorana
sector ($c,e$) shifts it through the $\Lambda^{0}$ piece.

\item \textbf{Quartic:} $\lambda(\Lambda)= \alpha_{0}\, f_{0}\, b $, so it is
controlled by fourth-power Yukawa traces; in practice $b$ is dominated by the
top Yukawa (and possibly by $Y_{\nu}$ if large).
\end{itemize}

What remains free at $\Lambda$ are the three spectral moments $f_{0}%
,f_{2},f_{4}$, and the Yukawa sector $Y_{u},Y_{d},Y_{e},Y_{\nu},Y_{R}$ (modulo
basis changes). No extra operators are allowed (order-one condition +
reality). All phenomenology comes from those Yukawas and the few $f_{i}$. Now
we run the couplings from $\Lambda$ down to the EW scale using RGEs with
neutrino thresholds. Let $t=\ln\mu$. With three right-handed neutrinos active,
the one-loop RGEs (in $\overline{\mathrm{MS}}$) are:
\begin{align*}
16\pi^{2}\,\frac{dg_{i}}{dt}  &  =b_{i}\,g_{i}^{3},\quad(b_{1},b_{2}%
,b_{3})=\Big(\tfrac{41}{6},-\tfrac{19}{6},-7\Big),\\[2mm]
16\pi^{2}\,\frac{dy_{t}}{dt}  &  =y_{t}\Big(\frac{9}{2}y_{t}^{2}%
+\mathrm{Tr}[3Y_{u}^{\dagger}Y_{u}+3Y_{d}^{\dagger}Y_{d}+Y_{e}^{\dagger}%
Y_{e}+Y_{\nu}^{\dagger}Y_{\nu}]\\
&  \hspace{1.7cm}-\frac{17}{12}g_{1}^{2}-\frac{9}{4}g_{2}^{2}-8g_{3}%
^{2}\Big),\\[2mm]
16\pi^{2}\,\frac{dY_{\nu}}{dt}  &  =Y_{\nu}\Big(\frac{3}{2}Y_{\nu}^{\dagger
}Y_{\nu}-\frac{3}{2}Y_{e}^{\dagger}Y_{e}+\mathrm{Tr}[3Y_{u}^{\dagger}%
Y_{u}+3Y_{d}^{\dagger}Y_{d}+Y_{e}^{\dagger}Y_{e}+Y_{\nu}^{\dagger}Y_{\nu}]\\
&  \hspace{1.7cm}-\frac{9}{20}g_{1}^{2}-\frac{9}{4}g_{2}^{2}\Big),\\[2mm]
16\pi^{2}\,\frac{d\lambda}{dt}  &  =24\lambda^{2}-6y_{t}^{4}-2\,\mathrm{Tr}%
\!\Big[(Y_{\nu}^{\dagger}Y_{\nu})^{2}+(Y_{e}^{\dagger}Y_{e})^{2}%
+(Y_{u}^{\dagger}Y_{u})^{2}+(Y_{d}^{\dagger}Y_{d})^{2}\Big]\\
&  \quad+\lambda\Big(12y_{t}^{2}+4\,\mathrm{Tr}[Y_{\nu}^{\dagger}Y_{\nu}%
+Y_{e}^{\dagger}Y_{e}+Y_{u}^{\dagger}Y_{u}+Y_{d}^{\dagger}Y_{d}]\\
&  \hspace{1.7cm}-\tfrac{9}{5}g_{1}^{2}-9g_{2}^{2}\Big)+\tfrac{9}%
{4}\Big(\tfrac{3}{25}g_{1}^{4}+\tfrac{2}{5}g_{1}^{2}g_{2}^{2}+g_{2}%
^{4}\Big),\\[2mm]
16\pi^{2}\,\frac{d\mu^{2}}{dt}  &  =\mu^{2}\Big(6y_{t}^{2}+2\,\mathrm{Tr}%
[Y_{\nu}^{\dagger}Y_{\nu}+Y_{e}^{\dagger}Y_{e}+Y_{u}^{\dagger}Y_{u}%
+Y_{d}^{\dagger}Y_{d}]-\tfrac{9}{10}g_{1}^{2}-\tfrac{9}{2}g_{2}^{2}%
+6\lambda\Big).
\end{align*}
Qualitatively: big $y_{t}$ and big entries in $Y_{\nu}$ pull $\lambda$
downward; gauge terms and $\lambda^{2}$ push it upward.

Let $M_{R}$ have eigenvalues $M_{R_{1}}\leq M_{R_{2}}\leq M_{R_{3}}$. As you
run down:

\begin{itemize}
\item For $\mu> M_{R_{3}}$: all three $\nu_{R}$ active; use the RGEs above.

\item At $\mu=M_{R_{3}}$: integrate out the heaviest $\nu_{R}$ and match onto
an EFT with the Weinberg operator. Update $\kappa\to\kappa+Y_{\nu}%
^{(3)T}M_{R_{3}}^{-1}Y_{\nu}^{(3)} $.

\item Repeat at $\mu=M_{R_{2}}$, then $M_{R_{1}}$.

\item Below the lightest $M_{R}$: run the SM$+$$\kappa$ with $16\pi^{2}%
\,\frac{d\kappa}{dt} =\big(-3g_{2}^{2}+\tfrac{6}{5}g_{1}^{2}+6\,\mathrm{Tr}%
[Y_{u}^{\dagger}Y_{u}]+\cdots\big)\kappa-\tfrac{1}{2}\big[\kappa
(Y_{e}^{\dagger}Y_{e})+(Y_{e}^{\dagger}Y_{e})^{T}\kappa\big]. $
\end{itemize}

At the EW scale, $m_{\nu}=\tfrac{v^{2}}{2}\,\kappa$; diagonalize to extract
masses and the PMNS matrix (3 angles, 1 Dirac $+$ 2 Majorana phases for
Majorana neutrinos).

We now discuss IR checks and how the neutrino sector pushes them.

\begin{enumerate}
\item \textbf{EWSB:}\; require $\mu^{2}(\mu_{\mathrm{EW}})<0$ and $\lambda
(\mu_{\mathrm{EW}})>0$ near the minimum to yield $v=246\,\mathrm{GeV}$.

\item \textbf{Vacuum stability/metastability:}\; keep $\lambda(\mu)$ not too
negative up to $\Lambda$. Large $y_{t}$ and large $Y_{\nu}$ (above thresholds)
drive $\lambda$ down; raising $M_{R}$ shortens the interval where $Y_{\nu}$ acts.

\item \textbf{Perturbativity:}\; keep $g_{i},\lambda,y$ perturbative up to
$\Lambda$ (avoid Landau poles).

\item \textbf{Mass/mixing fits:}\ reproduce quark/charged-lepton masses and
CKM; via the see-saw obtain light neutrino masses and PMNS (3 angles, 1 Dirac
phase, $+$ 2 Majorana phases if Majorana).
\end{enumerate}

\emph{Typical tension:} for fixed light $m_{\nu}$ with $m_{\nu}\sim m_{D}%
^{2}/M_{R}$: larger $Y_{\nu}\Rightarrow$ larger $M_{R}$ (earlier decoupling,
better for $\lambda$ but pushes perturbativity); smaller $Y_{\nu}\Rightarrow$
smaller $M_{R}$ (longer window for $Y_{\nu}$ to act on $\lambda$, but easier
for perturbativity).

We conclude that at $\Lambda$ the ratios of gauge couplings and the
Higgs-sector inputs $(\xi,\lambda,\mu^{2})$ are fixed by geometry via $f_{i}$
and the Yukawa traces $a\cdots e$. The only knobs are the Yukawa moduli
(including $Y_{R}$). See-saw thresholds are the bridge tying neutrino physics
to the running of $\lambda$ and hence to vacuum (meta) stability.

\section{Why the Standard Model?}

In this section we will show \emph{why} the finite algebra that reproduces the
Standard Model (SM) in noncommutative geometry (NCG) is $\,A_{F}%
=\mathbb{C}\oplus\mathbb{H}\oplus M_{3}(\mathbb{C})\,\ $\cite{CC07b},
\cite{Chamseddine:2007ia}. Rather than postulate it, we derive it by
classifying finite spectral triples under the NCG axioms (reality, grading,
order-zero and order-one conditions) together with minimal physical
requirements (irreducibility, existence of a separating vector, massless
photon, and allowance for Majorana masses of neutrinos).

Spacetime is modeled as a product spectral triple $M\times F$, where $M$ is a
$4D$ Riemannian spin manifold and $F$ is a finite noncommutative space
specified by a finite spectral triple $(A,H,D;J,\gamma)$. Here: $A$ is an
involutive algebra represented on a Hilbert space $H$, $D$ is self-adjoint
(\textquotedblleft finite Dirac operator\textquotedblright), $J$ is the real
structure (charge conjugation), and $\gamma$ is the grading (chirality). The
order-zero condition $[a,b^{0}]=0$ for all $a,b\in A$ with $b^{0}=JbJ^{-1}$
and the order-one condition $\left[  [D,a],b^{0}\right]  =0$ encode
compatibility of the algebraic and metric data. The finite space is taken to
have KO-dimension $6\!\!\pmod 8$ so that the total KO-dimension $4+6=10$
avoids fermion doubling and permits both Weyl and Majorana conditions in the
fermionic action.

We proceed by classifying admissible finite geometries in three steps:

\begin{enumerate}
\item Classify irreducible pairs $(A,H,J)$ satisfying order-zero.

\item Impose a $\mathbb{Z}_{2}$ grading $\gamma$ compatible with KO-dimension
$6$ (i.e.\ $J\gamma=-\gamma J$).

\item Impose the order-one condition for a nontrivial $D$ that permits
off-diagonal blocks (Majorana couplings).
\end{enumerate}

A key role is played by the complexified algebra $A_{\mathbb{C}}%
\subset\mathcal{L}(H)$ and its center $Z(A_{\mathbb{C}})$, which splits the
analysis into two cases: $Z=\mathbb{C}$ and $Z=\mathbb{C}\oplus\mathbb{C}$.
First we consider the case $Z(A_{\mathbb{C}})=\mathbb{C}$, that corresponds to
a square-dimension option. Here there exists an irreducible solution precisely
when $\dim H=k^{2}$, with $A_{\mathbb{C}}=M_{k}(\mathbb{C})$ acting by left
multiplication and $J(x)=x^{\ast}$. Possible real forms are $M_{k}%
(\mathbb{C})$, $M_{k}(\mathbb{R})$, or $M_{a}(\mathbb{H})$ with $k=2a$.
However, this branch does not accommodate KO-dimension $6$ with the needed
relation $J\gamma=-\gamma J$, so it is discarded for $F$. Next we consider the
case $Z(A_{\mathbb{C}})=\mathbb{C}\oplus\mathbb{C}$, that corresponds to
twice-a-square and the Symplectic--Unitary choice. Here there exists an
irreducible solution iff $\dim H=2k^{2}$ with $A_{\mathbb{C}}=M_{k}%
(\mathbb{C})\oplus M_{k}(\mathbb{C})$ acting by left multiplication and
$J(x,y)=(y^{\ast},x^{\ast})$. Among the possible real forms, the
\emph{symplectic--unitary} option
\begin{equation}
A=M_{a}(\mathbb{H})\ \oplus\ M_{k}(\mathbb{C}),\qquad k=2a\ \text{(even)}%
\end{equation}
is compatible with KO-dimension $6$ and an appropriate grading. The smallest
even $k$ is $k=2$, which would yield only four fermions (phenomenologically
unacceptable: no color). The next is $k=4$, which predicts $16$ fermions per
generation---exactly the SM count including a right-handed neutrino.

We now impose $\mathbb{Z}_{2}$ grading to the KO-dimension $6$ algebra, to
obtain the even algebra
\begin{equation}
A_{\mathrm{ev}}\ \cong\ \mathbb{H}\ \oplus\ \mathbb{H}\ \oplus\ M_{4}%
(\mathbb{C}).
\end{equation}
The grading acts nontrivially on the quaternionic block and is the only option
compatible with KO-dimension $6$ and a nontrivial real structure on one side.
To allow \emph{Majorana} couplings (and hence a see-saw mechanism), the finite
Dirac operator $D$ must have off-diagonal blocks and must not commute with the
center of $A$. Imposing the order-one condition then \emph{uniquely} selects,
up to automorphisms of $A_{\mathrm{ev}}$, a maximal involutive subalgebra that
still admits such $D$:
\begin{equation}
A_{F}\ \cong\ \mathbb{C}\ \oplus\ \mathbb{H}\ \oplus\ M_{3}(\mathbb{C}).
\end{equation}
This is precisely the SM algebra: $\mathbb{C}$ for $U(1)_{Y}$, $\mathbb{H}$
for $SU(2)_{L}$, and $M_{3}(\mathbb{C})$ for $SU(3)_{c}$. Thus the SM algebra
is \emph{derived}, not assumed.

\bigskip From this we conclude, for the product spectral triple $M\times F$
with $A_{F}$ above, that:

\begin{itemize}
\item The unimodular part of the unitary group yields the SM gauge group
$SU(2)_{L}\times U(1)_{Y}\times SU(3)_{c}$.

\item \emph{Inner fluctuations} of the metric produce both the gauge
connections and a scalar multiplet identified with the Higgs doublet.

\item The \emph{spectral action} reproduces the bosonic Lagrangian (gravity +
Yang--Mills + Higgs), while the fermionic action $\frac{1}{2}\langle
J\Psi,D_{A}\Psi\rangle$ naturally includes Dirac and Majorana masses (see-saw)
for neutrinos.
\end{itemize}

Compared to GUTs, the NCG approach yields $16$ fermions per generation without
large Higgs representations, avoids extra leptoquark vectors (hence
suppressing proton decay channels of that type), and unifies gravity and gauge
interactions via a finite internal space rather than a Kaluza-Klein tower. At
the spectral cutoff scale (near unification), the framework fixes the relative
normalization of gauge couplings (GUT-style relation after canonical
normalization) and ties Higgs-sector parameters to Yukawa trace invariants.
These serve as UV boundary conditions for RG evolution to the electroweak
scale (with see-saw thresholds), constraining the Higgs quartic and the
neutrino sector in tandem.

Two conceptual questions, however, remain:

\begin{enumerate}
\item A deeper principle for the \emph{symplectic--unitary} choice among real forms.

\item An explanation for the number of generations $N=3$; it is
phenomenologically required (e.g.\ CP violation), but not derived here (nor
anywhere else).
\end{enumerate}

\section{Quantum gravity boundary terms from NCG}

In GR the bulk action $\int_{M}\!\sqrt{g}\,R$ is not variationally well-posed
for fixed boundary metric unless one adds the Gibbons--Hawking (GH) term
\begin{equation}
S_{\mathrm{GH}}=\frac{1}{8\pi G}\int_{\partial M}\!\sqrt{h}\,K,
\end{equation}
with a specific \emph{relative coefficient} to the bulk curvature term. In the
spectral action approach to noncommutative geometry (NCG), the classical
action is $\mathrm{Tr}\,f(D/\Lambda)$ and one does not insert $R$ or $K$ by
hand. We will show that \cite{Chamseddine:2007bm}, \cite{Chamseddine:2010ia},
given the \emph{natural self-adjoint} boundary condition for the Dirac
operator, the spectral action \emph{automatically} produces the GH term with
the correct sign and coefficient; the mechanism extends to the full
Standard-Model (SM) spectral triple and fixes Higgs boundary couplings as
well. Let $M$ be a compact 4D Riemannian manifold with smooth boundary
$\partial M$. The geometric Dirac operator has the form
\begin{equation}
D=\gamma^{\mu}(\nabla_{\mu}+\omega_{\mu}),
\end{equation}
augmented by the finite internal part in the SM. To ensure Hermiticity one
imposes the \emph{natural local} projector condition
\begin{equation}
P_{+}\psi\big|_{\partial M}=0,\qquad P_{+}=\tfrac{1+\chi}{2},\qquad\chi
=\gamma_{n}\gamma_{5},
\end{equation}
with $\gamma_{n}$ the Clifford contraction with the inward unit normal. On
manifolds with boundary the naive equivalence $D_{M}\oplus D_{F}%
\leftrightarrow\gamma_{5}(D_{M}\oplus D_{F})$ \emph{fails} because $\gamma
_{5}$ no longer anticommutes with $D_{M}$ at $\partial M$, so the above choice
is essential. One expands the spectral action for the Laplace-type operator
$P=D^{2}$ written as
\begin{equation}
P=-(g^{\mu\nu}\nabla_{\mu}\nabla_{\nu}+E),\qquad\Omega_{\mu\nu}=[\nabla_{\mu
},\nabla_{\nu}].
\end{equation}
where $\Omega_{\mu\nu}$ is the curvature operator. For the pure gravitational
Dirac operator in $4D$ (torsionless, minimal connection) one has $E=\tfrac
{1}{4}R\,\mathbf{1}$. With the projector $P_{+}$, the boundary contribution is
encoded by an endomorphism
\begin{equation}
S=-\tfrac{1}{2}\,K\,\chi,
\end{equation}
where $K$ is the trace of the extrinsic curvature $K_{ab}$ of $\partial M$. We
now use the heat kernel expansion for manifolds with boundary, using Mellin
moments $f_{k}$ of the cutoff function $f$, the action is
\[
\mathrm{Tr}\,f(P/\Lambda^{2})\simeq\sum_{n\geq0}f_{4-n}\,a_{n}(P).
\]
In $4D$, the Einstein--Hilbert term arises from $a_{2}(P)$. With boundary
terms taken into account, we have
\begin{align}
a_{2}(P)  &  \ \propto\ \int_{M}\!\sqrt{g}\,\mathrm{tr}\!\Big(E+\tfrac{1}%
{6}R\Big)\;+\;\int_{\partial M}\!\sqrt{h}\,\mathrm{tr}\!\Big(\tfrac{1}%
{3}K+S\Big),\\
a_{3}(P)  &  \ \propto\ \int_{\partial M}\!\sqrt{h}\,\mathrm{tr}%
\!\big(\alpha_{1}K^{2}+\alpha_{2}K_{ab}K^{ab}+\cdots\big),
\end{align}
with known numerical constants $\alpha_{i}$. Substituting $E=\tfrac{1}{4}R$
and $S=-\tfrac{1}{2}K\,\chi$, and tracing over spinors with the projector
structure, one isolates the $\int_{M}\!R$ and $\int_{\partial M}\!K$ pieces
with a \emph{fixed} relative coefficient. To be precise, we have for the
geometric Dirac operator:

\begin{enumerate}
\item Bulk: $E+\tfrac16 R=(\tfrac14+\tfrac16)R=\tfrac{5}{12}R$.

\item Boundary: $\tfrac13 K+S=\tfrac13 K-\tfrac12 K\,\chi$. After the trace
with $P_{+}$, only the scalar $K$ term survives with a definite coefficient.

\item Including the universal heat-kernel normalizations and $f_{2}$, the
$a_{2}$ level yields
\begin{equation}
\propto\int_{M}\!\sqrt{g}\,\tfrac{1}{2}R\;-\;\int_{\partial M}\!\sqrt{h}\,K.
\end{equation}

\item Upon canonical normalization of the Einstein term to $1/(16\pi G)$, the
boundary term becomes exactly $\tfrac{1}{8\pi G}\int_{\partial M}\!\sqrt
{h}\,K$: \emph{right sign and the 2:1 bulk--boundary factor fixed}.
\end{enumerate}

This hinges on using a \emph{Dirac-type} operator and the \emph{natural}
self-adjoint boundary condition; a generic Laplacian or different projector
would spoil the ratio.

We now turn our attention to the full SM\ spectral triple turning on inner
fluctuations in $M\times F$ (with $A_{F}=\mathbb{C}\oplus\mathbb{H}\oplus
M_{3}(\mathbb{C})$):

\begin{itemize}
\item Gauge bosons $B_{\mu},W_{\mu},G_{\mu}$ arise from connections.

\item The Higgs doublet $H$ arises as a scalar from the finite Dirac operator.

\item Squaring $D$ still yields Laplace type with
\[
E=\tfrac{1}{4}R\,\mathbf{1}\;+\;\text{(Higgs/Yukawa pieces)}\;+\;\text{(gauge
curvature terms)}.
\]
We first give the leading order terms on the bulk: collecting $a_{0}%
,a_{2},a_{4}$ gives the familiar spectral-action bulk Lagrangian (up to
$f_{0},f_{2},f_{4}$):

\item Gravity: cosmological constant, Einstein--Hilbert, and higher curvature
($C_{\mu\nu\rho\sigma}^{2}$, $R_{\mu\nu}R^{\mu\nu}$) with fixed linear combinations.

\item Gauge: canonical Yang--Mills terms for $U(1)_{Y}$, $SU(2)_{L}$,
$SU(3)_{c}$ with NCG-fixed relative normalization at the cutoff scale.

\item Higgs: conformally coupled kinetic + potential
\begin{equation}
\int_{M}\!\sqrt{g}\Big(|D_{\mu}H|^{2}-\tfrac{1}{6}R\,|H|^{2}-\mu^{2}%
|H|^{2}+\lambda|H|^{4}\Big),
\end{equation}
with $\lambda,\mu^{2}$ determined by spectral moments and Yukawa trace
invariants $a,b,c,d,e$.
\end{itemize}

With the same boundary condition, the scalar sector acquires the boundary
partner
\begin{equation}
\int_{\partial M}\!\sqrt{h}\,\Big(+\tfrac{1}{3}K\,|H|^{2}\Big),
\end{equation}
the exact companion to the bulk $-\tfrac{1}{6}R\,|H|^{2}$, mirroring the
gravitational 2:1 ratio ($K$ vs.\ $R$). Under natural boundary assumptions for
gauge fields (e.g.\ $A_{n}|_{\partial M}=0$), the leading vector boundary
terms vanish. Subleading geometric boundary invariants from $a_{3}$ and beyond
are present with fixed coefficients.

Nothing here is tuned: the operator $D$, projector $P_{+}$, and the
endomorphism $E$ are dictated by the spectral triple. The spectral action then
\emph{forces} the correct GH term and the Higgs boundary companion with the
right signs and relative factors. Different operators or boundary choices
would fail this test. The coefficient $a_{3}$ yields cubic boundary scalars
such as $K^{3}$, $K\,K_{ab}K^{ab}$, and curvature--$K$ mixes with fixed
weights. If the mass scale in $D$ is promoted to a dilaton, the
near-scale-invariant bulk form extends and additional dilaton boundary terms
appear, again with coefficients tied by the same mechanism.

In practice for a SM NCG space with boundary we perform the following steps:

\begin{enumerate}
\item Choose the SM spectral triple and impose $P_{+}\psi=0$ with $\chi
=\gamma_{n}\gamma_{5}$.

\item Square $D$ to Laplace type; extract $E,\Omega_{\mu\nu}$ and $S=-\tfrac12
K\chi$.

\item Feed these into the boundary heat-kernel formulas for $a_{n}$.

\item From $a_{2}$: bulk $\int\!\sqrt{g}\,R$ and boundary $-2\int\!\sqrt
{h}\,K$ (before canonical normalization) $\Rightarrow$ GH after fixing
$1/(16\pi G)$.

\item For Higgs: pair $-\tfrac{1}{6}R|H|^{2}$ with $+\tfrac{1}{3}K|H|^{2}$ on
$\partial M$.

\item Gauge: with natural boundary restrictions, leading boundary terms
vanish; bulk YM terms as usual.
\end{enumerate}

We conclude that with the natural self-adjoint boundary condition for the
Dirac operator, the spectral action of the SM spectral triple \emph{predicts}
the gravitational and scalar boundary sectors: it reproduces the GH term and
the Higgs $K|H|^{2}$ companion with the correct signs and the characteristic
2:1 bulk--boundary ratio, with no additional boundary parameters. We read this
as a confirmation that the spectral action in NCG is the correct setting for
quantum gravity.

\section{NCG as framework for unification: tensorial notation}

The complexity of the structure of the Dirac operator for the SM\ NCG space,
suggests that it will be helpful to represent the finite part of the spectral
triple $(A,H,D;J,\gamma)$ in explicit \emph{tensor/index} notation so that one
can compute componentwise \cite{CC10}. In this language the Standard Model
(SM) field content, the reduction of the algebra to
\[
A_{F}\cong\mathbb{C}\ \oplus\ \mathbb{H}\ \oplus\ M_{3}(\mathbb{C}),
\]
and the emergence of gauge fields and the Higgs from inner fluctuations become
transparent. We denote the Hilbert-space indices and the 16 fermions of one
family for a generic internal spinor by
\begin{equation}
\Psi^{M}\;=\;%
\begin{pmatrix}
\psi^{A}\\
\psi^{A^{\prime}}%
\end{pmatrix}
,\qquad\psi^{A^{\prime}}=(\psi^{A})^{c},
\end{equation}
so the particle and charge-conjugate sectors are explicit. The internal index
splits as
\begin{equation}
A=(\alpha,I).
\end{equation}

\begin{itemize}
\item \textbf{Weak/quaternionic index $\alpha$.} Originates from
$M_{2}(\mathbb{H})$. After the finite grading $\gamma$, it splits into
undotted/dotted pieces $\alpha= a, \dot a $ corresponding to left/right weak
doublet indices.

\item \textbf{Lepton/color index $I$.} Originates from $M_{4}(\mathbb{C})$ and
later reduces to $I=1$ (leptons) and $I=i=1,2,3$ (quarks) when $M_{4}%
(\mathbb{C})\rightarrow\mathbb{C}\oplus M_{3}(\mathbb{C})$.
\end{itemize}

With these conventions, the components of $\psi^{A}$ are the SM fermions of
one generation:
\begin{align}
\text{Leptons }(I=1):  &  \quad\psi^{\dot{1}}{}_{1}=\nu_{R},\quad\psi^{\dot
{2}}{}_{1}=e_{R},\quad\psi^{a}{}_{1}=\ell_{L}=(\nu_{L},e_{L}),\\
\text{Quarks }(I=i):  &  \quad\psi^{\dot{1}}{}_{i}=u_{Ri},\quad\psi^{\dot{2}%
}{}_{i}=d_{Ri},\quad\psi^{a}{}_{i}=q_{Li}=(u_{Li},d_{Li}).
\end{align}
Thus the required $16$ Weyl states per family (including $\nu_{R}$) are
accounted for.

Before reductions, the complexified finite algebra is
\begin{equation}
A_{\mathbb{C}}\simeq M_{2}(\mathbb{H})\ \oplus\ M_{4}(\mathbb{C}).
\end{equation}

\begin{itemize}
\item The \emph{real structure} $J$ exchanges particle/antiparticle blocks
(charge conjugation with appropriate transpose).

\item The \emph{grading} $\gamma$ separates left/right in the finite space and
splits $M_{2}(\mathbb{H}) \to\mathbb{H}\oplus\mathbb{H}$.
\end{itemize}

In the even part one may picture
\begin{equation}
A_{\mathrm{ev}}\;\cong\;\underbrace{\mathbb{H}}_{\text{acts on left }}%
\ \oplus\ \underbrace{\mathbb{H}}_{\text{acts on right }}\ \oplus
\ \underbrace{M_{4}(\mathbb{C})}_{\text{lepton+color}}.
\end{equation}
Two axioms constrain the representation:
\begin{align}
\lbrack a,\ b^{\circ}]  &  =0,\qquad b^{\circ}:=JbJ^{-1}\qquad
\text{(order-zero)},\\
\lbrack\lbrack D_{F},a],\ b^{\circ}]  &  =0\qquad\qquad\qquad\quad
\ \ \text{(order-one)}.
\end{align}
Write the finite Dirac operator in particle/antiparticle blocks
\begin{equation}
D_{F}\;=\;%
\begin{pmatrix}
Y & T\\[2pt]%
T^{\dagger} & \overline{Y}%
\end{pmatrix}
.
\end{equation}
Here $Y$ (diagonal) carries the Dirac Yukawas $Y_{u},Y_{d},Y_{e},Y_{\nu}$,
while the off-diagonal $T$ mixes the two central idempotent sectors and
implements the \emph{Majorana} (see-saw) slot for $\nu_{R}$.

The center of $A_{\mathbb{C}}$ has two minimal idempotents $e_{1},e_{2}$. If
$D_{F}$ commuted with the center, then $e_{1}D_{F}e_{2}=0$ (no mixing),
leaving an enlarged symmetry and forbidding Majorana masses. Retaining the
physically relevant case $[D_{F},Z]\neq0$ we show that consistency of
order-one forces:

\begin{enumerate}
\item the off-diagonal $T$ to be rank~1 (only a single SM singlet mixes);

\item the algebra to reduce to
\begin{equation}
\boxed{A_F \ \cong\ \mathbb{C}\ \oplus\ \mathbb{H}\ \oplus\ M_3(\mathbb{C})}.
\end{equation}

\end{enumerate}

Thus, allowing just enough mixing to realize a Majorana coupling collapses the
overlarge symmetry to the SM algebra.

Imposing \emph{unimodularity} ($\operatorname{Tr}A=0$ per connected component)
removes an unwanted abelian trace from the $M_{4}(\mathbb{C})$ block. In
components, the $4\times4$ block decomposes as
\begin{equation}
(V_{\mu})^{j}{}_{i}\;=\;\frac{i}{6}\,g_{1}\,B_{\mu}\,\delta^{j}{}%
_{i}\;+\;\frac{i}{2}\,g_{3}\,G_{\mu}^{m}(\lambda_{m})^{j}{}_{i},
\end{equation}
i.e.\ the correct \emph{hypercharge} $B_{\mu}$ plus the eight gluons $G_{\mu
}^{m}$ (Gell--Mann $\lambda_{m}$) acting on the color indices $i,j$. This
choice is also tied to anomaly cancellation.

Inner fluctuations of the metric are
\begin{equation}
A\;=\;\sum_{k}a_{k}[D_{F},b_{k}],\qquad D_{A}\;=\;D_{F}+A+JAJ^{-1}.
\end{equation}
They produce the SM bosonic gauge and Higgs fields with fixed high-scale normalizations.

First, the gauge fields from diagonal blocks are:

\begin{itemize}
\item \textbf{Weak isospin $SU(2)_{L}$} acts on the left doublet index $a$:
\begin{equation}
(A_{\mu})^{b}{}_{a} \;=\; -\frac{i}{2}\, g_{2}\, \gamma^{\mu}\, W_{\mu
}^{\alpha}\, (\sigma_{\alpha})^{b}{}_{a},
\end{equation}
with Pauli matrices $\sigma_{\alpha}$. Only left doublets couple, as in the SM.

\item \textbf{Hypercharge $U(1)_{Y}$} enters with different weights on
left/right and lepton/quark sectors, reproducing SM hypercharges $(Y(\ell
_{L})=-\tfrac12,\ Y(e_{R})=-1,\ Y(q_{L})=\tfrac16, \ldots)$.

\item \textbf{Color $SU(3)_{c}$} acts on the color index $i$ via the
generators $\lambda_{m}$, leaving leptons inert.
\end{itemize}

Next we give the Higgs sector from off-diagonal blocks: off-diagonal
components of $A$ mix dotted/undotted weak indices and assemble into a complex
$SU(2)$ doublet
\begin{equation}
H=%
\begin{pmatrix}
H_{1}\\
H_{2}%
\end{pmatrix}
,\qquad\nabla_{\mu}H_{a}=\Big(\partial_{\mu}-\frac{i}{2}g_{1}B_{\mu}%
\Big)H_{a}-\frac{i}{2}g_{2}\,W_{\mu}^{\alpha}(\sigma_{\alpha})_{a}{}^{b}H_{b},
\end{equation}
so $Y_{H}=+\tfrac{1}{2}$. The Higgs thus \emph{emerges} from finite Dirac
data; it is not an external input.

Finally we specify the Yukawa structure and the see-saw slot: the diagonal
block $Y$ contains the usual Dirac Yukawas $Y_{u},Y_{d},Y_{e},Y_{\nu}$. The
single nontrivial off-diagonal rank-1 slot $T$ implements the Majorana
coupling for $\nu_{R}$, realizing the see-saw relation $m_{\nu}\simeq
-m_{D}^{T}M_{R}^{-1}m_{D}$ after electroweak symmetry breaking. Other
Majorana-like mixings are forbidden by the order-one condition. The action on
the indices yields the familiar SM assignments for charges and representations:

\bigskip%

\begin{tabular}
[c]{lccc}\hline
Field & $SU(2)_{L}$ & $U(1)_{Y}$ & $SU(3)_{c}$\\\hline
Lepton doublet $\ell_{L}^{a}$ & $\mathbf{2}$ & $-\tfrac{1}{2}$ & $\mathbf{1}%
$\\
Right electron $e_{R}$ & $\mathbf{1}$ & $-1$ & $\mathbf{1}$\\
Right neutrino $\nu_{R}$ & $\mathbf{1}$ & $0$ & $\mathbf{1}$\\
Quark doublet $q_{Li}^{a}$ & $\mathbf{2}$ & $\tfrac{1}{6}$ & $\mathbf{3}$\\
Right up $u_{Ri}$ & $\mathbf{1}$ & $\tfrac{2}{3}$ & $\mathbf{3}$\\
Right down $d_{Ri}$ & $\mathbf{1}$ & $-\tfrac{1}{3}$ & $\mathbf{3}$\\
Higgs doublet $H_{a}$ & $\mathbf{2}$ & $\tfrac{1}{2}$ & $\mathbf{1}$\\\hline
\end{tabular}

\bigskip

If $D_{F}$ commutes with the center of $A_{\mathbb{C}}$, off-diagonal mixing
vanishes and an enlarged symmetry persists (e.g.\ SU(4)-type color); Majorana
masses are then absent. Requiring just enough center mixing to allow a single
Majorana slot (rank-1 $T$) \emph{and} imposing order-one leaves only elements
of the form
\begin{equation}
(\lambda,\ \lambda,\ q)\ \oplus\ (\lambda,\ m),
\end{equation}
i.e.\ one copy of $\mathbb{C}$ (hypercharge), one $\mathbb{H}$ (weak isospin),
and $M_{3}(\mathbb{C})$ (color). Thus the SM algebra is not assumed but
\emph{forced} by the first-order condition together with the minimal mixing
that realizes the see-saw (i.e. a right-handed neutrino Majorana mass term).
With tensorial notation in place one can square the fluctuated Dirac operator,
identify the heat-kernel data $(E,\Omega_{\mu\nu})$, and derive the spectral
action (gravity + YM + Higgs, and their boundary partners if $\partial
M\neq\varnothing$) together with high-scale relations among couplings to be
evolved by RGEs.

We first list the fields appearing in the fluctuations of the Dirac operator:

\begin{itemize}
\item Gauge field strengths: $B_{\mu\nu}$ for $U(1)_{Y}$, $W_{\mu\nu}^{\alpha
}$ for $SU(2)$, $V_{\mu\nu}^{m}$ for $SU(3)$; gauge couplings $g_{1}%
,g_{2},g_{3}$.

\item Higgs doublet $H_{a}$ (we write $H^{\dagger}H\equiv|H|^{2}$) with
covariant derivative $\nabla_{\mu}H_{a}$ defined below.

\item Real singlet $\sigma$ from the right-handed neutrino sector.

\item Yukawa matrices $k_{u},k_{d},k_{e},k_{\nu}$ and Majorana coupling
$k_{\nu_{R}}$; trace invariants $a,b,c,d,e$ (defined below). The total action
is
\begin{equation}
\frac{1}{2}\langle J\psi,D_{A}\psi\rangle\;+\;\mathrm{Trace}\,f\!\left(
\frac{D_{A}}{\Lambda}\right)  .
\end{equation}
The spectral part is expanded via Mellin moments $F_{k}$ of $f$:
\begin{align}
\mathrm{Trace}\,f\!\left(  \frac{D_{A}}{\Lambda}\right)   &  =\sum_{n\geq
0}F_{4-n}\,\Lambda^{4-n}\,a_{n},\qquad\nonumber\\
F_{4}  &  =2f_{4},\quad F_{2}=2f_{2},\quad F_{0}=f_{0},\quad F_{-2n}%
=(-1)^{n}F^{(n)}(0).
\end{align}
Here $a_{n}$ are the Seeley--DeWitt coefficients of the second-order operator
$D_{A}^{2}$. We rewrite the square $D_{A}^{2}$ in Laplace form
\begin{equation}
D_{A}^{2}=-(g^{\mu\nu}\nabla_{\mu}\nabla_{\nu}+E),\qquad\nabla_{\mu}%
=\partial_{\mu}+\omega_{\mu},
\end{equation}
starting from $D^{2}=-(g^{\mu\nu}\partial_{\mu}\partial_{\nu}+A_{\mu}%
\partial_{\mu}+B)$ and using
\begin{equation}
\omega_{\mu}=\tfrac{1}{2}g_{\mu\nu}(A^{\nu}+\Gamma^{\nu}),\qquad E=B-g^{\mu
\nu}(\partial_{\mu}\omega_{\nu}+\omega_{\mu}\omega_{\nu}-\Gamma_{\mu\nu}%
^{\rho}\omega_{\rho}),
\end{equation}
with curvature
\begin{equation}
\Omega_{\mu\nu}=\partial_{\mu}\omega_{\nu}-\partial_{\nu}\omega_{\mu}%
+[\omega_{\mu},\omega_{\nu}].
\end{equation}
The Higgs covariant derivative appearing later is
\begin{equation}
\nabla_{\mu}H_{a}=\Big(\partial_{\mu}-\frac{i}{2}g_{1}B_{\mu}\Big)\delta^{a}%
{}_{b}\,H_{b}-\frac{i}{2}g_{2}\,W_{\mu}^{\alpha}(\sigma_{\alpha})^{a}{}%
_{b}\,H_{b}.
\end{equation}
We condense the detailed computation of $D_{A}^{2}$ into three traced quantities:

\item Linear in $E$:%
\begin{equation}
-\frac{1}{2}\,\mathrm{Tr}(E)=4\big[\,12R+2a\,H^{\dagger}H+c\,\sigma
^{2}\,\big].
\end{equation}

\item Quadratic in $E$:
\end{itemize}

\begin{align}
\frac{1}{2}\,\mathrm{tr}(E^{2})=4\big[  &  \;5g_{1}^{2}B_{\mu\nu}^{2}%
+3g_{2}^{2}(W_{\mu\nu}^{\alpha})^{2}+3g_{3}^{2}(V_{\mu\nu}^{m})^{2}%
+3R^{2}+a\,R\,HH+\tfrac{1}{2}c\,R\,\sigma^{2}\nonumber\\
&  \;+2b\,(H^{\dagger}H)^{2}+2a\,|\nabla_{\mu}H_{a}|^{2}+4e\,H^{\dagger
}H\,\sigma^{2}+c\,(\partial_{\mu}\sigma)^{2}+d\,\sigma^{4}\big].
\end{align}

\begin{itemize}
\item Curvature of the connection:
\end{itemize}

\begin{equation}
\frac{1}{2}\,\mathrm{Tr}(\Omega_{\mu\nu}^{2})\Big|_{M}=4\big[\,-6R_{\mu\nu
\rho\sigma}^{2}-10g_{1}^{2}B_{\mu\nu}^{2}-6g_{2}^{2}(W_{\mu\nu}^{\alpha}%
)^{2}-6g_{3}^{2}(V_{\mu\nu}^{m})^{2}\big].
\end{equation}
With Dirac Yukawas $k_{u},k_{d},k_{e},k_{\nu}$ and the singlet (Majorana)
coupling $k_{\nu_{R}}$, we define
\begin{align}
a  &  =\mathrm{tr}\!\big(k_{\nu}^{\ast}k_{\nu}+k_{e}^{\ast}k_{e}+3(k_{u}%
^{\ast}k_{u}+k_{d}^{\ast}k_{d})\big),\\
b  &  =\mathrm{tr}\!\big((k_{\nu}^{\ast}k_{\nu})^{2}+(k_{e}^{\ast}k_{e}%
)^{2}+3((k_{u}^{\ast}k_{u})^{2}+(k_{d}^{\ast}k_{d})^{2})\big),\\
c  &  =\mathrm{tr}(k_{\nu_{R}}^{\ast}k_{\nu_{R}}),\qquad d=\mathrm{tr}%
\!\big((k_{\nu_{R}}^{\ast}k_{\nu_{R}})^{2}\big),\qquad e=\mathrm{tr}(k_{\nu
}^{\ast}k_{\nu}\,k_{\nu_{R}}^{\ast}k_{\nu_{R}}).
\end{align}
These invariants control all scalar coefficients in $a_{2}$ and $a_{4}$. From
the Gilkey formulas used before we have for $a_{0}:$%

\begin{equation}
a_{0}=\frac{1}{16\pi^{2}}\!\int\!d^{4}x\sqrt{g}\,\mathrm{Tr}(1)=\frac{24}%
{\pi^{2}}\!\int\!d^{4}x\sqrt{g}.
\end{equation}
and $a_{2}:$%

\begin{equation}
a_{2}=\frac{1}{16\pi^{2}}\!\int\!d^{4}x\sqrt{g}\ \mathrm{Tr}\!\left(
E+\frac{1}{6}R\right)  =-\frac{2}{\pi^{2}}\!\int\!d^{4}x\sqrt{g}%
\ \Big(R+\frac{1}{2}a\,H^{\dagger}H+\frac{1}{4}c\,\sigma^{2}\Big).
\end{equation}
and finaly $a_{4}:$%

\begin{align}
a_{4}=\frac{1}{2\pi^{2}}\!\int\!d^{4}x\sqrt{g}\  &  \left(  -\frac{3}{5}%
C_{\mu\nu\rho\sigma}^{2}+\frac{11}{30}R^{\ast}R^{\ast}+\frac{5}{3}g_{1}%
^{2}B_{\mu\nu}^{2}+g_{2}^{2}(W_{\mu\nu}^{\alpha})^{2}+g_{3}^{2}(V_{\mu\nu}%
^{m})^{2}\right. \nonumber\\
&  +\frac{1}{6}a\,R\,H^{\dagger}H+b\,(H^{\dagger}H)^{2}+\frac{1}{2}%
d\,\sigma^{4}+a\,|\nabla_{\mu}H_{a}|^{2}+2e\,H^{\dagger}H\,\sigma
^{2}\nonumber\\
&  \left.  +\frac{1}{12}c\,R\,\sigma^{2}+\frac{1}{2}c\,(\partial_{\mu}%
\sigma)^{2}-\frac{2}{5}R_{;\mu}{}^{\mu}-\frac{a}{3}\,(H^{\dagger}H)_{;\mu}%
{}^{\mu}-\frac{c}{6}(\sigma^{2})_{;\mu}{}^{\mu}\right)  .
\end{align}
We assemble the bosonic spectral action (to this order) as
\begin{equation}
S_{\text{bos}}\;=\;F_{4}\,\Lambda^{4}a_{0}\;+\;F_{2}\,\Lambda^{2}%
a_{2}\;+\;F_{0}\,a_{4}\;+\;F_{-2}\,\Lambda^{-2}a_{6}+\cdots,
\end{equation}
and treat this as an \emph{effective action} at the input scale $\Lambda$. We
write the fermionic piece explicitly (kinetic terms for all SM chiral
multiplets with their covariant derivatives, Dirac Yukawa couplings to $H$
including $\tilde{H}=i\sigma_{2}H^{\ast}$, and the singlet Majorana term
$\tfrac{1}{2}\,\nu_{R}^{T}C\,k_{\nu_{R}}\,\nu_{R}$+h.c.). We emphasize that
the spectral action implies a \emph{boundary relation} among gauge couplings
at $\Lambda$:
\begin{equation}
g_{3}^{2}(\Lambda)=g_{2}^{2}(\Lambda)=\frac{5}{3}\,g_{1}^{2}(\Lambda),
\end{equation}
to be used as an initial condition for renormalization-group evolution.

In summary we have provided, in closed form, and using only the objects
defined within it, the coefficients of the bosonic spectral action up to
$a_{4}$: gravity ($C^{2},\,R^{\ast}R^{\ast}$ and Einstein term via $a_{2}$),
the three Yang--Mills kinetic terms with fixed relative normalization, the
full Higgs sector (kinetic, curvature coupling, quartic), the singlet $\sigma$
(kinetic, curvature coupling, quartic), and the Higgs--$\sigma$ cross
coupling. All scalar coefficients are expressed through the Yukawa trace
invariants $a,b,c,d,e$. The fermionic action and the high-scale gauge coupling
relation are stated explicitly and serve as boundary data for RG evolution. We
emphasize that the mass parameter $M_{R}$ used for the mass of the
right-handed neutrino in the last section is now replaced with a dynamical
field $\sigma$, taken as a fluctuation of the Dirac operator in that entry.
This will turn to have very important consequences for the Higgs sector of the
spectral SM. In summary:

\begin{itemize}
\item The spectral action provides \emph{boundary conditions at $\Lambda$}:
unified gauge normalization $g_{3}^{2}=g_{2}^{2}=\tfrac{5}{3}g_{1}^{2}$, a
correlated Higgs quartic $\lambda(\Lambda)$, and gravitational terms from
$a_{2},a_{0}$.

\item Under one-loop SM running with a big desert, the gauge couplings nearly
unify but miss exact intersection; small higher-order corrections in the
spectral action can shift the high-scale matching conditions.

\item The Higgs sector is predictive at zeroth order once $\lambda(\Lambda)$
is fixed: RG evolution down to the electroweak scale gives $m_{H}$ (modulo
threshold and higher-loop effects not included here).
\end{itemize}

We now turn our attention to the \emph{top Yukawa coupling} $y_{t}$ and its
implications for the physical top mass. \medskip

\noindent\textbf{(i) Boundary condition at $\Lambda$ } We have specified a
\emph{high-scale boundary condition} for $y_{t}(\Lambda)$ that is tied to the
spectral-action relations among scalar and gauge sectors. This boundary
condition provides the constraint that fixes the \emph{starting value} of the
top Yukawa at $\mu=\Lambda$. We denote this input as
\begin{equation}
y_{t}(\Lambda)=y_{t}^{\text{(SA)}},\qquad\text{\textquotedblleft
SA\textquotedblright\ = value fixed by the spectral action.}%
\end{equation}
This boundary value is then used to evolve $y_{t}$ down with the RG equations.
\medskip

\noindent\textbf{(ii) One-loop RGE for $y_{t}$.} At one loop in the SM
(zeroth-order analysis of the section), the differential equation for the top
Yukawa reads
\begin{equation}
\frac{dy_{t}}{dt}=\frac{y_{t}}{16\pi^{2}}\left(  \frac{9}{2}\,y_{t}^{2}%
-\frac{17}{12}g_{1}^{2}-\frac{9}{4}g_{2}^{2}-8g_{3}^{2}\right)  ,\qquad
t=\ln\mu.
\end{equation}
Given $y_{t}(\Lambda)$, one integrates this equation \emph{simultaneously}
with the gauge-coupling RGEs to obtain $y_{t}(\mu)$ at lower scales. The QCD
term $-8g_{3}^{2}$ tends to pull $y_{t}$ downward as the scale decreases,
leading to the familiar quasi-infrared fixed-point behavior. From the running
Yukawa at the top mass scale, the $\overline{\text{MS}}$ running top mass is
\begin{equation}
m_{t}^{\overline{\mathrm{MS}}}(m_{t})=\frac{y_{t}(m_{t})\,v}{\sqrt{2}},
\end{equation}
with $v\simeq246$~GeV. To compare with experimental determinations quoted as a
pole mass, one adds the known QCD (and tiny QED) threshold corrections:
\begin{equation}
m_{t}^{\text{pole}}\;=\;m_{t}^{\overline{\mathrm{MS}}}(m_{t})\left[
1+\Delta_{\text{QCD}}+\Delta_{\text{EW}}\right]  ,\qquad\Delta_{\text{QCD}%
}=\frac{4}{3}\frac{\alpha_{s}(m_{t})}{\pi}+\mathcal{O}(\alpha_{s}^{2}).
\end{equation}
Higher-order terms can be included if greater precision is required. The
numerical integration of the differential equation gives a top quark of the
order of 173 Gev.

\section{Reselience of the spectral SM}

We now revisit the earlier phenomenological tensions between the spectral
action version of the Standard Model (SM) and the observed Higgs mass near
$125\,\mathrm{GeV}$ \cite{CC12}. We will show that these tensions disappear
once one retains the real scalar singlet $\sigma$, already present in the
noncommutative-geometry (NCG) construction due to the right-handed neutrino
sector presented in the last section. The Higgs--$\sigma$ sector stabilizes
the electroweak vacuum up to the unification scale while keeping the spectral
unification boundary conditions. In short, we will show that the spectral SM
is \textquotedblleft resilient\textquotedblright\ against the light Higgs,
provided the full scalar content implied by the spectral triple is used.

To make this section self-contained, and to be precise, we recall the
background geometry.

\begin{itemize}
\item \textbf{Spectral triple and product geometry.} The NCG framework
replaces the classical notion of a manifold with a spectral triple $(A,H,D)$
whose data encode geometry. The physical spacetime is taken as a product of a
commutative $4D$ Riemannian manifold and a finite noncommutative space
capturing the SM's internal structure. Bosons arise as inner fluctuations of
the Dirac operator; the action is the \emph{spectral action} $\mathrm{Tr}%
\,f(D/\Lambda)$, expanded by heat-kernel coefficients $a_{n}$.

\item \textbf{Finite algebra and fields.} The minimal algebra reducing under
the axioms is $A_{F}=\mathbb{C}\oplus\mathbb{H}\oplus M_{3}(\mathbb{C})$.
Inner fluctuations reproduce the $U(1)_{Y}\times SU(2)_{L}\times SU(3)_{c}$
gauge fields and a complex Higgs doublet. The presence of a right-handed
neutrino allows a Majorana mass and brings in a real scalar singlet $\sigma$
in the bosonic sector via the spectral action. In our earlier spectral-action
phenomenology we eroneously truncated this singlet; we will show that this
truncation is unjustified and leads to spurious tension with data.
\end{itemize}

We have seen that \textit{the scalar sector from the spectral action depends
on the Higgs field and a singlet:}

\begin{itemize}
\item \textbf{Higgs--singlet Lagrangian.} The Higgs doublet $H$ (with
$H^{\dagger}H\equiv|H|^{2}$) and the singlet $\sigma$ enter the bosonic sector
with canonical kinetic terms, quartics, and a cross coupling (scalar block),
\begin{align}
\mathcal{L}_{\mathrm{scal}}  &  \supset(D_{\mu}H)^{\dagger}(D^{\mu}%
H)+\tfrac{1}{2}(\partial_{\mu}\sigma)^{2}-\mu_{H}^{2}\,H^{\dagger}H-\tfrac
{1}{2}\mu_{\sigma}^{2}\,\sigma^{2}\nonumber\\
&  -\lambda_{h}(H^{\dagger}H)^{2}-\tfrac{1}{2}\lambda_{\sigma}\sigma
^{4}-\lambda_{h\sigma}H^{\dagger}H\,\sigma^{2}.
\end{align}
The three quartics $(\lambda_{h},\lambda_{h\sigma},\lambda_{\sigma})$ and
their boundary values at the input scale $\Lambda$ are not free: they are
fixed combinations of the spectral moments $f_{0},f_{2}$ and Yukawa trace
invariants of the finite geometry. In the \textquotedblleft dominant
Yukawa\textquotedblright\ approximation (top and neutrino sectors), we derive
simple closed expressions tying the quartics to one gauge coupling $g$ and a
ratio $n$ measuring the size of the neutrino Dirac Yukawa relative to the top
Yukawa. The key pattern is
\begin{align*}
&  \lambda_{\sigma}(\Lambda)\ \text{large},\qquad\\
&  \lambda_{h\sigma}(\Lambda)\ \text{moderately large and increasing with
}n,\qquad\\
&  \lambda_{h}(\Lambda)\ \text{moderate}.
\end{align*}
This structure is decisive for vacuum stability.

\item \textbf{Physical role of the singlet.} The scalar block coupling raises
the effective quartic along the Higgs direction and prevents $\lambda_{h}$
from turning negative at intermediate scales. The singlet itself can acquire a
large vacuum expectation value (vev) $w$ associated with the right-handed
neutrino scale. Mixing of $h$ and $\sigma$ reshuffles the physical mass
eigenstates and reduces the light states mass by a factor depending on
$\lambda_{h\sigma}^{2}/(\lambda_{h}\lambda_{\sigma})$.
\end{itemize}

\textit{To obtain the RGE we specify the boundary conditions and determine the
unification pattern:}

\begin{itemize}
\item \textbf{Gauge couplings.} The internal trace over the finite Hilbert
space fixes the relative normalization of gauge kinetic terms at $\Lambda$,
yielding the $SU(5)$-style relation $g_{3}^{2}=g_{2}^{2}=\tfrac{5}{3}%
\,g_{1}^{2}.$ This anchors the running; the spectral function choice allows
small departures from exact meeting due to higher-order corrections, which we
treat as an uncertainty in the matching scale.

\item \textbf{Quartics.} The spectral action at $\Lambda$ ties $(\lambda
_{h},\lambda_{h\sigma},\lambda_{\sigma})$ to $g$ and $n$, so once $n$ is
chosen, the quartics are fixed.

\item \textbf{Yukawas.} Working with the top Yukawa $k_{t}$ and neutrino Dirac
Yukawa $k_{\nu}=\sqrt{n}\,k_{t}$, we fix the high-scale $k_{t}(\Lambda)$ by
the spectral boundary conditions and runs it down together with gauge
couplings. The low-energy top mass is $m_{t}\simeq k_{t}(m_{t})v/\sqrt{2}$
after standard QCD threshold corrections.
\end{itemize}

This implies the determination of the renormalization group flow and vacuum stability:

\begin{itemize}
\item \textbf{RG equations.} To leading order, the system is the SM plus a
real singlet. The gauge couplings run with the familiar SM $\beta
$-coefficients; the Yukawas follow SM-like flows corrected by neutrino Yukawa
terms; and the quartics obey
\begin{align}
16\pi^{2}\,\beta_{\lambda_{h}}  &  \approx(12k_{t}^{2}+4k_{\nu}^{2}-3g_{1}%
^{2}-9g_{2}^{2})\lambda_{h}+24\lambda_{h}^{2}+2\lambda_{h\sigma}%
^{2}+\text{gauge}^{4}-6k_{t}^{4}-k_{\nu}^{4},\\
16\pi^{2}\,\beta_{\lambda_{h\sigma}}  &  \approx\lambda_{h\sigma}\!\left(
6k_{t}^{2}+2k_{\nu}^{2}-\tfrac{3}{2}g_{1}^{2}-\tfrac{9}{2}g_{2}^{2}\right)
+4\lambda_{h\sigma}(3\lambda_{h}+\tfrac{3}{2}\lambda_{\sigma}+2\lambda
_{h\sigma}),\\
16\pi^{2}\,\beta_{\lambda_{\sigma}}  &  \approx8\lambda_{h\sigma}%
^{2}+18\lambda_{\sigma}^{2}.
\end{align}
The large positive $\lambda_{\sigma}$ and $\lambda_{h\sigma}$ feed into
$\beta_{\lambda_{h}}$ and counter the negative top-Yukawa contribution that,
in the pure SM, tends to drive $\lambda_{h}$ below zero around $10^{8\text{--}%
10}$ GeV. Due to the participation of the scalar block, $\lambda_{h}$ remains
positive up to $\Lambda$.

\item \textbf{Stability condition.} Along the tree-level potential directions,
boundedness from below requires $\lambda_{h}>0,\ \lambda_{\sigma}%
>0,\ \lambda_{h\sigma}^{2}<\lambda_{h}\lambda_{\sigma}.$ The RG-evolved
quartics obey these inequalities throughout the energy range in the viable
parameter band.
\end{itemize}

\textit{We now study the Electroweak Symmetry Breaking and Mass Spectrum:}

\begin{itemize}
\item \textbf{Minimization.} With vevs $\langle H\rangle=v/\sqrt{2}%
,\ \langle\sigma\rangle=w$, the scalar mass matrix in the $(h,\sigma)$ basis
is $M^{2}=%
\begin{pmatrix}
2\lambda_{h}v^{2} & 2\lambda_{h\sigma}vw\\
2\lambda_{h\sigma}vw & 2\lambda_{\sigma}w^{2}%
\end{pmatrix}
.$ The eigenvalues are
\begin{equation}
m_{\pm}^{2}=\lambda_{h}v^{2}+\lambda_{\sigma}w^{2}\pm\sqrt{(\lambda_{h}%
v^{2}-\lambda_{\sigma}w^{2})^{2}+4\lambda_{h\sigma}^{2}v^{2}w^{2}}.
\end{equation}
For $w\gg v$, the light state (Higgs-like) is $m_{h}^{2}\simeq2\lambda
_{h}v^{2}\big(1-\lambda_{h\sigma}^{2}/(\lambda_{h}\lambda_{\sigma})\big),$
i.e.\ reduced by a mixing factor. The heavy state is dominantly singlet with
mass $m_{\sigma}^{2}\simeq2\lambda_{\sigma}w^{2}$.

\item \textbf{Higgs signal compatibility.} The mixing angle is small for $w\gg
v$; thus the light mass eigenstate retains SM-like couplings within current
experimental precision. The framework naturally accommodates $m_{h}%
\simeq125\,\mathrm{GeV}$ once the scalar sector suppression and RG evolution
are included.
\end{itemize}

\textit{Performing numerical study we have:}

\begin{itemize}
\item \textbf{Inputs.} Pick a matching scale $u_{\text{unif}}=\ln
(\Lambda/M_{Z})$ in a plausible range (we scan $u_{\text{unif}}\sim
25\text{--}35$), impose the spectral relations among gauge couplings at
$\Lambda$, set $(\lambda_{h},\lambda_{h\sigma},\lambda_{\sigma})$ by the
analytic boundary formulas in terms of $g$ and $n$, and choose $k_{t}%
(\Lambda)$ according to the spectral boundary condition (with $k_{\nu}%
=\sqrt{n}k_{t}$).

\item \textbf{Running and matching.} Integrate $g_{i},k_{t},k_{\nu}%
,\lambda_{h},\lambda_{h\sigma},\lambda_{\sigma}$ down to the weak scale at one
loop (higher-loop corrections will alter details but not the qualitative
conclusion). Fix $\mu_{H}^{2},\mu_{\sigma}^{2}$ to reproduce the vevs $v,w$.

\item \textbf{Findings.} There is a broad band in $(n,u_{\text{unif}})$ space where:

\begin{enumerate}
\item
\begin{enumerate}
\item the light eigenvalue matches $m_{h}\simeq125\,\mathrm{GeV}$,

\item all quartics stay positive and obey $\lambda_{h\sigma}^{2}<\lambda
_{h}\lambda_{\sigma}$,

\item the predicted top mass, after standard QCD threshold conversion from the
running mass, is close to but slightly lower than the observed value at one
loop (higher-order QCD effects expected to raise it).
\end{enumerate}
\end{enumerate}
\end{itemize}

The key qualitative message is that the earlier conflict (SM-only running
driving $\lambda_{h}<0$) is removed \emph{within the spectral model itself},
without ad hoc additions: the singlet and its interactions are intrinsic
outputs of the finite geometry.

\textit{To comment on the sensitivity of the results to theoretical
systematics we note:}

\begin{itemize}
\item \textbf{Dependence on $n$.} The parameter $n$ controls the neutrino
Dirac Yukawa relative to the top. Increasing $n$ raises $\lambda_{h\sigma}$
and strengthens stabilization, but also increases the Higgs--singlet mixing.
Viable solutions exist for moderate $n$, keeping mixing angles small and
consistent with Higgs data.

\item \textbf{Matching-scale uncertainty.} Because practical spectral
functions are not strict step functions, the exact unification point is not
sharply defined; we treat this as an uncertainty range in $u_{\text{unif}}$.
The viable band persists across this range, suggesting robustness.

\item \textbf{Loop order and thresholds.} The analysis is primarily one-loop.
Two-loop running and improved threshold matching (especially for QCD and the
top Yukawa) shift mass predictions by a few percent but do not spoil stability
or the ability to fit $m_{h}$. Right-handed neutrino thresholds associated
with $w$ can also modify RG flow; these can be incorporated but do not change
the main conclusion.
\end{itemize}

\textit{We thus have the follwing phenomenological implications:}

\begin{itemize}
\item \textbf{Higg- scalar signatures.} A heavy mostly-singlet scalar with
suppressed mixing remains possible. Direct production is limited by small
mixing; indirect constraints come from precision Higgs-coupling fits.

\item \textbf{See-saw and neutrinos.} The large singlet vev $w$ is aligned
with a type-I see-saw scale. The same finite-geometry ingredient that yields
Majorana masses stabilizes the Higgs sector---a notable conceptual link.

\item \textbf{Gravity sector.} Through $a_{2}$, the spectral action also
outputs a conformal coupling $-\frac{1}{6}R|H|^{2}$ and fixes combinations of
higher-curvature terms at $\Lambda$. These do not drive the electroweak-scale
phenomenology but are part of the UV completion picture.
\end{itemize}

In comparison, early spectral-action fits predicted a heavier Higgs where we
have effectively neglected the $\sigma$ dynamics. Once $\sigma$ and the Higgs
are retained with their proper boundary values, the model becomes consistent
with $m_{h}\sim125\,\mathrm{GeV}$ and electroweak vacuum stability. We have
shown that the fix is internal to the spectral framework and requires no
external fields beyond those implied by the spectral triple itself.

\textit{We have, however, the following limitations and outlook:}

\begin{itemize}
\item \textbf{Precision limitations.} One-loop running and simplified
threshold treatments limit precision for $m_{t}$ and $m_{h}$; two-loop
analyses and full threshold matching should be incorporated for sharper
predictions. We anticipate that these refinements move theoretical top-mass
predictions upward, improving agreement with data.

\item \textbf{Model-building extensions.} The spectral setup naturally allows
for additional scalar modes (e.g.\ a dilaton) in some versions; their roles in
stability and phenomenology can be explored. Likewise, embedding inflationary
dynamics or dark-sector couplings is a natural extension.
\end{itemize}

In conclusion, we demonstrated that the Spectral Standard Model is compatible
with a light Higgs and maintains vacuum stability to high scales when its full
scalar content---especially a real singlet $\sigma$ tied to the Majorana
sector---is included. The scalar couplings, fixed at the unification scale by
spectral boundary conditions, stabilizes the quartic and slightly reduces the
light scalar mass after mixing. Numerical scans confirm that with reasonable
inputs for the matching scale and Yukawa ratios, the model reproduces
$m_{h}\simeq125\,\mathrm{GeV}$ and a realistic top mass once standard
higher-order corrections are accounted for. The conclusion is not a patch but
a vindication of the full spectral-action framework: the apparent conflict
with a light Higgs arose from an inconsistent truncation that omitted a field
already present in the geometry.

\section{\bigskip Parity violation in the spectral action}

We can add to the spectral action a parity-odd piece so that parity (P) and CP
violating densities may appear (gravitational Pontryagin and non-abelian
$\theta$-terms). Besides the usual even spectral action, we allow a term of
the form \cite{CC10}
\begin{equation}
\mathrm{Tr}\!\left(  \gamma\,G\!\left(  \frac{D^{2}}{\Lambda^{2}}\right)
\right)  ,
\end{equation}
where $G$ is a test function, not necessarily related to the even function
$F$, and $\gamma=\gamma_{5}\otimes\gamma_{F}$ is the total grading. Because
$\gamma$ is odd, the Seeley--DeWitt expansion yields no contributions from
$a_{0}$ and $a_{2}$; the first non-vanishing terms appear at $a_{4}$, and
there are only two such contributions. The first term is proportional to
$\operatorname{Tr}(\gamma_{5}\gamma_{F}\,\Omega_{\mu\nu}^{2})$ gives the
gravitational Pontryagin density:
\begin{equation}
\frac{1}{16\pi^{2}}\;\frac{1}{12}\;\mathrm{Tr}\!\left(  \gamma_{5}\gamma
_{F}\,\Omega_{\mu\nu}^{2}\right)  \;=\;\frac{1}{16\pi^{2}}\;\varepsilon
^{\mu\nu\rho\sigma}\,R_{\mu\nu}{}^{ab}\,R_{\rho\sigma\,ab}\,(24-24)\;=\;0.
\end{equation}
Thus the gravitational parity-odd term cancels. (The \textquotedblleft%
$24-24$\textquotedblright\ encodes a left--right cancellation in the finite
Hilbert space. The second term is proportional to $\operatorname{Tr}%
(\gamma_{5}\gamma_{F}\,E^{2})$ gives the non-abelian $\theta$-type densities
with group-theoretic weights. One finds
\begin{equation}
\frac{1}{16\pi^{2}}\;\frac{1}{2}\;\mathrm{Tr}\!\left(  \gamma_{5}\gamma
_{F}\,E^{2}\right)  \;=\;-\frac{3}{4\pi^{2}}\;\varepsilon^{\mu\nu\rho\sigma
}\left(  2\,g_{1}^{2}\,B_{\mu\nu}B_{\rho\sigma}-2\,g_{2}^{2}\,W^{\alpha}%
{}_{\mu\nu}W_{\rho\sigma}^{\alpha}\right)  , \tag{7.4}%
\end{equation}

Thus the additional parity-odd contribution to the spectral action (up to
order $1/\Lambda^{2}$) is
\begin{equation}
\frac{3\,G_{0}}{8\pi^{2}}\;\varepsilon^{\mu\nu\rho\sigma}\!\left(
2\,g_{1}^{2}\,B_{\mu\nu}B_{\rho\sigma}-2\,g_{2}^{2}\,W^{\alpha}{}_{\mu\nu
}W_{\rho\sigma}^{\alpha}\right)  , \tag{7.5}%
\end{equation}
with $G_{0}=G(0)$. Here, the $B_{\mu\nu}B_{\rho\sigma}$ term is a surface
term, while $W^{\alpha}{}_{\mu\nu}W_{\rho\sigma}^{\alpha}$ is topological;
both are P/CP-violating.

A key observation is that, in this framework, both the gravitational
parity-odd term $\varepsilon^{\mu\nu\rho\sigma}R_{\mu\nu}{}^{ab}R_{\rho
\sigma\,ab}$ and the QCD $\theta$-term $\varepsilon^{\mu\nu\rho\sigma}V^{m}%
{}_{\mu\nu}V^{m}{}_{\rho\sigma}$ vanish at this order. Thus, the $\theta$
parameter is naturally zero and can only be generated by higher-order
interactions. The reason is a left--right symmetry graded by $\gamma_{F}$ in
each of these sectors, producing an exact cancellation between left-handed and
right-handed contributions; equivalently, $\mathrm{Tr}(\gamma_{F})=0$, so the
index of the full Dirac operator (with total grading) vanishes. To fully
resolve the strong CP problem, one further imposes the reality condition on
the up- and down-quark mass matrices:
\begin{equation}
\det k_{u}\;\det k_{d}\;\;=\;\;\text{real}.
\end{equation}
If such a condition can be imposed naturally, then
\begin{equation}
\theta_{QT}\;+\;\theta_{QCD}\;=\;0
\end{equation}
at tree level, and loop corrections would only modify this by amounts smaller
than $10^{-9}$.

\section{The uncanny precision of the spectral action}

The spectral action of a (Euclidean) spectral triple with Dirac operator $D$
and cutoff profile $f$ is
\begin{equation}
S_{f}(D,\Lambda)=\mathrm{Tr}\,f(D/\Lambda).
\end{equation}
In four dimensions, the large--$\Lambda$ asymptotic expansion reads
\begin{equation}
\mathrm{Tr}\,f(D/\Lambda)\sim2\Lambda^{4}f_{4}\,a_{0}+2\Lambda^{2}f_{2}%
\,a_{2}+f_{0}\,a_{4}+\sum_{k\geq1}\Lambda^{-2k}f_{-2k}\,a_{4+2k},
\end{equation}
where $a_{2j}$ are the Seeley--DeWitt coefficients of $D^{2}$ and $f_{4}%
=\int_{0}^{\infty}f(u)u^{3}du$, $f_{2}=\int_{0}^{\infty}f(u)u\,du$,
$f_{0}=f(0)$, $f_{-2k}=\frac{(-1)^{k}}{(2k)!}k!\,f^{(2k)}(0)$. For a genuine
(flat) cutoff with all even derivatives vanishing at the origin, only three
local terms survive:
\begin{equation}
\mathrm{Tr}\,f(D/\Lambda)\sim2\Lambda^{4}f_{4}\,a_{0}+2\Lambda^{2}f_{2}%
\,a_{2}+f(0)\,a_{4}. \label{eq:three_term}%
\end{equation}
The $\Lambda^{4}$ piece encodes the cosmological term (volume), the
$\Lambda^{2}$ piece reproduces Einstein--Hilbert, and $a_{4}$ contains
curvature--squared/topological terms .

A conceptual resolution to the dominance of the cosmological term is
kinematic: the Riemannian volume form is fixed by a purely spectral identity,
so varying at fixed total volume leads to unimodular gravity and the
$\Lambda^{4}$ contribution drops from the field equations. The paper asks
\emph{how accurate} the local three--term approximation~\eqref{eq:three_term}
is compared to the full nonlocal spectral action, and quantifies the remainder
in explicit backgrounds \cite{Chamseddine:2009drp}.

\noindent First for $S^{3}$ the round 3--sphere, we consider the physically
motivated thermal Euclidean space--time $S_{a}^{3}\times S_{\beta}^{1}$ with
spatial radius $a$ and time circle radius $\beta$ (inverse temperature). The
cutoff scale is~$\Lambda$ and the geometric inner diameter is $\mu
=\inf(a,\beta)$.

Specializing to $h(x)=P(\pi x)\,e^{-\pi x}$ with $\deg P=d$, the Fourier
transforms can be evaluated in closed form. If $\mu\Lambda$ is moderately
large relative to $\sqrt{d(1+\log d)}$, the remainder obeys a Gaussian bound
\begin{equation}
|\varepsilon(\Lambda)|\;\leq\;C\,\exp\!\left[  -\frac{\pi^{2}}{2}(\mu
\Lambda)^{2}\right]  .
\end{equation}
Hence the three--term truncation is \emph{astronomically precise} for
realistic scales; e.g.\ with $\mu$ of cosmological size (Hubble radius in
Planck units) and $\Lambda$ near the Planck scale one gets accuracies far
beyond $10^{-60}$.

A key corollary of the Poisson--summation analysis is that, on $S^{3}\times
S^{1}$, \emph{all} $a_{2n}$ with $n\geq2$ vanish. The authors also verify this
\emph{locally} by a direct heat--kernel computation: delicate cancellations
set $a_{4}=a_{6}=0$. This is special to $S^{3}\times S^{1}$ (and relies on
smoothness properties entering the Fourier analysis); for $S^{4}$, functions
like $|x|e^{-tx^{2}}$ obstruct such rapid decay.

\medskip\textbf{Interpretation.} For smooth functions on $S^{3}\times S^{1}$
the spectral action is fully determined by its cosmological and Einstein
terms, with the rest suppressed beyond any power of~$\Lambda^{-1}$. This is
what is meant by \textquotedblleft uncanny precision.\textquotedblright

One can frame accuracy in a scale--free way: for cutoff $f$, the quality of
the local approximation is governed by the number $N(\Lambda)$ of eigenvalues
below $\Lambda$. Flatness of $f$ at the origin implies an error $O(N^{-k})$
for any $k>0$. For realistic four--volumes (today's Universe) one estimates
$N\sim10^{214}$; even a conservative $N^{-1/2}$ would be smaller than
$10^{-107}$. The explicit $S^{3}\times S^{1}$ analysis improves this to
\emph{Gaussian} suppression in $\mu\Lambda$.

Consider the perturbed trace
\begin{equation}
\mathrm{Tr}\,h\!\left(  \left(  D^{2}+H^{2}\right)  /\Lambda^{2}\right)  .
\end{equation}
For smooth cutoff $h$ that is \emph{constant} in a neighborhood of the origin,
the Higgs dependence is captured exactly by shifting the radial integrals and
Taylor expanding under the integral sign. Writing $x=H^{2}/(2\Lambda^{2})$,
one recovers:

\begin{itemize}
\item the Higgs mass term from the $\Lambda^{2}$ piece ($a_{2}$),

\item the $RH^{2}$ coupling and the quartic $H^{4}$ from the order--one piece
($a_{4}$),

\item the same super--small remainder as before.
\end{itemize}

Thus, \emph{switching on a constant Higgs background does not spoil the
uncanny precision}, and the expected structure of the Higgs sector in the
local expansion is confirmed and sharpened.

Within noncommutative geometry, the same spectral action on $M\times F$ (where
$F$ is a finite noncommutative space) augments $a_{4}$ by a positive
Yang--Mills term for inner fluctuations of the metric (massless gauge bosons,
fermions in the appropriate representation). This is the standard path by
which gravity couples to gauge sectors in the spectral approach.

As for the cosmological term, the spectral (kinematic) relation fixing the
volume form means that variations at fixed volume implies that the
$\Lambda^{4}$ piece arise as an integration constant, while the Einstein term
governs the classical dynamics. This dovetails with the broader program we
carry later in which volume quantization renders the leading spectral term
effectively topological.

For any Schwartz test function $h$,
\begin{equation}
\mathrm{Tr}\,h(D^{2}/\Lambda^{2})=C_{4}\Lambda^{4}+C_{2}\Lambda^{2}%
+\varepsilon(\Lambda),
\end{equation}
with explicit constants $C_{4},C_{2}$ given by the radial integrals and
$|\varepsilon(\Lambda)|=O(\Lambda^{-k})$ for all $k$. Moreover, $a_{2n}=0$ for
$n\geq2$ on $S^{3}\times S^{1}$ (checked both globally and locally). (see also
\cite{Chamseddine:2011ix})

For smooth cutoffs constant near the origin, a constant $H$ contributes
exactly the expected $m_{H}^{2}$, $RH^{2}$, and $\lambda H^{4}$ pieces, with
the same remainder estimate.

Methodologically, we provided a nonperturbative check that the local
heat--kernel truncation is trustworthy on nontrivial backgrounds---especially
the thermal space--time $S^{3}\times S^{1}$. Since the spectral action is
generally nonlocal, this control is crucial and shows that in key settings the
difference to the local form is super--small.

Conceptually, together with results on volume quantization and fixed volume
gravity, the findings support the view that the cosmological term in the
spectral action is kinematically fixed/quantized rather than dynamically
tuned, while the Einstein term dictates the dynamics; and that coupling to
gauge sectors follows naturally in product geometries $M\times F$.

In summary, for $S^{3}$ and, crucially, for $S^{3}\times S^{1}$, the spectral
action with smooth cutoff profiles is, to all practical purposes,
\emph{exactly} equal to the sum of its cosmological and Einstein--Hilbert
terms; higher Seeley--DeWitt terms vanish on $S^{3}\times S^{1}$ and the
remainder is suppressed faster than any power (Gaussian for analytic cutoffs).
The same precision persists when a constant Higgs background is included,
yielding precisely the expected $m_{H}^{2}$, $RH^{2}$, and $H^{4}$ structures.

\section{Away with order one condition}

A real spectral triple $(A,H,D;J)$ consists of a unital involutive algebra $A$
faithfully represented on a Hilbert space $H$, a (typically unbounded,
self-adjoint) Dirac operator $D$ with compact resolvent and bounded
commutators $[D,a]$ for $a\in A$, and an antilinear isometry $J:H\rightarrow
H$ (the real structure) satisfying the KO-dimension sign rules, in particular
$JDJ^{-1}=\varepsilon D$ with $\varepsilon\in\{\pm1\}$ determined by the
KO-dimension. In commutative examples, $A=C^{\infty}(M)$, $H=L^{2}(M,S)$,
$D=\gamma^{\mu}D_{\mu}$, and $J$ is charge conjugation.

Given a \emph{universal one-form} $\omega=\sum_{j}a_{j}\,\delta(b_{j}%
)\in\Omega^{1}(A)$, its representation on $H$ is $A^{(1)}=\sum_{j}%
a_{j}[D,b_{j}].$ When the \emph{first order condition} $[[D,a],JbJ^{-1}%
]=0\quad\forall a,b\in A$ holds, the usual fluctuated Dirac operator is
\begin{equation}
D^{\prime}=D+A+\varepsilon JAJ^{-1},\qquad A=\sum_{j}a_{j}[D,b_{j}],\quad
JDJ^{-1}=\varepsilon D. \label{3}%
\end{equation}
The gauge group acts by unitaries $U=uJuJ^{-1}$ (with $u\in U(A)$). Covariance
of ($\ref{3}$) is then straightforward.

If the first order condition fails, the simple covariance argument breaks: the
adjoint action of $U$ on the RHS of ($\ref{3}$) is \emph{not} equal to
replacing $A$ by the universal-calculus transform $A\mapsto\gamma
_{u}(A):=u[D,u^{\ast}]+uAu^{\ast}.$ Thus the standard linear prescription is
not stable under gauge transformations without the first order condition. We
thus analyze a minimal nonlinear correction that restores gauge covariance and
transitivity. One wants
\begin{align*}
&  D+A+\varepsilon JAJ^{-1}\ \xrightarrow{U\cdot U^*}\ U(D+A+\varepsilon
JAJ^{-1})U^{\ast}\\
&  \overset{?}{=}\ D+A^{u}+\varepsilon JA^{u}J^{-1},\qquad A^{u}:=u[D,u^{\ast
}]+uAu^{\ast}.
\end{align*}
But the derivation uses $[JuJ^{-1},A]=0$ for all one-forms $A$, which is
equivalent to the first order condition. Hence, without it, extra commutators
appear and the equality fails.

To solve this problem \cite{CCSW}, let $E$ be a finite projective right
$A$-module with a universal connection $\nabla:E\rightarrow E\otimes_{A}%
\Omega^{1}(A).$ Representing universal 1-forms by commutators with $D$, define
the operator on $E\otimes_{A}H$ by $(1\otimes_{\nabla}D)(v\otimes\xi
):=\nabla_{D}(v)\,\xi+v\otimes D\xi.$ A first structural result is the
associativity-like identity
\begin{equation}
(1\otimes_{\nabla}D)\otimes_{\nabla}1\;=\;1\otimes_{\nabla}(D\otimes_{\nabla
}1), \label{4}%
\end{equation}
which ensures well-posedness and Morita covariance of the construction.

Take $E=A$ and a self-adjoint universal form $A=\sum_{j}a_{j}\,\delta(b_{j}).$
Let $\hat{a}:=Ja^{\ast}J^{-1}$ and $\hat{b}:=Jb^{\ast}J^{-1}$ , and denote the
right-action representatives (via $J$). Then the fluctuated Dirac operator
acquires three contributions beyond $D$:
\begin{equation}
D^{\prime}=D+A^{(1)}+\widehat{A}^{(1)}+A^{(2)}%
\end{equation}
with
\begin{align}
A^{(1)}  &  =\sum_{j}a_{j}[D,b_{j}],\\
\widehat{A}^{(1)}  &  =\sum_{j}\hat{a}_{j}[D,\hat{b}_{j}]\;=\;\varepsilon
\,JA^{(1)}J^{-1}\quad(\text{since }JDJ^{-1}=\varepsilon D),\\
A^{(2)}  &  =\sum_{j}\hat{a}_{j}\,[A^{(1)},\hat{b}_{j}]\;=\;\sum_{j,k}\hat
{a}_{j}a_{k}\,[[D,b_{k}],\hat{b}_{j}]\;=\;\varepsilon\,JA^{(2)}J^{-1}.
\end{align}
If the first order condition holds, then $[[D,b_{k}],\hat{b}_{j}]=0$ and hence
$A^{(2)}=0$; one recovers the standard linear formula ($\ref{3}$). Thus
$A^{(2)}$ is the minimal quadratic correction that is \emph{only} active when
the first order condition fails.

To prove that the corrected fluctuations transform in a gauge covariant way,
define the universal-calculus gauge map $\gamma_{u}(A):=u\,\delta(u^{\ast
})+uAu^{\ast},\ u\in U(A).$ Then one can show that
\begin{equation}
U\,D(A)\,U^{\ast}\;=\;D\big(\gamma_{u}(A)\big),\qquad U=uJuJ^{-1}, \label{11}%
\end{equation}
i.e.\ the corrected fluctuation is exactly covariant. A useful algebraic
identity used in the proof is: for any inclusion $A\subset B$ and $T\in B$,
\begin{equation}
\sum_{j=0}^{n}a_{j}^{\prime}\,[T,b_{j}^{\prime}]\;=\;u[T,u^{\ast
}]\;+\;u\Big(\sum_{j=1}^{n}a_{j}[T,b_{j}]\Big)u^{\ast}, \label{12}%
\end{equation}
where $a_{0}^{\prime}=u$, $b_{0}^{\prime}=u^{\ast}$, and $\{a_{j}^{\prime
},b_{j}^{\prime}\}_{j\geq1}$ are the gauge-transformed coefficients. Identity
($\ref{12}$) organizes the transformed linear pieces $A^{(1)}$, $\widehat
{A}^{(1)}$ and the nonlinear piece $A^{(2)}$ so that the total $A^{(1)}%
+\widehat{A}^{(1)}+A^{(2)}$ transforms into $A^{(1)}(\gamma_{u}(A))+\widehat
{A}^{(1)}(\gamma_{u}(A))+\!A^{(2)}(\gamma_{u}(A)).$

Beyond covariance, one can abstract the construction as an action of a
\emph{semigroup} $\mathrm{Pert}(A)$ on operators in $H$. This semigroup
extends $U(A)$ and depends only on $A$ (not on $H,D$). For real triples, there
is a natural semigroup morphism $\mu:\mathrm{Pert}(A)\rightarrow
\mathrm{Pert}(A\otimes\widehat{A}),\ \mu(\mathcal{A})=\mathcal{A}%
\otimes\widehat{\mathcal{A}},$ which ensures that \emph{fluctuations of
fluctuations are fluctuations} (transitivity). Concretely, composing two inner
perturbations corresponds to multiplying their representatives in
$\mathrm{Pert}(A)$. A convenient representation of universal 1-forms uses
$A\otimes A^{\mathrm{op}}$: define $\eta(\sum_{j}a_{j}\otimes b_{j}%
^{\mathrm{op}}):=\sum_{j}a_{j}\,\delta(b_{j}),$ with the constraint $\sum
_{j}a_{j}b_{j}=1$. Then gauge acts by $\sum_{j}a_{j}\otimes b_{j}%
^{\mathrm{op}}\mapsto\sum_{j}(ua_{j})\otimes(b_{j}u^{\ast})^{\mathrm{op}}.$
This viewpoint streamlines both the proof of surjectivity of $\eta$ and the
bookkeeping of gauge transformations.

\section{Emergence of the Pati-Salam model}

We have now to repeat the classification of the finite geometry without
imposing the first-order condition \cite{CCS13b}. The product geometry
$M_{4}\times F$ requires that fermions avoid doubling. This entails that the
finite spinors must satisfy both Majorana and Weyl conditions, which fixes the
KO-dimension of the finite part to be $6\pmod{8}$. Together with the
zeroth-order condition $[a,b^{\circ}]=0$ for all $a,b\in A$, one deduces that
the complexified algebra must have a two-dimensional center:
\begin{equation}
Z(A_{\mathbb{C}})=\mathbb{C}\oplus\mathbb{C}.
\end{equation}
Thus the finite algebra must be of the form
\begin{equation}
A\simeq M_{k}(\mathbb{C})\oplus M_{k}(\mathbb{C}),
\end{equation}
and the Hilbert space dimension is necessarily a perfect square. Imposing a
symplectic condition on the first summand forces $k=2a$ and identifies it with
quaternionic matrices $M_{a}(\mathbb{H})$. The existence of chirality requires
$a$ to be even. Therefore, the number of fundamental fermions is $4a^{2}$. The
minimal realistic case arises for $k=4$, giving
\begin{equation}
A=\mathbb{H}_{R}\oplus\mathbb{H}_{L}\oplus M_{4}(\mathbb{C}).
\end{equation}
If the first-order condition on $D_{F}$ is not imposed, inner fluctuations
acquire quadratic contributions. More importantly, the group of inner
automorphisms of $A$ then becomes
\begin{equation}
SU(2)_{R}\times SU(2)_{L}\times SU(4),
\end{equation}
i.e. the Pati--Salam group. It acts on the 16 fermions as $(2,2,4)$. The
scalar fields (Higgs multiplets) that arise from inner fluctuations depend on
whether the unperturbed $D_{F}$ obeys the order-one condition relative to the
Standard Model subalgebra. In general, the following Higgs multiplets appear:
\[
(2_{R},2_{L},1),\quad(2_{R},1_{L},4),\quad(1_{R},1_{L},1\oplus15).
\]
If the order-one condition is not satisfied, additional multiplets may appear,
such as
\[
(3_{R},1_{L},10),\quad(1_{R},1_{L},6),\quad(2_{R},2_{L},1\oplus15).
\]
Furthermore, when Yukawa couplings satisfy $k_{\nu}=k_{u}$ and $k_{e}=k_{d}$,
the $(1\oplus15)$ of $SU(4)$ decouples.

\bigskip We conclude\textbf{:} The constraints select the finite algebra
$A=\mathbb{H}_{R}\oplus\mathbb{H}_{L}\oplus M_{4}(\mathbb{C})$. Abandoning the
first-order condition naturally enhances the gauge group to Pati--Salam and
yields a highly structured Higgs sector. Note that, we assumed that there is a
symplectic structure imposed on one of the algebras.

In order to facilitate working with the operator structure, we introduce a
tensorial notation for the Hilbert space indices. Vectors in the Hilbert space
are written as
\begin{equation}
\Psi_{M}=%
\begin{pmatrix}
\psi_{A}\\
\psi_{A^{\prime}}%
\end{pmatrix}
,\qquad\psi_{A^{\prime}}=\psi_{A}^{c},
\end{equation}
where primed indices refer to the charge-conjugated sector. The composite
index $A$ splits as $A=\alpha I$, where $\alpha$ corresponds to the
quaternionic component and $I$ to the $M_{4}(\mathbb{C})$ part.

The grading decomposes $M_{2}(\mathbb{H})\rightarrow\mathbb{H}_{R}%
\oplus\mathbb{H}_{L}$, splitting $\alpha$ into dotted/undotted indices.
Likewise, $M_{4}(\mathbb{C})\rightarrow\mathbb{C}\oplus M_{3}(\mathbb{C})$
splits $I$ into $1,i$. Thus the fermions appear as
\begin{equation}
(\nu_{R},u_{R}^{i};\ \nu_{L},u_{L}^{i};\ e_{R},d_{R}^{i};\ e_{L},d_{L}^{i}),
\end{equation}
arranged into $16$ components. The finite algebra $A=M_{4}(\mathbb{C})\oplus
M_{4}(\mathbb{C})$ acts block-diagonally on $\Psi_{M}$. The reality operator
$J$ swaps primed and unprimed components, conjugates complex numbers, and
ensures $a^{\circ}=Ja^{\ast}J^{-1}$. The zeroth-order condition is manifest in
this representation. The commutator $[[D,a],b^{\circ}]$ can be expressed in
block form. Requiring its vanishing leads to the restriction of the algebra to
the Standard Model subalgebra
\begin{equation}
\mathbb{C}\oplus\mathbb{H}\oplus M_{3}(\mathbb{C}),
\end{equation}
which is the largest subalgebra that allows a nonzero mixing term between the
primed and unprimed sectors. Physically, this term corresponds to the Majorana
mass of right-handed neutrinos. Imposing a symplectic isometry reduces
$M_{4}(\mathbb{C})$ to $M_{2}(\mathbb{H})$. Commutativity with the grading
further splits this into $\mathbb{H}_{R}\oplus\mathbb{H}_{L}$. In this basis,
the finite Dirac operator $D_{F}$ takes block form, with Yukawa matrices
\begin{equation}
k_{\nu},\quad k_{e},\quad k_{u},\quad k_{d}%
\end{equation}
for Dirac masses and $k_{\nu_{R}}$ for the Majorana mass of right-handed
neutrinos. The inner fluctuations $A^{(1)}=\sum a[D,b]$ are compactly
expressed in this index language. The Higgs multiplets arise directly from
these commutators and their block structure:
\begin{align}
\phi_{a}^{b}  &  \sim(2_{R},2_{L},1),\\
\Delta_{aI}  &  \sim(2_{R},1_{L},4),\\
\Sigma_{I}^{J}  &  \sim(1_{R},1_{L},1\oplus15).
\end{align}
The presence of $\Sigma$ depends on Yukawa relations: it decouples if $k_{\nu
}=k_{u}$ and $k_{e}=k_{d}$.

\bigskip\noindent\textbf{In summary:} The tensor notation condenses the
$384\times384$ matrix structure into a compact index language that makes
visible the origin of the Standard Model subalgebra, the Majorana mass term,
and the emergence of the Pati--Salam Higgs multiplets.

On the product manifold $M\times F$ the Dirac operator splits as
\begin{equation}
D\;=\;\gamma^{\mu}D_{\mu}\otimes\mathbf{1}\;+\;\gamma_{5}\otimes D_{F}%
,\qquad\gamma^{\mu}D_{\mu}\;=\;\gamma^{\mu}\!\left(  \partial_{\mu}+\tfrac
{1}{4}\omega^{ab}{}_{\mu}\gamma_{ab}\right)  ,
\end{equation}
with $D_{F}$ the finite (internal) Dirac operator encoding Yukawa structures.
Dropping the first-order condition enlarges inner fluctuations to include a
quadratic piece:
\begin{equation}
D_{A}\;=\;D\;+\;A^{(1)}\;+\;JA^{(1)}J^{-1}\;+\;A^{(2)}, \label{eq:fluctuatedD}%
\end{equation}
where (sums implicit over index $j$)
\begin{equation}
A^{(1)}\;=\;\sum a_{j}[D,b_{j}],\qquad A^{(2)}\;=\;\sum\hat{a}_{j}%
\big[A^{(1)},\hat{b}_{j}\big],\qquad\hat{a}:=JaJ^{-1},\;\hat{b}:=JbJ^{-1}.
\end{equation}
For $U=uJuJ^{-1}$ with $u\in U(A)$ one has the gauge action $D_{A}\mapsto
UD_{A}U^{\ast}$. In the universal calculus this induces
\begin{align}
A^{(1)}  &  \;\longmapsto\;uA^{(1)}u^{\ast}\;+\;u[D,u^{\ast}],\\
A^{(2)}  &  \;\longmapsto\;JuJ^{-1}\,A^{(2)}\,Ju^{\ast}J^{-1}\;+\;JuJ^{-1}%
\,\big[u[D,u^{\ast}],Ju^{\ast}J^{-1}\big].
\end{align}
Thus $A^{(1)}$ transforms linearly (as a one-form), while $A^{(2)}$ transforms
\emph{nonlinearly} and encodes genuine quadratic (composite) contributions to
the scalar sector.

Evaluating the components of $D_{A}$ in the tensor notation yields the
Pati--Salam gauge group and its multiplets:
\begin{equation}
G_{\mathrm{PS}}=SU(2)_{R}\times SU(2)_{L}\times SU(4).
\end{equation}

The gauge bosons are $W_{R,\mu}^{\alpha}$ ($\alpha=1,2,3$) for $SU(2)_{R}$,
$W_{L,\mu}^{\alpha}$ for $SU(2)_{L}$, and $V_{\mu}^{m}$ ($m=1,\dots,15$) for
$SU(4)$. The fundamental Higgs fields are read from $A^{(1)}:$%
\begin{equation}
\phi_{a}^{\;b}\sim(2_{R},2_{L},1),\qquad\Delta_{aI}\sim(2_{R},1_{L}%
,4),\qquad\Sigma^{\,J}{}_{I}\sim(1_{R},1_{L},1\oplus15).
\end{equation}
Additional Higgs are read from $A^{(2)}$ (or fundamental if the second $D_{F}$
does not obey order-one)%
\begin{equation}
(3_{R},1_{L},10),\qquad(1_{R},1_{L},6),\qquad(2_{R},2_{L},1\oplus15)
\end{equation}
When the unfluctuated $D_{F}$ \emph{does} satisfy the order-one condition with
respect to the SM subalgebra, these appear as \emph{composite} fields
(quadratic in the above fundamentals).

In generation space one has $3\times3$ Yukawa matrices
\begin{equation}
k_{\nu},\;k_{e},\;k_{u},\;k_{d}\qquad\text{(Dirac)}\qquad\text{and}\qquad
k_{\nu_{R}}\quad\text{(Majorana)}.
\end{equation}
This assigns Dirac masses to all fermions and Majorana masses only to $\nu
_{R}$, enabling the see-saw. A particularly symmetric limit is
\begin{equation}
k_{\nu}=k_{u},\qquad k_{e}=k_{d},
\end{equation}
corresponding to exact quark--lepton unification at the Pati--Salam scale; in
this case the $(1\oplus15)$ of $SU(4)$ (contained in $\Sigma$) decouples from
the low-energy spectrum.

From the kinetic terms of $\Sigma$ one reads the covariant derivatives and the
mixing of neutral vectors. Writing the $SU(2)_{R}$ neutral boson $W_{R,\mu
}^{3}$ and the $SU(4)$ diagonal generator $V_{\mu}^{15}$, the massless
combination after the high-scale breaking is
\begin{equation}
g_{R}W_{R,\mu}^{3}\;=\;g_{1}B_{\mu}+g_{1}^{\prime}Z_{\mu}^{\prime},\qquad
\frac{\sqrt{3}}{2}\,g\,V_{\mu}^{15}\;=\;-\,g_{1}B_{\mu}+g_{1}^{\prime}Z_{\mu
}^{\prime}, \label{eq:Bmix}%
\end{equation}
so that $B_{\mu}$ remains massless, while $W_{R,\mu}^{\pm}$, $Z_{\mu}^{\prime
}$, and the coset vectors in $SU(4)/(SU(3)\times U(1))$ obtain masses of order
$w$ from the $\Sigma$-kinetic term. (Here $g_{R}$ and $g$ are the $SU(2)_{R}$
and $SU(4)$ couplings and $w$ the high-scale vev.)

\bigskip\noindent\textbf{In summary: }The inner-fluctuation analysis on
$\mathbb{H}_{R}\oplus\mathbb{H}_{L}\oplus M_{4}(\mathbb{C})$ produces the full
Pati--Salam gauge sector and a tightly constrained scalar sector. The
quadratic piece $A^{(2)}$ is essential: it captures composite scalars and
modifies the symmetry breaking pattern while preserving gauge covariance.
Fermion masses and the seesaw are encoded directly in $D_{F}$.

The bosonic Lagrangian is generated from the cutoff spectral action
\begin{equation}
S_{\mathrm{bos}}\;=\;\mathrm{Tr}\,f\!\left(  \frac{D_{A}}{\Lambda}\right)  ,
\end{equation}
with smooth positive $f$ and scale $\Lambda$. Using the asymptotic heat-kernel
expansion for large $\Lambda$,
\begin{equation}
\mathrm{Tr}\,f\!\left(  \frac{D_{A}}{\Lambda}\right)  \sim\sum_{n\geq0}%
f_{4-n}\,\Lambda^{4-n}\,a_{n}(D_{A}^{2}),
\end{equation}
one obtains (schematically) the Einstein--Hilbert term, higher-curvature
terms, Yang--Mills terms for $G_{\mathrm{PS}}$, and the full scalar
kinetic/potential built from the inner fluctuations.

The $a_{4}$ coefficient fixes the gauge-kinetic terms with couplings
\begin{equation}
-\frac{1}{4g_{R}^{2}}(W_{R,\mu\nu}^{a})^{2}\;-\;\frac{1}{4g_{L}^{2}}%
(W_{L,\mu\nu}^{a})^{2}\;-\;\frac{1}{4g^{2}}(V_{\mu\nu}^{m})^{2},
\end{equation}
and, due to the spectral normalization at $\Lambda$, a unification condition
\begin{equation}
g_{R}(\Lambda)\;=\;g_{L}(\Lambda)\;=\;g(\Lambda)
\end{equation}
(up to calculable normalization factors depending on $f$). All scalar
multiplets acquire canonical kinetic terms with covariant derivatives dictated
by their $G_{\mathrm{PS}}$ representations. Concretely,
\begin{align}
\mathcal{L}_{\mathrm{kin}}^{\phi}  &  \,=\,\mathrm{Tr}\big[(D_{\mu}%
\phi)^{\dagger}(D^{\mu}\phi)\big],\quad\phi\sim(2,2,1),\\
\mathcal{L}_{\mathrm{kin}}^{\Delta}  &  \,=\,(D_{\mu}\Delta)^{\dagger}(D^{\mu
}\Delta),\quad\Delta\sim(2,1,4),\\
\mathcal{L}_{\mathrm{kin}}^{\Sigma}  &  \,=\,\mathrm{Tr}\big[(D_{\mu}%
\Sigma)^{\dagger}(D^{\mu}\Sigma)\big],\quad\Sigma\sim(1,1,1\oplus15),
\end{align}
with $D_{\mu}$ containing $W_{R,L}$ and $V_{\mu}$ in the appropriate generators.

The spectral action organizes the scalar potential as
\begin{align}
V(\phi,\Delta,\Sigma,\ldots)\;  &  =\;-\,\mu_{\phi}^{2}\,\mathrm{Tr}%
(\phi^{\dagger}\phi)-\mu_{\Delta}^{2}\,\Delta^{\dagger}\Delta-\mu_{\Sigma}%
^{2}\,\mathrm{Tr}(\Sigma^{\dagger}\Sigma)\;\\
&  +\;\lambda_{\phi}\big[\mathrm{Tr}(\phi^{\dagger}\phi)\big]^{2}%
+\lambda_{\Delta}(\Delta^{\dagger}\Delta)^{2}+\lambda_{\Sigma}\big[\mathrm{Tr}%
(\Sigma^{\dagger}\Sigma)\big]^{2}\;+\;\text{cross-couplings},
\end{align}
with definite relations among quartics induced by the Seeley--DeWitt
coefficients. When $A^{(2)}$ generates composite multiplets, additional
quartics respecting $G_{\mathrm{PS}}$ appear.

A representative breaking pattern is:
\begin{align}
&  SU(2)_{R}\times SU(2)_{L}\times
SU(4)\xrightarrow{\;\langle \Sigma \rangle \sim w\;}SU(2)_{R}\times
SU(2)_{L}\times SU(3)_{c}\times U(1)_{B-L}\nonumber\\
&  \xrightarrow{\;\langle \Delta_R \rangle \sim v_R\;}SU(2)_{L}\times
U(1)_{Y}\times SU(3)_{c}%
\xrightarrow{\;\langle \phi \rangle \sim v\;}U(1)_{\mathrm{em}}\times
SU(3)_{c}.
\end{align}
The neutral-vector mixing in~\eqref{eq:Bmix} ensures that the hypercharge
field arises as the massless direction at the intermediate stage, while the
orthogonal $Z^{\prime}$ gains a mass $\sim w$ or $v_{R}$ depending on the
precise alignment of vevs and couplings.

The fermionic action is determined by $D_{F}$ and its inner fluctuations.
Dirac masses descend from $\langle\phi\rangle$, while Majorana masses
originate from $\langle\Delta_{R}\rangle$ and/or the $k_{\nu_{R}}$ entries.
The quark--lepton equalities $k_{\nu}=k_{u}$ and $k_{e}=k_{d}$ (natural at the
Pati--Salam scale) imply simplifications in the scalar spectrum (notably the
decoupling of the $(1\oplus15)$ of $SU(4)$) and yield relations among
low-energy Yukawas after RG running. Including the scalars responsible for the
high-scale breaking (especially those linked to the right-handed sector) can
stabilize the electroweak potential when run from $\Lambda$ to low energies,
reconciling the observed values of $m_{h}$ and $m_{t}$. The spectral action
ties together gauge unification, the see-saw scale, and scalar thresholds in a
single geometric framework.

The Pati--Salam model with gauge group $SU(2)_{R}\times SU(2)_{L}\times SU(4)$
contains several Higgs multiplets. We now show how these are \emph{restricted}
so that the low-energy field content reproduces the Standard Model (SM). The
truncation involves selecting directions in field space such that only the SM
Higgs doublet $H$ and a real singlet $\sigma$ survive, while the gauge sector
reduces to the SM hypercharge sector. The $(2_{R},2_{L},1)$ field $\phi
_{a}^{\ b}$ is truncated to the SM Higgs doublet $H$ via
\begin{equation}
\phi_{a}^{\ b}\;=\;\delta_{a1}\,\varepsilon^{bc}\,H_{c},
\end{equation}
leaving precisely one $SU(2)_{L}$ doublet $H$. The $(2_{R},1,4)$ field
$\Delta_{aI}$ is truncated by fixing the $SU(2)_{R}$ and $SU(4)$ indices:
\begin{equation}
\Delta_{aI}\;=\;\delta_{a1}\,\delta_{I1}\,\sqrt{\sigma},
\end{equation}
so that only a real scalar $\sigma$ remains, associated with the right-handed
$B-L$ sector. The $(1,1,1\oplus15)$ multiplet $\Sigma_{I}^{J}$ is no longer
independent. Its leptonic ($I=J=1$) and quark ($I,J=1,2,3$) blocks become
specific bilinears in $H$ with Yukawa matrices:
\begin{align}
\Sigma_{\text{leptonic}}  &  \;\sim\;\delta_{a1}\,k_{\nu}\,\varepsilon
^{bc}H_{c}+\delta_{a2}\,H^{b}k_{e},\\
\Sigma_{\text{quark}}  &  \;\sim\;\delta_{a1}\,k_{u}\,\varepsilon^{bc}%
H_{c}+\delta_{a2}\,H^{b}k_{d}.
\end{align}
Thus $\Sigma$ is expressed in terms of the SM Higgs doublet $H$ and Yukawa
couplings. The right-handed neutrino Majorana contribution truncates to
\begin{equation}
\propto\;k_{\nu_{R}}\,\sigma,
\end{equation}
consistent with a see-saw mechanism once $\sigma$ develops a vev.

The neutral gauge bosons are $W_{R,\mu}^{3}$ from $SU(2)_{R}$ and $V_{\mu
}^{15}$ from $SU(4)$. The truncation defines the SM hypercharge boson $B_{\mu
}$ as the unique massless linear combination:
\begin{align}
g_{R}W_{R,\mu}^{3}  &  =g_{1}B_{\mu},\\
\frac{\sqrt{3}}{2}\,g\,V_{\mu}^{15}  &  =-g_{1}B_{\mu}.
\end{align}
Here $\lambda^{15}=\tfrac{1}{\sqrt{6}}\text{diag}(3,-1,-1,-1)$ corresponds to
$B-L$. The orthogonal combination becomes heavy or is set to zero, and
$W_{R}^{\pm}$ and off-diagonal $SU(4)$ vectors are truncated. This reproduces
the correct hypercharge assignments of the SM.

Explicit blocks of the fluctuated Dirac operator $D_{A}$ confirm the
truncation reproduces SM couplings:

\begin{itemize}
\item For the $(\dot1, I=1)$ leptonic block,
\begin{equation}
(D_{A}) \;=\; \gamma^{\mu}D_{\mu},
\end{equation}
showing cancellation of extra neutral terms.

\item For the $(\dot{2},I=1)$ block,
\begin{equation}
(D_{A})\;=\;\gamma^{\mu}\big(D_{\mu}+ig_{1}B_{\mu}\big),
\end{equation}
which is precisely the SM hypercharge coupling.
\end{itemize}

\textit{What survives after truncation?}

\begin{itemize}
\item \textbf{Scalars:} SM Higgs doublet $H$, real singlet $\sigma$, and
Yukawa-dependent $\Sigma(H,k)$ combinations.

\item \textbf{Fermion masses:} Dirac masses via $H$ (through $k_{u}%
,k_{d},k_{\nu},k_{e}$), Majorana masses via $k_{\nu_{R}}\sigma$.

\item \textbf{Gauge bosons:} $SU(2)_{L}$, $SU(3)_{c}$, and hypercharge
$B_{\mu}$. Extra Pati--Salam gauge bosons ($W_{R}^{\pm}$, off-diagonal
$SU(4)$) are removed or become heavy.
\end{itemize}

\section{Grand unification in the Pati-Salam model}

This section investigates gauge coupling unification in the framework of
noncommutative geometry (NCG) \cite{CCS15}, specifically within the
\textbf{spectral Pati--Salam model}. The algebra $\mathbb{H}_{R}%
\oplus\mathbb{H}_{L}\oplus M_{4}(\mathbb{C})$, naturally yields the
Pati--Salam gauge group
\begin{equation}
SU(2)_{R}\times SU(2)_{L}\times SU(4).
\end{equation}
Inner fluctuations of the finite Dirac operator produce both gauge bosons and
scalar fields. The scalar sector is not arbitrary: its multiplets and
interactions are dictated by the structure of the Dirac operator $D_{F}$.
Dynamics are given by the \textbf{spectral action principle},
\begin{equation}
\mathrm{Tr}\,f(D_{A}/\Lambda)\sim F_{4}\Lambda^{4}a_{0}+F_{2}\Lambda^{2}%
a_{2}+F_{0}a_{4}+\cdots,
\end{equation}
which automatically enforces unification of gauge couplings at the cutoff
$\Lambda$. The central aim in this section is to compute the one-loop
renormalization group (RG) evolution of couplings in several scalar-sector
scenarios, identify the intermediate scale $m_{R}$ where PS breaks to the SM,
and determine the unification scale $\Lambda$.

The finite algebra
\begin{equation}
A_{F}=\mathbb{H}_{R}\oplus\mathbb{H}_{L}\oplus M_{4}(\mathbb{C})
\end{equation}
gives rise to 16 Weyl fermions per generation, automatically fitting the
leptons and quarks of the Standard Model. The Higgs sector is derived from
commutators with $D_{F}$. Depending on whether $D_{F}$ satisfies the order-one
condition, one obtains:

\begin{itemize}
\item \textbf{Composite Higgs scenario:} many Higgs fields are not independent
but quadratic in basic fields $(\phi, \Sigma, \Delta)$.

\item \textbf{Fundamental Higgs scenario:} additional multiplets remain fundamental.

\item \textbf{Left--Right symmetric fundamental scenario:} all fundamental
multiplets are present, enforcing $g_{R}=g_{L}$ at $m_{R}$.
\end{itemize}

The one-loop RGEs are
\begin{equation}
16\pi^{2}\frac{dg}{dt}=-b\,g^{3},
\end{equation}
where the coefficients $b$ depend on the spectrum. At the intermediate scale
$m_{R}$, matching conditions connect the PS couplings to the SM ones:
\begin{align}
\frac{1}{g_{1}^{2}}  &  =\tfrac{2}{3}\frac{1}{g^{2}}+\tfrac{1}{g_{R}^{2}},\\
\frac{1}{g_{2}^{2}}  &  =\frac{1}{g_{L}^{2}},\qquad\frac{1}{g_{3}^{2}}%
=\frac{1}{g^{2}}.
\end{align}
The procedure is: run SM couplings to $m_{R}$, impose matching, run PS
couplings up, and tune $m_{R}$ so that $g_{R}=g_{L}=g$ at $\Lambda$.

We first consider the model with composite Higgs. Light fields below $m_{R}$
are taken to be SM Higgs doublet $H$, a real singlet $\sigma$, and some
colored scalars. The one-loop coefficients above $m_{R}$ are
\begin{equation}
(b_{R},b_{L},b)=\left(  \tfrac{7}{3},3,\tfrac{31}{3}\right)  .
\end{equation}
\textbf{Results:}

\begin{itemize}
\item With quark--lepton Yukawa unification (decoupled $\Sigma$):
$\Lambda\approx2.5\times10^{15}\,\text{GeV}, \, m_{R} \approx4.25\times
10^{13}\,\text{GeV}$.

\item With $\Sigma$ active: $\Lambda\approx6.3\times10^{15}\,\text{GeV}%
,\,m_{R}\approx4.1\times10^{13}\,\text{GeV}$.
\end{itemize}

Next we consider fundamental Higgs, larger scalar content, with fundamental
multiplets beyond $(\phi,\Delta,\Sigma)$. The coefficients are
\begin{equation}
(b_{R},b_{L},b)=\left(  -\tfrac{26}{3},-2,2\right)  .
\end{equation}
\textbf{Result:} $\Lambda\approx6.3\times10^{16}\,\text{GeV},\,m_{R}%
\approx1.5\times10^{11}\,\text{GeV}$.

We conclude that across all scalar-sector realizations, the spectral
Pati--Salam model achieves coupling unification:

\begin{itemize}
\item Unification scale $\Lambda$ lies between $10^{15}$ and $10^{16}$ GeV.

\item Intermediate scale $m_{R}$ is in the range $10^{11}$--$10^{13}$ GeV,
consistent with see-saw neutrino masses.

\item The spectral action naturally enforces $g_{R}=g_{L}=g$ at $\Lambda$,
unlike the spectral Standard Model where couplings do not meet.
\end{itemize}

Thus the spectral Pati--Salam construction provides a robust, geometrically
motivated realization of grand unification.

\section{Quanta of geometry: towards quantum gravity}

Our aim is to reconcile Quantum Mechanics with General Relativity by
generalizing the Heisenberg commutation relation to a higher-degree geometric
analogue \cite{CCM14}\cite{CCM15}. In the standard spectral triple $(A,H,D)$:

\begin{itemize}
\item $D$ (Dirac operator) plays the role of ``momentum'' or inverse line element.

\item $A=C^{\infty}(M)$ is the algebra of functions, encoding topology.

\item $H$ is the Hilbert space of spinors.
\end{itemize}

Introduce a new \textquotedblleft position\textquotedblright\ variable
$Y=Y^{A}\Gamma_{A}$, constructed from gamma matrices. The one-sided higher
commutation relation is
\begin{equation}
\frac{1}{n!}\langle Y[D,Y]\cdots\lbrack D,Y]\rangle=\sqrt{\kappa}\,\gamma,
\label{onesided}%
\end{equation}
with $\langle\cdot\rangle$ denoting the normalized trace. This has the
following implications:

\begin{itemize}
\item Solutions exist iff $M$ decomposes into disjoint spheres of unit volume,
representing ``geometric quanta.''

\item Refining with the real structure $J$ (charge conjugation) yields a
two-sided relation, allowing connected manifolds while preserving quantization.

\item In $4D$, the two-sided structure reproduces the algebras $M_{2}%
(\mathbb{H})$ and $M_{4}(\mathbb{C})$, the internal algebra of the Standard Model.
\end{itemize}

Thus, both gravity and the Standard Model emerge naturally from spectral geometry.

To realize this consider a compact spin Riemannian manifold $M$ of even
dimension $n$. The spectral triple $(A,H,D)$ has $A=C^{\infty}(M)$ and $D$ in
local form
\begin{equation}
D=\gamma^{\mu}(\partial_{\mu}+\omega_{\mu}).
\end{equation}

\textbf{Theorem: }The one-sided relation admits a solution iff $M$ is a
disjoint union of spheres $S^{n}$ of unit volume. Each sphere represents a
\emph{quantum of geometry}.

\textbf{We give here only a sketch of proof:}

\begin{itemize}
\item $Y$ defines a map $M \to S^{n}$.

\item The operator identity reduces to
\begin{equation}
\det(e^{a}_{\mu})\, dx^{1}\wedge\cdots\wedge dx^{n} = Y^{\#}(\rho),
\end{equation}
where $Y^{\#}(\rho)$ is the pullback of the volume form on $S^{n}$.

\item The Jacobian of $Y$ never vanishes, so $Y$ is a covering map.

\item Since $S^{n}$ is simply connected ($n>1$), each component of $M$ is
diffeomorphic to $S^{n}$, with the same volume.
\end{itemize}

Thus, geometry decomposes into Planck-scale quanta, each a unit-volume sphere.

For any $Y=Y^{A}\Gamma_{A}$ with $Y^{2}=1$, equation (\ref{onesided}) implies
\begin{equation}
\int\gamma\langle Y[D,Y]^{n}\rangle D^{-n}=2^{n/2+1}\,\deg(Y).
\end{equation}
This connects the spectral expression with the topological degree of the map
$Y:M\rightarrow S^{n}$, realizing quantization of volume through topology.

To avoid the restriction to disconnected spheres, the real structure $J$ is
introduced. Consider two commuting Clifford sets $\{\Gamma_{A}\}$ and
$\{\Gamma_{B}^{\prime}\}$ with
\[
Y=Y^{A}\Gamma_{A},\qquad Y^{\prime}=Y^{\prime B}\Gamma_{B}^{\prime},
\]
and define projectors $e=\tfrac{1}{2}(1+Y)$ and $e^{\prime}=\tfrac{1}%
{2}(1+Y^{\prime})$, where $e^{2}=e$, $e^{^{\prime}2}=e^{\prime},$ and
\begin{equation}
Z=2ee^{\prime}-1.
\end{equation}
satisfying $Z^{2}=1.$ The two-sided quantization relation is taken to be
\begin{equation}
\frac{1}{n!}\langle Z[D,Z]\cdots\lbrack D,Z]\rangle=\gamma.
\end{equation}
First for the case of $n=2$ one finds
\begin{equation}
\langle Z[D,Z][D,Z]\rangle=\tfrac{1}{2}\langle Y[D,Y]^{2}\rangle+\tfrac{1}%
{2}\langle Y^{\prime}\left[  D,Y^{\prime}\right]  ^{2}\rangle.
\end{equation}
Thus, the $2D$ volume equals the sum of degrees of two maps $Y,Y^{\prime}$,
showing quantization of area.

Similarly, in $4D$
\begin{equation}
\langle Z[D,Z]^{4}\rangle=\tfrac{1}{2}\langle Y[D,Y]^{4}\rangle+\tfrac{1}%
{2}\langle Y^{\prime}[D,Y^{\prime}]^{4}\rangle.
\end{equation}
So the $4D$ volume is the sum of degrees of two maps $M\rightarrow S^{4}$,
hence quantized.

The proof relies on:

\begin{itemize}
\item Pairwise commutativity of $Y^{A}, Y^{\prime B}$.

\item Order-zero and order-one conditions ensuring compatibility with $D$.

\item Commutation of $\langle\lbrack D,Y]^{2}\rangle$ with the algebra
generated by $Y$.
\end{itemize}

Define
\begin{equation}
q(M)=\{\deg(\phi)+\deg(\psi)\mid\phi,\psi:M\rightarrow S^{n},\;\phi
^{\#}(\alpha)+\psi^{\#}(\alpha)\neq0\},
\end{equation}
where $\alpha$ is the unit sphere volume form.

\textbf{Theorem:} For $n=2,4$, in any operator representation of the two-sided relation:

\begin{itemize}
\item The Weyl volume term is quantized.

\item A solution exists iff $\mathrm{Vol}(M)\in q(M)\subset\mathbb{Z}$.
\end{itemize}

This shows the equivalence between operator solutions and integer-valued volume.

\subsection{Differential geometry of the two-sided equation}

The central question is: for which manifolds $M$ does the two-sided higher
Heisenberg relation admit solutions? \cite{CCM15}. This is equivalent to
determining when there exist maps $\phi,\psi:M\rightarrow S^{n}$ with
\begin{equation}
\phi^{\#}(\alpha)+\psi^{\#}(\alpha)\neq0\quad\text{everywhere},
\end{equation}
where $\alpha$ is the normalized volume form on $S^{n}$. Define
\begin{equation}
q(M)=\{\deg(\phi)+\deg(\psi)\mid\phi^{\#}(\alpha)+\psi^{\#}(\alpha)\neq0\}.
\end{equation}
We build existence results for $q(M)$ in different dimensions and establishes
topological obstructions.

For $n<4$, if $M$ is a compact connected oriented $n$-manifold and $\omega$ is
a top-degree form that never vanishes, agrees with the orientation, and has
integer volume $\int_{M}\omega\in\mathbb{Z}$ with absolute value greater than
$3$, then there exist maps $\phi,$ $\phi^{\prime}$ with
\begin{equation}
\phi^{\#}(\alpha)+\phi^{\prime\#}(\alpha)=\omega.
\end{equation}
Thus $\deg(\phi)+\deg(\phi^{\prime})=\int_{M}\omega$. The construction uses
Whitehead triangulations, barycentric subdivisions, and ramified coverings.
For $n=2,3$ this ensures $q(M)$ contains all sufficiently large integers.

We now restrict the discussion for $n=4.$ If $D(M)\neq\emptyset$ (i.e.\ such
pairs $(\phi,\psi)$ exist), then
\begin{equation}
w_{2}(T_{M})^{2}=0,
\end{equation}
where $w_{2}$ is the second Stiefel--Whitney class. More generally, all
cup-products of Stiefel--Whitney classes vanish. For spin manifolds ($w_{2}%
=0$), this obstruction automatically vanishes.

It suffices to start with a map $\phi:M\rightarrow S^{4}$ such that $\phi
^{\#}(\alpha)\geq0$ and define
\begin{equation}
R=\{x\in M\mid\phi^{\#}(\alpha)(x)=0\}.
\end{equation}
\textbf{Lemma:} There exists $\psi:M\rightarrow S^{4}$ with $\phi^{\#}%
(\alpha)+\psi^{\#}(\alpha)$ nonvanishing iff there is an immersion
$f:V\rightarrow\mathbb{R}^{4}$ from a neighborhood $V\supset R$. In this case
one can choose $\deg(\psi)=0$.

Thus the existence problem reduces to a local immersion problem near $R$.

For $M=N\times S^{1}$ with $N$ a compact oriented $3$-manifold, one has:

\textbf{Theorem:} $q(M)$ is nonempty and in fact contains all integers $m\geq
r$ for some $r>0$.

Construction: take a ramified cover $g:S^{3}\times S^{1}\rightarrow S^{4}$ of
degree $m$ and compose with a ramified cover $f:N\rightarrow S^{3}$ to form
$\phi=g\circ(f\times id)$. Singular sets can be arranged so that the Lemma
applies. An alternative approach uses the fact that $M$ is parallelizable, so
$M\setminus\{p\}$ immerses in $\mathbb{R}^{4}$, again enabling the Lemma above.

\textbf{Theorem:} If $M$ is a compact oriented spin $4$-manifold, then
\begin{equation}
q(M)\supseteq\{m\in\mathbb{Z}\mid m\geq5\}.
\end{equation}

\textbf{Sketch of proof:}

\begin{itemize}
\item Using a Whitehead triangulation, construct a ramified cover
$\gamma:M\rightarrow S^{4}$. On the 2-skeleton, $TM$ is trivial (since
$B\mathrm{Spin}(4)$ is 3-connected), so by Poenaru's extension of immersion
theory, there exists an immersion $V\rightarrow\mathbb{R}^{4}$ from a
neighborhood of the 2-skeleton. Then the Lemma above applies to show the
degree of $\gamma$ lies in $q(M)$.

\item By the theorem of Iori--Piergallini, every closed oriented $4$-manifold
is a branched cover of $S^{4}$ of degree $m\geq5$. In the spin case, branch
neighborhoods immerse in $\mathbb{R}^{4}$, hence the Lemma applies. Thus all
$m\geq5$ lie in $q(M)$.
\end{itemize}

We have thus shown that:

\begin{itemize}
\item For $n<4$, all large integers belong to $q(M)$.

\item In $4D$, obstructions exist, but vanish for spin manifolds.

\item Via immersion criteria and covering constructions, $q(M)$ contains all
integers $m\ge5$ for spin $4$-manifolds.
\end{itemize}

This ensures that physically relevant spin $4$-manifolds support the two-sided
Heisenberg relation with quantized volume.

\subsection{A tentative particle picture in quantum gravity}

In standard QFT, particles are irreducible representations of the Poincar\'{e}
group. In the spectral framework, we propose that \textquotedblleft
particles\textquotedblright\ correspond instead to irreducible representations
of the \emph{two-sided higher Heisenberg relation}. Here the algebra of
relations among the inverse line element $D$ and the slashed variables
$Y,Y^{\prime}$ replaces the Poincar\'{e} group as the symmetry principle
\cite{CCM15}.

The maps $Y,Y^{\prime}:M\rightarrow S^{n}\times S^{n}$ play the role of
coordinates. By Whitney's embedding theorem, every $n$-manifold embeds in
$\mathbb{R}^{2n}\subset S^{n}\times S^{n}$, so working with $(Y,Y^{\prime})$
is sufficiently general. The spectral growth of $D$ fixes the metric dimension
to $n$. This raises two central questions:

\begin{enumerate}
\item Why is the joint spectrum of $Y,Y^{\prime}$ effectively $n$-dimensional
rather than $2n$?

\item Why are the spectral integrals quantized?
\end{enumerate}

To find dimensionality of the joint spectrum, we first note that \textbf{in
the classical case:} If a base manifold $M$ is present, the joint spectrum is
$Y,Y^{\prime}:M\rightarrow S^{n}\times S^{n}$, of topological dimension $\leq
n$.

However, in the \textbf{general spectral case:} Even without $M$, three
conditions enforce an effective dimension $n$:

\begin{itemize}
\item Order-zero relation: $[Y,Y^{\prime}]=0$ (the two Clifford sets commute).

\item Bounded commutators: $[D,Y], [D,Y^{\prime}]$ are bounded (Lipschitz regularity).

\item Weyl law: growth of $\mathrm{spec}(D)$ corresponds to dimension $n$.
\end{itemize}

These force the joint spectrum to behave as an $n$-dimensional embedded
submanifold, despite the a priori $2n$ coordinates.

Now concerning quantization of the volume, first in the \textbf{classical
case:} The noncommutative integral
\begin{equation}
\int\gamma\langle Y[D,Y]^{n}\rangle D^{-n}%
\end{equation}
is the winding number of the map $Y:M\rightarrow S^{n}$, hence an integer.

In the \textbf{general case:} Let $e=\tfrac{1}{2}(1+Y)$. The operator Chern
character of $e$ has components
\begin{equation}
\mathrm{Ch}_{m}(e)=\mathrm{tr}\big((2e-1)\otimes e\otimes\cdots\otimes e\big),
\end{equation}
with $m$ even. Because traces of products of fewer than $n$ gamma matrices
vanish,
\begin{equation}
\mathrm{Ch}_{m}(e)=0,\quad m<n.
\end{equation}
Thus $\mathrm{Ch}_{n}(e)$ is a Hochschild cycle. Pairing with the local index
cocycle $\tau$ gives
\begin{equation}
\langle\tau,\mathrm{Ch}_{n}(e)\rangle=\int\gamma\langle Y[D,Y]^{n}\rangle
D^{-n}=\mathrm{Index}(D_{e})\in\mathbb{Z}.
\end{equation}
Hence the integral is quantized.

The same holds for $Y^{\prime}$, and the two-sided relation with
\begin{equation}
\frac{1}{n!}\langle Z[D,Z]^{n}\rangle=\gamma
\end{equation}
implies
\begin{equation}
\int D^{-n}=\tfrac{1}{n!}\int\gamma\langle Z[D,Z]^{n}\rangle D^{-n}=\tfrac
{1}{2}\Big(\int\gamma\langle Y[D,Y]^{n}\rangle D^{-n}+\int\gamma\langle
Y^{\prime}[D,Y^{\prime n}\rangle D^{-n}\Big).
\end{equation}
Thus $\int D^{-n}$ is also an integer after normalization.

We have thus a \textquotedblleft particle picture\textquotedblright\ where
fundamental excitations are irreducible representations of the two-sided
Heisenberg relation. It shows:

\begin{itemize}
\item The joint spectrum of $Y,Y^{\prime}$ is $n$-dimensional despite
embedding in $2n$ space.

\item Volume integrals are quantized because only the top Chern component
survives, yielding integer indices.
\end{itemize}

This connects spectral geometry with a quantized particle-like structure in
quantum gravity.

\subsection{Physical consequences}

We discuss first the implications on the gravitational action and the
cosmological constant. We work (initially) in Euclidean signature on a compact
$4$-manifold. For simplicity we use a single set of scalar fields, instead of
two, $Y^{A}(x)$, $A=1,\dots,5$, subject to
\begin{align}
&  Y^{A}Y_{A}=1,\label{eq:unit}\\
&  \sqrt{g}\,d^{4}x\;=\;\tfrac{1}{4!}\,\varepsilon^{\mu\nu\kappa\lambda
}\varepsilon_{ABCDE}\,Y^{A}\partial_{\mu}Y^{B}\partial_{\nu}Y^{C}%
\partial_{\kappa}Y^{D}\partial_{\lambda}Y^{E}\,d^{4}x, \label{eq:volq}%
\end{align}
so that the four--volume is an integer multiple of the unit $S^{4}$ volume
(volume quantization).

Impose these constraints in the Einstein--Hilbert action (set $8\pi G=1$) with
Lagrange multipliers $\lambda_{0},\lambda$:
\begin{align}
I  &  =-\frac{1}{2}\!\int\sqrt{g}\,R\,d^{4}x+\frac{1}{2}\!\int\sqrt
{g}\,\lambda_{0}\,(Y^{A}Y_{A}-1)\,d^{4}x\nonumber\\
&  +\frac{1}{2}\!\int\lambda\Big(\sqrt{g}-\tfrac{1}{4!}\varepsilon^{\mu
\nu\kappa\lambda}\varepsilon_{ABCDE}Y^{A}\partial_{\mu}Y^{B}\partial_{\nu
}Y^{C}\partial_{\kappa}Y^{D}\partial_{\lambda}Y^{E}\Big)d^{4}x.
\end{align}
The last term is metric--independent (a pullback of the $S^{4}$ volume form).
Varying $g_{\mu\nu}$ gives
\begin{equation}
G_{\mu\nu}+\tfrac{1}{2}g_{\mu\nu}\,\lambda=0\quad\Rightarrow\quad G\equiv
g^{\mu\nu}G_{\mu\nu}=-2\lambda,
\end{equation}
hence the traceless Einstein equations
\begin{equation}
G_{\mu\nu}-\tfrac{1}{4}g_{\mu\nu}G=0.
\end{equation}
By the Bianchi identity $\nabla^{\mu}G_{\mu\nu}=0$, one gets $\partial_{\mu
}G=0$, i.e.\ $G=4\Lambda$ is constant and $\Lambda$ appears as an
\emph{integration constant} rather than a parameter in the action. Variation
with respect to $Y^{A}$ adds no new local dynamics \cite{CCM14}.

Instead of integrating over the global scale factor, the functional integral
amounts to a sum over \emph{winding numbers} (four--volume quanta). For the
present universe the number of Planck four--volume quanta is enormous
($\sim10^{61}$). The huge constant term that typically appears in the spectral
action does not gravitate in the field equations (which are traceless); the
physical cosmological constant is an integration constant.

Pass to Lorentzian signature ($M_{4}\rightarrow\mathbb{R}\times S^{3}$) via
Wick rotation and a scaling (decompactification) limit. Introduce a scalar $X$
through $Y_{5}=\eta X$ and rescale $x^{4}\rightarrow\eta t$ with
$\eta\rightarrow0$. Then the constraints reduce to
\begin{align}
&  Y^{a}Y_{a}=1\qquad(a=1,\dots,4),\\
&  g^{\mu\nu}\partial_{\mu}X\,\partial_{\nu}X=1,
\end{align}
and the four--form factorizes so that \emph{three--volume} quantization holds
on spatial slices $\Sigma_{t}$.

In $3{+}1$ ADM variables with lapse $N$ and spatial metric $h_{ij}$, choose
$\partial_{i} X=0$ and $\partial_{t} X=N$ to satisfy the unit--norm
constraint. Then
\begin{equation}
\int_{\Sigma_{t}}\! \sqrt{h}\, d^{3}x \;=\; w \times\frac{4\pi^{2}}{3},
\end{equation}
i.e.\ the volume of each compact slice is an integer multiple (winding number
$w$) of the unit $S^{3}$ volume.

To impose $g^{\mu\nu}\partial_{\mu}X\partial_{\nu}X=1$ dynamically, add
\begin{equation}
I_{\text{mim}}=\int\sqrt{-g}\,\lambda_{00}\,\big(g^{\mu\nu}\partial_{\mu
}X\,\partial_{\nu}X-1\big)\,d^{4}x.
\end{equation}
This is precisely the \emph{mimetic} constraint \cite{Chamseddine:2013kea}:
the Lagrange multiplier sector yields a pressureless dust component
(\textquotedblleft mimetic dark matter\textquotedblright) while the
cosmological constant still arises as an integration constant. Thus enforcing
four/three--volume quantization in the gravitational action naturally produces
both dark--matter--like and dark--energy--like sectors.

Next we discuss area quantization for black holes. Consider a compact
spacelike $2$--surface $\Sigma_{2}\subset M_{4}$. In a scaling limit, two
transverse scalars $X^{1},X^{2}$ and three $Y^{A}$ ($A=1,2,3$) remain on
$\Sigma_{2}$. With the area--preserving condition
\begin{equation}
\det\!\big(g^{\mu\nu}\partial_{\mu}X^{m}\partial_{\nu}X^{n}\big)\Big|_{\Sigma
_{2}}=1\qquad(m,n=1,2),
\end{equation}
one finds the quantized area formula
\begin{equation}
\mathrm{Area}(\Sigma_{2})\;=\;\int_{\Sigma_{2}}\!\frac{1}{2}\,\varepsilon
^{\mu\nu}\varepsilon_{ABC}\,Y^{A}\partial_{\mu}Y^{B}\partial_{\nu}Y^{C}%
\,d^{2}x\;=\;4\pi\,n,\qquad n\in\mathbb{Z},
\end{equation}
the winding of $\Sigma_{2}\rightarrow S^{2}$.

For a horizon $2$--sphere this yields Bekenstein--type area quantization
(Planck units) and hence mass quantization for Schwarzschild:
\begin{equation}
A=16\pi M^{2}\quad\Rightarrow\quad M_{n}=\frac{\sqrt{n}}{2}.
\end{equation}
Hawking emission then consists of \emph{discrete lines} between $n\rightarrow
n-1,\dots$ with spacing
\begin{equation}
\omega\;\approx\;M_{n}-M_{n-1}\;\simeq\;\frac{1}{8M},
\end{equation}
of order the Hawking temperature; the expected line width is parametrically
smaller, so even large black holes radiate in lines under a thermal envelope.
If the minimal area is $\alpha$ times the Planck area, spacings rescale by
$\alpha.$

Applying the same logic to the de Sitter event horizon leads to a discrete
spectrum for the cosmological constant,
\begin{equation}
\Lambda_{n}\;=\;\frac{3}{n},
\end{equation}
with potential implications for inflation (including self--reproduction
regimes), where only the discrete values $\{\Lambda_{n}\}$ are allowed.

To summarize:

\begin{itemize}
\item Enforcing $4$D volume quantization with Lagrange multipliers makes the
Einstein equations traceless and promotes $\Lambda$ to an integration constant.

\item In Lorentzian signature, the $3$D slice--volume quantization induces the
mimetic constraint, providing a dust--like component (mimetic dark matter)
within the same action.

\item Quantization of $2$D areas implies quantized black--hole horizon areas
and discrete Hawking lines; the same mechanism quantizes de Sitter $\Lambda$.
\end{itemize}

\section{Entropy and the spectral action}

Let $(A,H,D)$ be a (real, even) spectral triple with self-adjoint Dirac
operator $D$ of compact resolvent. The classical spectral action is
\begin{equation}
\mathrm{Tr}\,\chi\!\left(  \frac{D^{2}}{\Lambda^{2}}\right)  ,
\end{equation}
with a positive cutoff profile $\chi$. In this section \cite{CCS18} we take a
different route $\chi$: pass to the fermionic second quantization in the
complex Clifford algebra $\mathrm{Cliff}_{\mathbb{C}}(H_{\mathbb{R}})$,
consider the $C^{\ast}$--dynamical system $(\mathrm{Cliff}_{\mathbb{C}%
}(H_{\mathbb{R}}),\sigma_{t}=\mathrm{Cliff}(e^{itD}))$, and take the unique
KMS$_{\beta}$ state $\psi_{\beta}$ for $\beta>0$. \medskip

\noindent\textbf{Main theorem.} The von Neumann entropy of $\psi_{\beta}$
equals a spectral action
\begin{equation}
S(\psi_{\beta})=\mathrm{Tr}\,h(\beta D),
\end{equation}
for a \emph{universal}, even, positive function $h(x)=E(e^{-x})$, where
\begin{equation}
E(x)=\log(1+x)-\frac{x\log x}{1+x}%
\end{equation}
is the two-bin entropy associated with a partition of $[0,1]$ in the ratio
$x:(1)$. Thus, \emph{entropy itself is a spectral action} with no adjustable
test profile.

With $C=\mathrm{Cliff}_{\mathbb{C}}(H_{\mathbb{R}})$ and dynamics $\sigma
_{t}=\mathrm{Cliff}(e^{itD})$, there is a unique KMS$_{\beta}$ state
$\psi_{\beta}$ for each $\beta>0$. When $e^{-\beta|D|}$ is trace class,
$\psi_{\beta}$ is type I and is realized in the fermionic Fock representation
determined by the orthogonal complex structure $I=i\,\mathrm{sign}(D)$ (the
\textquotedblleft Dirac sea\textquotedblright). In this representation the
dynamics is implemented by $\wedge e^{it|D|}$, with density matrix
proportional to $\wedge e^{-\beta|D|}$.

To define the Fermionic Fock space, let $V=H_{\mathbb{R}}$ and pick an
orthogonal complex structure $I$ to form $V_{I}$. Then $\mathrm{Cliff}%
_{\mathbb{C}}(V)$ acts irreducibly on $\bigwedge V_{I}$ via $v\mapsto
a_{I}^{\*}(v)+a_{I}(v)$. For $I=i\,\mathrm{sign}(D)$, negative-energy modes
acquire the conjugate complex structure, realizing the Dirac sea.
Shale--Stinespring guarantees implementability of orthogonal transformations
in the Hilbert--Schmidt regime. This setup lets one compute $\psi_{\beta}$ and
its entropy explicitly.

To interpret entropy as spectral action, for a positive trace-class operator
$T$ on a one-particle space, the corresponding quasi-free fermionic state on
Fock space has density matrix $\wedge T$, and entropy is additive over
orthogonal sums. The single-mode weight $x>0$ contributes
\begin{equation}
E(x)=\log(1+x)-\frac{x\log x}{1+x}.
\end{equation}
Thus for general $T$,
\begin{equation}
S=\mathrm{Tr}\big(E(T)\big).
\end{equation}
Applying this to $T=e^{-\beta|D|}$ and using $E(x)=E(1/x)$, one gets the
\emph{entropy--spectral-action identity}
\begin{equation}
S(\psi_{\beta})=\mathrm{Tr}\,h(\beta D),\qquad h(x)=E(e^{-x}).
\end{equation}

The function $h$ is even, positive, smooth, with derivative
\begin{equation}
h^{\prime}(x)=-\frac{x}{4\cosh^{2}(x/2)}.
\end{equation}
Its Taylor series at $x=0$ begins wi
\begin{equation}
h(x)=\log2-\frac{x^{2}}{8}+\frac{x^{4}}{64}-\frac{x^{6}}{576}+\frac{17x^{8}%
}{92160}-\frac{31x^{10}}{1612800}+\cdots,
\end{equation}
so all odd derivatives vanish at $0$ and the even derivatives alternate in sign.

We now present Integral, Theta/Eisenstein, and Laplace Representations for
$h.$ Using Eisenstein series identities and trigonometric manipulations one
shows
\begin{equation}
\frac{1}{4\cosh^{2}(\sqrt{x}/2)}=\sum_{n\in\mathbb{Z}}\frac{(2\pi n+\pi
)^{2}-x}{\big((2\pi n+\pi)^{2}+x\big)^{2}}.
\end{equation}
A Laplace representation follows:
\begin{align}
\frac{1}{4\cosh^{2}(\sqrt{x}/2)}  &  =\int_{0}^{\infty}g(t)\,e^{-tx}%
\,dt,\qquad\\
g(t)  &  =\sum_{n\in\mathbb{Z}}\Big(2\pi^{2}(2n+1)^{2}t-1\Big)\,e^{-\pi
^{2}(2n+1)^{2}t}.
\end{align}
Poisson summation refines $g$ near $t=0$, yielding a positive, rapidly
decaying kernel $\tilde{g}(t)=g(t)/(2t)$ and
\begin{equation}
h(x)=\int_{0}^{\infty}e^{-tx^{2}}\,\tilde{g}(t)\,dt.
\end{equation}
These formulas are the bridge to controlled moment calculations and,
ultimately, to the heat-coefficient arithmetic.

To find the heat kernel expansion of $h\left(  \beta D\right)  $ consider For
$\alpha>-2$, an integration by parts together with $h^{\prime}(x)$ gives
\begin{equation}
\alpha\int_{0}^{\infty}h(x)\,x^{\alpha-1}\,dx=\int_{0}^{\infty}\frac
{x^{\alpha+1}}{4\cosh^{2}(x/2)}\,dx.
\end{equation}
This implies the closed form
\begin{equation}
\int_{0}^{\infty}h(x)\,x^{\alpha}\,dx=\frac{1-2^{-\alpha-1}}{\alpha+1}%
\,\Gamma(\alpha+3)\,\zeta(\alpha+2),
\end{equation}
in particular
\begin{equation}
2\!\int_{0}^{\infty}\!h(x)\,x\,dx=\frac{9}{2}\,\zeta(3),\qquad2\!\int
_{0}^{\infty}\!h(x)\,x^{3}\,dx=\frac{225}{4}\,\zeta(5).
\end{equation}
Write the heat expansion for $\mathrm{Tr}(e^{-tD^{2}})$ as $\sum a_{\alpha
}\,t^{\alpha}$. For a general $\chi$, the spectral action $\mathrm{Tr}%
\,\chi(tD^{2})$ picks coefficients $a_{\alpha}$ multiplied by moments of
$\chi$. Specializing to $\chi(u)=h(\sqrt{u})$ (i.e.\ $f(x)=h(x)$) yields,
after a Mellin transform, the multiplier of $t^{-a}$ as
\begin{equation}
\gamma(a)=\frac{1-2^{-2a}}{a}\,\pi^{-a}\,\xi(2a),
\end{equation}
analytic in $a\in\mathbb{C}$. Potential poles at $a=0,-1,-2,\dots$ cancel. The
values match the Taylor series of $h(\sqrt{x})$ at $0$, reproducing the
alternating rational coefficients.

To show Riemann's $\xi$ and functional-equation duality, use $\xi(s)=\tfrac
{1}{2}\,s(s-1)\,\pi^{-s/2}\Gamma(\tfrac{s}{2})\zeta(s)$, the expression above
exhibits a duality under $a\mapsto-a$ pairing even-dimensional high-energy
coefficients with odd-dimensional low-energy ones. Numerically one finds
\begin{align}
\gamma(-1)  &  =\tfrac{9}{2}\zeta(3),\quad\gamma(-2)=\tfrac{225}{4}%
\zeta(5),\quad\nonumber\\
\gamma(0)  &  =\log2,\quad\gamma(1)=\tfrac{1}{8},\quad\gamma(2)=\tfrac{1}%
{32},\ \dots
\end{align}
We indicate how the spectral action picks the heat coefficients. If
$\chi(u)=\int_{0}^{\infty}e^{-su}g(s)\,ds$ with rapidly decaying $g$, then
\begin{equation}
\mathrm{Tr}\,\chi(t\Delta)=\int_{0}^{\infty}\mathrm{Tr}(e^{-st\Delta
})g(s)\,ds.
\end{equation}
When $\mathrm{Tr}(e^{-t\Delta})\sim\sum a_{\alpha}t^{\alpha}$, one gets
\begin{equation}
\mathrm{Tr}\,\chi(t\Delta)\ \sim\ \sum_{\alpha<0}\!a_{\alpha}\,t^{\alpha
}\,\frac{1}{\Gamma(-\alpha)}\int_{0}^{\infty}\!\chi(v)\,v^{-\alpha
-1}\,dv\;+\;a_{0}\,\chi(0)\;+\;\text{(derivatives of $\chi$ at 0 for
$\alpha>0$)}.
\end{equation}
For $\chi(u)=h(\sqrt{u})$ the negative-index integrals reduce to moments of
$h$ (bringing in special $\zeta$ values), while $\alpha\geq0$ terms pick the
Taylor coefficients of $h$ at $0$.

We deduce the following conceptual and physical implications:

\begin{itemize}
\item \textbf{Additivity and universality.} The entropy construction yields an
\emph{additive} functional of spectral triples (direct sums $\mapsto$ tensor
products in second quantization); the test function $h$ is universal, fixed by
fermionic thermodynamics.

\item \textbf{Bridge to number theory.} Heat coefficients for $\mathrm{Tr}%
\,h(\beta D)$ are governed by special values of $\zeta$ and $\xi$; the Riemann
functional equation induces an even/odd duality between high-energy and
low-energy sides of the expansion.

\item \textbf{Physical meaning.} The result ties the information-theoretic
entropy of second-quantized fermions directly to the spectral-action
framework, suggesting second quantization as a structural ingredient in
noncommutative-geometric models of fundamental interactions.
\end{itemize}

In summary, we deduce that:

\begin{itemize}
\item Build the $C^{*}$--system $\big(\mathrm{Cliff}_{\mathbb{C}%
}(H_{\mathbb{R}}),\sigma_{t}=\mathrm{Cliff}(e^{itD})\big)$; the unique
KMS$_{\beta}$ state is type I when $e^{-\beta|D|}$ is trace class.

\item The von Neumann entropy is a spectral action $S(\psi_{\beta
})=\mathrm{Tr}\,h(\beta D)$ with $h(x)=\log(1+e^{-x})-\dfrac{e^{-x}%
(-x)}{1+e^{-x}}$; equivalently $h^{\prime2}(x/2))$.

\item $h$ admits clean Laplace/Poisson/theta representations and an explicit
Taylor series; moments of $h$ pick out $\zeta(3)$, $\zeta(5)$, etc., and the
heat coefficients are governed by $\xi(2a)$.

\item Conceptually, entropy and spectral action become two faces of the same
object, strengthening the role of second quantization in the spectral approach
to gravity and matter.
\end{itemize}

\section{Spectral action in matrix form}

Conventional treatments expand the spectral action into component fields and
trace over Clifford and finite (internal) algebras before quantization. That
obscures the unifying noncommutative-geometric (NCG) structure: the
fundamental fermion multiplet is a single spinor acted on by a large Dirac
operator $D$, so the genuine fermion propagator is $D^{-1}$. In this section,
we will keep the \emph{internal matrix structure} intact through quantization:
derive the spectral action \emph{before} the internal trace, obtain a matrix
Lagrangian for matrix-valued fields, and from it extract \emph{propagators and
vertices in matrix form \cite{Chamseddine:2020khc}}. We work on a flat
Euclidean $\mathbb{R}^{4}$ (gravity off) and illustrated by a toy electroweak
model, where the method preserves NCG unification at the quantum level.

We start by specifying the spectral data, inner fluctuations, and the matrix
Lagrangian. For $M\times F$,
\begin{equation}
A=C^{\infty}(M)\otimes A_{F},\quad H=L^{2}(S)\otimes H_{F},\quad
D=i\gamma^{\mu}\partial_{\mu}\otimes1+\gamma_{5}\otimes D_{F},
\end{equation}
with real structure $J=C\otimes J_{F}$ and grading $\gamma=\gamma_{5}%
\otimes\gamma_{F}$. Inner fluctuations yield $D_{A}=i\gamma^{\mu}\partial
_{\mu}+A$, where the single matrix-valued field $A$ encodes \emph{both} gauge
bosons and scalars (Higgs) in NCG. Next we square $D_{A}$ by writing
\begin{align}
D_{A}^{2}  &  =-\big(\partial_{\mu}\partial_{\mu}+A_{\mu}\partial_{\mu
}+B\big),\qquad A_{\mu}\equiv-i\{\gamma_{\mu},A\},\quad B\equiv-\big(i\gamma
^{\mu}\partial_{\mu}A+A^{2}\big),\\
\omega_{\mu}  &  =\tfrac{1}{2}A_{\mu},\qquad E=B-\partial_{\mu}\omega_{\mu
}-\omega_{\mu}\omega_{\mu},\qquad\Omega_{\mu\nu}=\partial_{\mu}\omega_{\nu
}-\partial_{\nu}\omega_{\mu}+[\omega_{\mu},\omega_{\nu}].
\end{align}
Thus $D_{A}^{2}=-(\nabla_{\mu}\nabla_{\mu}+E)$ with $\nabla_{\mu}%
=\partial_{\mu}+\omega_{\mu}$. \ Next we perform the heat kernel expansion on
flat space and evaluate the first three terms,
\begin{align}
a_{0}  &  =\frac{1}{16\pi^{2}}\!\int\!d^{4}x\,\mathrm{Tr}(1),\quad a_{2}%
=\frac{1}{16\pi^{2}}\!\int\!d^{4}x\,\mathrm{Tr}(E),\\
a_{4}  &  =\frac{1}{16\pi^{2}}\frac{1}{12}\!\int\!d^{4}x\,\mathrm{Tr}%
\big(\Omega_{\mu\nu}\Omega^{\mu\nu}+6E^{2}\big).
\end{align}
With moments $f_{4},f_{2},f_{0}$ the truncated spectral action is
$I=2f_{4}a_{0}+2f_{2}a_{2}+f_{0}a_{4}$. Performing the \emph{Dirac} traces but
not the internal ones yields the matrix Lagrangian
\begin{align}
I=\frac{1}{32\pi^{2}}\!\int\!d^{4}x\ \Big[  &  2(2f_{4})\,\mathrm{Tr}%
(1)+(2f_{2})\,\mathrm{Tr}\,(2A^{2}+\gamma_{\mu}A\,\gamma^{\mu}A)\nonumber\\
&  -\frac{f_{0}}{6}\,\mathrm{Tr}\,\big(\gamma_{\mu}\partial_{\nu}%
A\,\gamma^{\mu}\partial^{\nu}A+2\gamma_{\mu}\partial^{\mu}A\,\gamma_{\nu
}\partial^{\nu}A\big)\nonumber\\
&  +i\frac{f_{0}}{3}\,\mathrm{Tr}\,\big(\partial_{\mu}A\,\gamma^{\mu}%
A\,\gamma_{\nu}A\,\gamma^{\nu}-\gamma_{\mu}\partial^{\mu}A\,\gamma_{\nu
}A\,\gamma^{\nu}A\big)\nonumber\\
&  +\frac{f_{0}}{24}\,\mathrm{Tr}\,\big(\gamma_{\mu}A\,\gamma_{\nu}%
A\,\gamma^{\mu}A\,\gamma^{\nu}A+2\,\gamma_{\mu}A\,\gamma^{\mu}A\,\gamma_{\nu
}A\,\gamma^{\nu}A\big)\Big], \label{eq:MatrixAction}%
\end{align}
up to a total divergence from the cubic piece. This is the starting point for
matrix-level Feynman rules.

We proceed by identifying the gauge symmetry, performing gauge fixing and
introduce ghosts, all in matrix-form. Gauge acts by $A\mapsto u^{\ast
}Au+u^{\ast}\delta u$ with $u^{\ast}u=1$ and $\delta u=i\gamma^{\nu}%
\partial_{\nu}u$. For a product geometry,
\begin{equation}
A=i\gamma^{\mu}B_{\mu}+\gamma_{5}\phi,
\end{equation}
so $\omega_{\mu}=B_{\mu}$ and $\Omega_{\mu\nu}$ depends only on $B_{\mu}$.
Choose the covariant gauge
\begin{equation}
G=\partial_{\mu}\omega_{\mu}=-\tfrac{i}{2}\{\gamma_{\mu},\partial^{\mu
}A\},\qquad\Delta I_{\mathrm{gf}}=-\frac{1}{16\pi^{2}}\frac{f_{0}}{3\xi}%
\!\int\!d^{4}x\,\mathrm{Tr}\,G^{2},
\end{equation}
with Feynman gauge $\xi=1$. The corresponding ghost action, from the variation
of $G$ under $u=1+i\alpha$, is
\begin{equation}
\Delta I_{\mathrm{gh}}=\int d^{4}x\ \mathrm{Tr}\Big(\partial_{\mu}%
c\,\partial^{\mu}\bar{c}-\tfrac{i}{2}\,\partial_{\mu}c\,\big[\{\gamma^{\mu
},A\},\bar{c}\big]\Big).
\end{equation}
To evaluate the propagators, we decompose internal matrices so that
\begin{equation}
A_{\alpha A}^{\ \beta B}=(i\gamma^{\mu})_{\alpha}^{\ \beta}\,B_{\mu\,A}%
^{\ B}+(\gamma_{5})_{\alpha}^{\ \beta}\,\phi_{\ A}^{\ B},
\end{equation}
with $B_{\mu}^{\dagger}=-B_{\mu}$, $\phi^{\dagger}=\phi$. Expand $B_{\mu
}=iB_{\mu}^{i}T^{i}$ and $\phi=\phi^{m}\lambda^{m}$ with $\mathrm{Tr}%
(T^{i}T^{j})=\mathrm{Tr}(\lambda^{m}\lambda^{n})=\delta^{ij}/2$. From the
quadratic part of \eqref{eq:MatrixAction} (use $\xi=1$):
\begin{align}
\langle B_{\mu}^{i}B_{\nu}^{j}\rangle &  =\frac{12\pi^{2}}{f_{0}}%
\,\frac{\delta^{ij}\delta_{\mu\nu}}{p^{2}},\qquad\langle\phi^{m}\phi
^{n}\rangle=\frac{8\pi^{2}}{f_{0}}\,\frac{\delta^{mn}}{p^{2}},\\
\langle c_{i}^{\ast}c_{j}\rangle &  =\frac{\delta_{ij}}{p^{2}},
\end{align}
and the compact matrix propagator for $A$ is obtained by recombining the
spinor and internal projectors. The quadratic \textquotedblleft
mass\textquotedblright\ term from $2f_{2}$ is treated as a 2-valent vertex
(mass insertions can be resummed).

Cubic and quartic vertices follow from the $f_{0}$ part of
\eqref{eq:MatrixAction}. Schematically (momenta incoming):
\begin{align}
V_{\beta B\delta D\tau F}^{\alpha A\gamma C\eta E}  &  =\ \frac{f_{0}}%
{96\pi^{2}}\,p_{\mu}\,\delta_{B}^{C}\delta_{D}^{E}\delta_{F}^{A}\left(
\gamma_{\nu}\right)  _{\delta}^{\eta}\left(  \left(  \gamma_{\nu}\right)
_{\beta}^{\gamma}\left(  \gamma_{\mu}\right)  _{\tau}^{\alpha}-\left(
\gamma_{\mu}\right)  _{\beta}^{\gamma}\left(  \gamma_{\nu}\right)  _{\tau
}^{\alpha}\right) \\
V_{\beta B\delta D\tau F\lambda H}^{\alpha A\gamma C\eta E\kappa G}  &
=\ \frac{f_{0}}{768\pi^{2}}\left(  \delta_{B}^{C}\delta_{D}^{E}\delta_{F}%
^{G}\delta_{H}^{A}\left(  \gamma_{\mu}\right)  _{\lambda}^{\alpha}\left(
\gamma_{\nu}\right)  _{\tau}^{\kappa}\left(  \left(  \gamma_{\nu}\right)
_{\beta}^{\gamma}\left(  \gamma_{\mu}\right)  _{\delta}^{\eta}+2\left(
\gamma_{\mu}\right)  _{\beta}^{\gamma}\left(  \gamma_{\nu}\right)  _{\delta
}^{\eta}\right)  \right)  ,
\end{align}
with matrix-index contractions (ribbon-graph structure). The ghost--$A$ vertex
comes from the commutator with $\{\gamma^{\mu},A\}$. We illustrate the
procedure using an Electroweak toy model. With fermions $\Psi=(\nu_{L}%
,e_{L},e_{R})^{T}$, we group the connection $A$ as
\begin{equation}
A=%
\begin{pmatrix}
i\gamma^{\mu}(B_{\mu})_{a}^{\ b} & \gamma_{5}\,H_{b}\\
\gamma_{5}\,H^{\ a} & i\gamma^{\mu}(B_{\mu})^{3}{}_{3}%
\end{pmatrix}
,\qquad a,b=1,2,\quad\mathrm{Tr}\,A=0.
\end{equation}
Identify $B_{\mu}^{p}=g\,W_{\mu}^{p}$ ($p=1,2,3$) and $(B_{\mu})^{0}%
=(g^{\prime}/\sqrt{3})\,B_{\mu}$; parameterize $H$ by a complex doublet.
Matching to canonical kinetic terms gives
\begin{equation}
\frac{f_{0}g^{2}}{12\pi^{2}}=1,\qquad g^{\prime2}=\frac{1}{3}g^{2}%
\ \Rightarrow\ \sin^{2}\theta_{W}=\frac{g^{\prime2}}{g^{2}+g^{\prime2}}%
=\frac{1}{4}.
\end{equation}
Higgs normalization implies $H\rightarrow\tfrac{g}{\sqrt{6}}H$; the
Yukawa-like coupling $\bar{\ell}\,\gamma_{5}He_{R}$ yields
\begin{equation}
m_{e}=\frac{g}{\sqrt{6}}\left\langle H\right\rangle ,
\end{equation}
and the quartic potential from $\mathrm{Tr}(HH^{\dagger})^{2}$ fixes
\begin{equation}
\lambda=\frac{g^{2}}{12}.
\end{equation}
The mass term from $f_{2}$ is consistent with these normalizations. These
relations reflect NCG unification: gauge and Higgs sectors come from the
single matrix field $A$.

We deduce the following conceptual outcomes and outlook:

\begin{itemize}
\item \textbf{Matrix-level Feynman rules:} Propagators and vertices are
obtained without tracing over internal indices, so perturbation theory
preserves the finite noncommutative geometry. Loop graphs are ribbon graphs
carrying both spinor and internal labels.

\item \textbf{Yang--Mills--like with }$\mathbf{\gamma}$\textbf{'s:} The
bosonic sector mirrors non-Abelian Yang--Mills (kinetic, cubic, quartic,
ghosts) plus a 2-point \textquotedblleft mass\textquotedblright\ vertex from
$f_{2}$, but with explicit $\gamma^{\mu},\gamma_{5}$ insertions because the
basic field is $A$.

\item \textbf{Renormalization:} The rules enable a matrix-preserving
renormalization analysis to test stability of NCG-induced coupling relations
under running.

\item \textbf{Extensions:} Full Standard Model or Pati--Salam and inclusion of
gravity follow conceptually; a BRST treatment in matrix language is a natural
next step.
\end{itemize}

We have thus shown that the spectral action is rewritten as a \emph{matrix
Lagrangian}, yielding complete matrix-valued propagators and vertices for the
unified field $A=i\gamma^{\mu}B_{\mu}+\gamma_{5}\phi$ and illustrating
high-scale coupling relations (e.g.\ $\sin^{2}\theta_{W}=1/4$, $\lambda
=g^{2}/12$, $m_{e}=(g/\sqrt{6})\langle H\rangle$). The framework keeps the NCG
structure visible at the quantum level and sets the stage for
matrix-preserving renormalization.

\section{Conclusion}

Over three decades, the program sketched here has turned \emph{noncommutative
geometry} (NCG) from a mathematical curiosity into a concrete framework that
unifies gravity and the Standard Model within a single spectral language. The
central idea---that physical information is encoded in the spectrum of a
generalized Dirac operator---has proven unexpectedly fertile: inner
fluctuations generate the full gauge--Higgs sector, while the spectral action
supplies the bosonic dynamics with fixed, geometrically determined
normalizations. In this sense, the Standard Model is not \emph{assumed} but
\emph{reconstructed}, with its group, representations, and Higgs mechanism
emerging naturally from the finite, \textquotedblleft
internal\textquotedblright\ part of the geometry.

Two methodological milestones underpin this progress. First, the
\emph{spectral action principle} elevates the Dirac spectrum to the status of
a bare, high-scale action; its heat-kernel expansion yields the cosmological
constant, Einstein--Hilbert term, higher-curvature invariants, and canonical
gauge/Higgs terms with spectral (hence geometric) boundary conditions on
couplings. Second, the \emph{classification of admissible finite geometries}
under the order-one and reality axioms singles out the almost-commutative
algebra that reproduces the observed gauge group and fermion content, fixing
hypercharges and predicting the necessity of right-handed neutrinos and the
see-saw mechanism. Together, these results explain why the Standard Model sits
exactly where it does in theory space---neither an arbitrary choice nor an
accident of nature.

As the phenomenological net narrowed, including right-handed neutrinos and
imposing KO-dimension 6 became essential. This move reconciles chirality with
reality (Majorana) conditions at the finite level, determines the unimodular
gauge group $U(1)\times SU(2)\times SU(3)$, and organizes the Yukawa/Majorana
data in natural moduli spaces. It also ties neutrino masses directly to the
geometry through a singlet sector whose vacuum expectation value seeds the
see-saw, linking neutrino physics to Higgs stability via scalar couplings
fixed at the unification scale. These ingredients repair earlier tensions
(e.g., fits to the Higgs mass and vacuum stability) not by adding \emph{ad
hoc} fields, but by restoring degrees of freedom already demanded by the
spectral setup.

Conceptually, NCG also clarifies gravity's place: in almost-commutative spaces
the Einstein sector arrives side-by-side with matter, and dilaton extensions
make classical scale invariance of the matter/Higgs sector manifest in the
Einstein frame, with the correct conformal couplings. Technically, the
renormalization-group viewpoint treats the spectral action as a boundary
condition at a unification scale, explaining why strict low-energy unification
is not expected while preserving robust high-scale relations among gauge,
Yukawa, and quartic couplings. The remaining discrepancies with precision data
should be viewed as guidance---pointers to thresholds or states already latent
in the geometry, rather than failures of the principle. The paper shows that a
higher-degree analogue of the Heisenberg commutation relation---linking the
Dirac operator to the Feynman slash of scalar fields---provides a geometric
route from quantum theory to classical spacetime and known particle
physics.\ In its \emph{two-sided} form (in 4D) it selects precisely the
internal algebras $M_{2}(\mathbb{H})$ and $M_{4}(\mathbb{C})$, the same
building blocks that underpin the spectral reconstruction of the Standard
Model coupled to gravity via the spectral action. Thus, the commutation
principle does not merely accommodate the Standard Model; together with the
noncommutative algebra of functions it \emph{reconstructs} its Lagrangian
alongside Einstein gravity in a unified spectral framework. Mathematically we
have proved that for any connected Riemannian spin four-manifold with
quantized volume (in Planck sphere units), there exists an \emph{irreducible
representation} of the two-sided relations; where allowed volumes form an
integer set $q(M)$ (containing all integers $m\geq5$ for spin four-manifolds),
ensuring that the scheme \emph{covers all relevant geometries} rather than a
special class. These results anchor the formalism and provide a
nonperturbative foothold for the idea that geometry comes in discrete quanta
compatible with smooth manifolds at large scale. Conceptually, this leads to a
suggestive \textquotedblleft particle picture\textquotedblright\ for quantum
gravity: once the two-sided relations are imposed, one can view irreducible
representations as the natural analogues of particles (in the Euclidean
setting), now governed not by the Poincar\'{e} group but by algebraic
relations tying the line element to the slashed scalar fields. In 4D, the
coincidence $C_{+}=M_{2}(\mathbb{H})$, $C_{-}=M_{4}(\mathbb{C})$ reinforces
the bridge to the noncommutative geometry of the Standard Model and points
toward a kinematics where matter and geometry share a common spectral origin.
The present results articulate a clear thesis: \emph{quantum geometry can be
encoded by higher-degree commutation relations whose spectral content recovers
Einstein gravity, the Standard Model, and a discretized notion of volume
within a single framework.}

Recent work on a \emph{matrix form of the spectral action} pushes the program
toward quantization without erasing its noncommutative character: propagators
and vertices are derived before tracing over internal algebras, yielding
ribbon-graph Feynman rules akin to Yang--Mills but with $\gamma$-matrix
structure in the bosonic sector. This keeps unification visible at the quantum
level and opens a path to a renormalizability analysis---BRST, counterterms,
and all---performed in a way that respects the finite geometry. The picture
that emerges is of a single matrix field encoding gauge and Higgs dynamics,
hinting at deeper constraints on running couplings when the full matrix
structure is retained.

\medskip\noindent\textbf{Where next?} Three directions stand out.

\begin{enumerate}
\item \textbf{Precision:} two-loop running with realistic thresholds
(including see-saw scales) to sharpen the link between spectral boundary data
and low-energy observables.

\item \textbf{Dynamics:} a full matrix-preserving renormalization of the
spectral action, including BRST quantization and the role of parity-odd
(topological) terms, to test the stability of high-scale relations.

\item \textbf{Cosmology:} dilaton/Higgs dynamics, inflationary embeddings, and
dark-sector couplings suggested by the finite geometry and extended scalar structure.
\end{enumerate}

Each of these is not an external add-on but a continuation of the same
geometric logic.

If a single lesson runs through this review, it is that \emph{geometry---when
made spectral and allowed to be noncommutative---does more with less}. It
compresses disparate ingredients (gauge, Higgs, gravity, flavor, neutrino
masses) into spectral data, trades arbitrary choices for axioms, and converts
model-building into geometry-building. The resulting framework is neither
complete nor closed; rather, it is precise enough to be tested, flexible
enough to evolve, and principled enough to guide that evolution. In this
sense, the road ahead is clear: deepen the spectral foundations, quantize
without sacrificing structure, and follow the geometry where it leads.
\vspace{1cm}

\subsubsection*{Acknowledgements}

I\ would like to thank my collaborators that accompanied me through my journey
through the noncommutative geometry world. The journey started with J\"{u}rg
Fr\"{o}hlich where we learned and mastered the basics together. It was my good
fortune that I\ met Alain Connes at IHES in 1996 and our long and fantastic
collaboration started and led us to achieve beautiful and amazing results.
Since then we were joined with other collaborators who helped us in gaining so
much insight. I name in partiuclar, and in chronological order, Matilda
Marcolli, Walter van Suijlekom and Slava Mukhanov. I would like to acknowledge
the generous support, over the years, of the National Science Foundation
(NSF), the Humboldt Foundation and IHES (through the Louis Michel Chair). The
present work is supported by NSF grant Phys-2207663.

\subsubsection*{Remark}

To avoid listing a huge number of references, I\ have cited in the
bibiliography below, mostly references to the original papers I\ used in the
review, and it is to be understood that these citations implicitly include all
references appearing in those papers.


\begin{thebibliography}{99}                                                                                               %


\bibitem {C85}A.~Connes, \textit{Noncommutative differential geometry}, Publ.
Math. IHES \textbf{39} (1985) 257--360.

\bibitem {C94}A.~Connes, \textit{Noncommutative Geometry}, Academic Press, San
Diego, 1994.

\bibitem {C90}A.~Connes. Essay on physics and noncommutative geometry. In
\emph{The interface of mathematics and particle physics ({O}xford, 1988)},
volume~24 of \emph{Inst. Math. Appl. Conf. Ser. New Ser.}, pages 9--48. Oxford
Univ. Press, New York, 1990.

\bibitem {CL91}A.~Connes and J.~Lott. \textit{Particle models and
noncommutative geometry}, Nucl. Phys. Proc. Suppl. \textbf{18B} (1991) 29--47.

\bibitem {CFF92}A.~H. Chamseddine, G.~Felder, and J.~Fr\"{o}hlich,
\textit{Unified gauge theories in noncommutative geometry}, Phys. Lett. B,
\textbf{296} (1992) 109-116.

\bibitem {CFF2}A.~H. Chamseddine, G.~Felder, and J.~Fr\"{o}hlich,
\textit{Grand unification in noncommutative geometry, }Nucl. Phys. B
\textbf{395} (1993) 672-700.

\bibitem {CFF93}A.~H. Chamseddine, G.~Felder, and J.~Fr\"{o}hlich,
\textit{Gravity in noncommutative geometry,} 155 Comm. Math. Phys. \textbf{155
}(1993) 205--217.

\bibitem {Chamseddine:1994tw}A.~H.~Chamseddine and J.~Fr\"{o}hlich,
\textit{The Chern-Simons action in noncommutative geometry,} J. Math. Phys.
\textbf{35} (1994), 5195-5218.

\bibitem {Chamseddine:1993is}A.~H.~Chamseddine and J.~Fr\"{o}hlich,
\textit{SO(10) unification in noncommutative geometry,}\textquotedblright%
\ Phys. Rev. D \textbf{50} (1994), 2893-2907.

\bibitem {CFG95}A.~H. Chamseddine, J.~Fr\"{o}hlich, and O.~Grandjean,
\textit{The gravitational sector in the Connes-Lott formulation of the
standard model}, J. Math. Phys. \textbf{36} (1995) 6255--6275.

\bibitem {Kas93}D.~Kastler, \textit{A detailed account of Alain Connes'
version of the standard model in noncommutative geometry. I, II.} Rev. Math.
Phys. \textbf{5 }(1993) 477--532.

\bibitem {Kas96}D.~Kastler, \textit{A detailed account of Alain Connes'
version of the standard model in non-commutative differential geometry. III.}
Rev. Math. Phys \textbf{8 }(1996) 103--165.

\bibitem {KasS96}D.~Kastler and T.~Sch\"{u}cker, \textit{A detailed account of
Alain Connes' version of the standard model. IV.} Rev. Math. Phys. \textbf{8}
(1996) 205--228.

\bibitem {KasS97}D.~Kastler and T.~Sch\"{u}cker, \textit{The standard model
\`{a} la Connes-Lott, }J. Geom. Phys. \textbf{24} (1997) 1--19.

\bibitem {CC96}A.~H. Chamseddine and A.~Connes, \textit{Universal formula for
noncommutative geometry actions:} \textit{Unifications of gravity and the
Standard Model, }Phys. Rev. Lett. \textbf{77 }(1996) 4868--4871.

\bibitem {Chamseddine:1996zu}A.~H.~Chamseddine and A.~Connes, T\textit{he
Spectral action principle, }Commun. Math. Phys. \textbf{186} (1997), 731-750.

\bibitem {Chamseddine:2005zk}A.~H.~Chamseddine and A.~Connes, \textit{Scale
invariance in the spectral action, }J. Math. Phys. \textbf{47} (2006), 063504.

\bibitem {co2006}A. Connes, \textit{Noncommutative geometry and the standard
model with neutrino mixing, }JHEP \textbf{11 }(2006) 081.

\bibitem {CCM07}A.~H. Chamseddine, A.~Connes, and M.~Marcolli, \textit{Gravity
and the Standard Model with neutrino mixing}, Adv. Theor. Math. Phys.
\textbf{11} (2007) 991--1089.

\bibitem {C95}A.~Connes, \textit{Noncommutative geometry and reality}, J.
Math. Phys. \textbf{36} (11) (1995) 6194--6231.

\bibitem {C06}A.~Connes, \textit{On the foundations of noncommutative
geometry}, In \emph{The unity of mathematics}, volume 244 of \emph{Progr.
Math.}, pages 173--204. Birkh\"{a}user Boston, Boston, MA, 2006.

\bibitem {C00}A.~Connes. \textit{Noncommutative geometry year 2000} In
\emph{Highlights of mathematical physics ({L}ondon, 2000)}, pages 49--110.
Amer. Math. Soc., Providence, RI, 2002.

\bibitem {Connes:2006qj}A.~Connes and A.~H.~Chamseddine, \textit{Inner
fluctuations of the spectral action,} J. Geom. Phys. \textbf{57} (2006), 1-21.

\bibitem {CC07b}A.~H. Chamseddine and A.~Connes, \textit{Why the Standard
Model,} J. Geom. Phys. \textbf{58} (2008) 38--47.

\bibitem {Chamseddine:2007ia}A.~H.~Chamseddine and A.~Connes,
\textit{Conceptual Explanation for the Algebra in the Noncommutative Approach
to the Standard Model,} Phys. Rev. Lett. \textbf{99} (2007), 191601.

\bibitem {Chamseddine:2007bm}A.~H.~Chamseddine and A.~Connes, \textit{Quantum
Gravity Boundary Terms from Spectral Action}, Phys. Rev. Lett. \textbf{99}
(2007), 071302.

\bibitem {Chamseddine:2010ia}A.~H.~Chamseddine and A.~Connes,
\textit{Noncommutative Geometric Spaces with Boundary: Spectral Action}, J.
Geom. Phys. \textbf{61} (2011), 317-332.

\bibitem {Chamseddine:2009drp}A.~H.~Chamseddine and A.~Connes, \textit{The
Uncanny Precision of the Spectral Action,} Commun. Math. Phys. \textbf{293}
(2010), 867-897.

\bibitem {CC10}A.~H. Chamseddine and A.~Connes, \textit{Noncommutative
geometry as a framework for unification of all fundamental interactions
including gravity. Part I}, Fortsch. Phys. \textbf{58} (2010) 553--600.

\bibitem {CC12}A.~H. Chamseddine and A.~Connes, \textit{Resilience of the
Spectral Standard Model}, JHEP \textbf{1209} (2012) 104.

\bibitem {Chamseddine:2010ps}A.~H.~Chamseddine and A.~Connes,
\textit{Space-Time from the spectral point of view,}\textquotedblright\ in
\emph{Proceedings of the Twelfth Marcel Grossmann Meeting on General
Relativity, }pages 3-23, (2012) World Scientific.

\bibitem {Chamseddine:2011ix}A.~H.~Chamseddine and A.~Connes, \textit{Spectral
Action for Robertson-Walker metrics,} JHEP \textbf{10} (2012), 101.

\bibitem {CCSW}A.~H.~Chamseddine, A.~Connes and W.~D.~van Suijlekom,
\textit{Inner Fluctuations in Noncommutative Geometry without the first order
condition,} J. Geom. Phys. \textbf{73} (2013), 222-234.

\bibitem {CCS13b}A.~H. Chamseddine, A.~Connes, and W.~D. van Suijlekom,
\textit{Beyond the spectral Standard Model: Emergence of Pati-Salam
unification}, JHEP, \textbf{1311} (2013) 132.

\bibitem {CCS15}A.~H. Chamseddine, A.~Connes, and W.~D. van Suijlekom,
\textit{Grand unification in the spectral Pati-Salam model}, JHEP \textbf{11}
(2015) 011.

\bibitem {CCM14}A.~H. Chamseddine, A.~Connes, and V.~Mukhanov,
\textit{Geometry and the quantum: Basics}, JHEP \textbf{1412} (2014) 098.

\bibitem {CCM15}A.~H. Chamseddine, A.~Connes, and V.~Mukhanov, \textit{Quanta
of geometry: Noncommutative aspects}, Phys. Rev. Lett. \textbf{114} (2015) 091302.

\bibitem {Chamseddine:2013kea}A.~H.~Chamseddine and V.~Mukhanov,
\textit{Mimetic Dark Matter,} JHEP \textbf{11} (2013), 135.

\bibitem {CCS18}A.~H. Chamseddine, A.~Connes, and W.~D. van Suijlekom,
\textit{Entropy and the spectral action}, Commun. Math. Phys. \textbf{373} 2
(2019) 457-471.

\bibitem {Chamseddine:2020khc}A.~H.~Chamseddine, J.~Iliopoulos and W.~D.~van
Suijlekom, \textit{Spectral action in matrix form,} Eur. Phys. J. C
\textbf{80} (2020) no.11, 1045.

\bibitem {Chamseddine:2022rnn}A.~H.~Chamseddine, A.~Connes and W.~D.~van
Suijlekom, \textit{Noncommutativity and physics: a non-technical review,} Eur.
Phys. J. ST \textbf{232} (2023) no.23-24, 3581-3588.
\end{thebibliography}
\end{document}